\documentclass[lettersize,journal]{IEEEtran}
\usepackage{amsmath,amsfonts}
\usepackage{algorithmic}
\usepackage{algorithm}
\usepackage{array}
\usepackage[caption=false,font=normalsize,labelfont=sf,textfont=sf]{subfig}
\usepackage{textcomp}
\usepackage{stfloats}
\usepackage{xurl}
\usepackage{verbatim}
\usepackage{graphicx}

\usepackage{cite}
\usepackage{svg}

\usepackage{adjustbox}
\usepackage{tikz}
\usepackage{pgfplots}
\usepackage{amsmath}
\usepackage{adjustbox}
\usetikzlibrary{shapes, patterns, positioning, backgrounds}
\usepackage{import}

\usepackage[table]{xcolor}
\usepackage{tabularx}
\usepackage{makecell} 
\definecolor{rowgray}{HTML}{E4E4E4}
\newcolumntype{Y}{>{\raggedright\arraybackslash}X} 

\pgfplotsset{compat=1.17}

\usepackage{tikz}

\usepackage{xcolor}
\usepackage{color}

\usepackage[numbers]{natbib}
\usepackage{xstring}

\begin{document}

\title{
Application-Specific Power Side-Channel Attacks and Countermeasures: A Survey
}

\author{Sahan~Sanjaya~\IEEEmembership{Student Member,~IEEE,},
        Aruna~Jayasena,~\IEEEmembership{Member,~IEEE,}
        and~Prabhat~Mishra,~\IEEEmembership{Fellow,~IEEE,}\vspace{-0.1in}

\IEEEcompsocitemizethanks{%
\IEEEcompsocthanksitem S. Sanjaya and P. Mishra are with the Department of Computer \& Information Science \& Engineering, University of Florida, USA.%
\ Email: \{ssanjaya, prabhat\}@ufl.edu. A. Jayasena is with the Department of Computer Science and Engineering, University of Tennessee, Chattanooga, USA.%
\ Email: aruna@tennessee.edu
}}


\markboth{Proceedings of the IEEE,~Vol.~XXX, No.~YY, ZZZZ~2024}%
{Shell \MakeLowercase{\textit{et al.}}: A Sample Article Using IEEEtran.cls for IEEE Journals}

\maketitle


\begin{abstract}


Side-channel attacks try to extract secret information from a system by analyzing different side-channel signatures, such as power consumption, electromagnetic emanation, thermal dissipation, acoustics, time, etc. Power-based side-channel attack is one of the most prominent side-channel attacks in cybersecurity, which rely on data-dependent power variations in a system to extract sensitive information. 
While there are related surveys, they primarily focus on power side-channel attacks on cryptographic implementations. In recent years, power-side channel attacks have been explored in diverse application domains, including key extraction from cryptographic implementations, reverse engineering of machine learning models, user behavior data exploitation, and instruction-level disassembly. In this paper, we provide a comprehensive survey of power side-channel attacks and their countermeasures in different application domains. Specifically, this survey aims to classify recent power side-channel attacks and provide a comprehensive comparison based on application-specific considerations.

\end{abstract}

\begin{IEEEkeywords}
Power Side-Channel, Cryptographic Implementations, Machine Learning, Disassembler, User Behavior Data, Survey, Side-Channel Attacks, Hardware Security, Countermeasures
\end{IEEEkeywords}

\section{Introduction}
\label{sec:introduction}

With the rapid advancement of technology, electronic devices have become an integral part of everyday life. From simple personal devices, like smartphones and wearables, to complex systems in critical infrastructure, these devices process and store a significant amount of sensitive information. As a result, the security of electronic devices is of critical importance, specially in the face of an increasingly sophisticated threat landscape. While traditional cryptographic measures are designed to protect data during transmission and storage, they often overlook the potential risks posed by physical side channels, which can leak information unintentionally through observable side-channels, such as timing, electromagnetic emanation, power consumption, acoustics, etc.

\begin{figure}[htp]
    \begin{center}
        \includegraphics[width=0.5\textwidth]{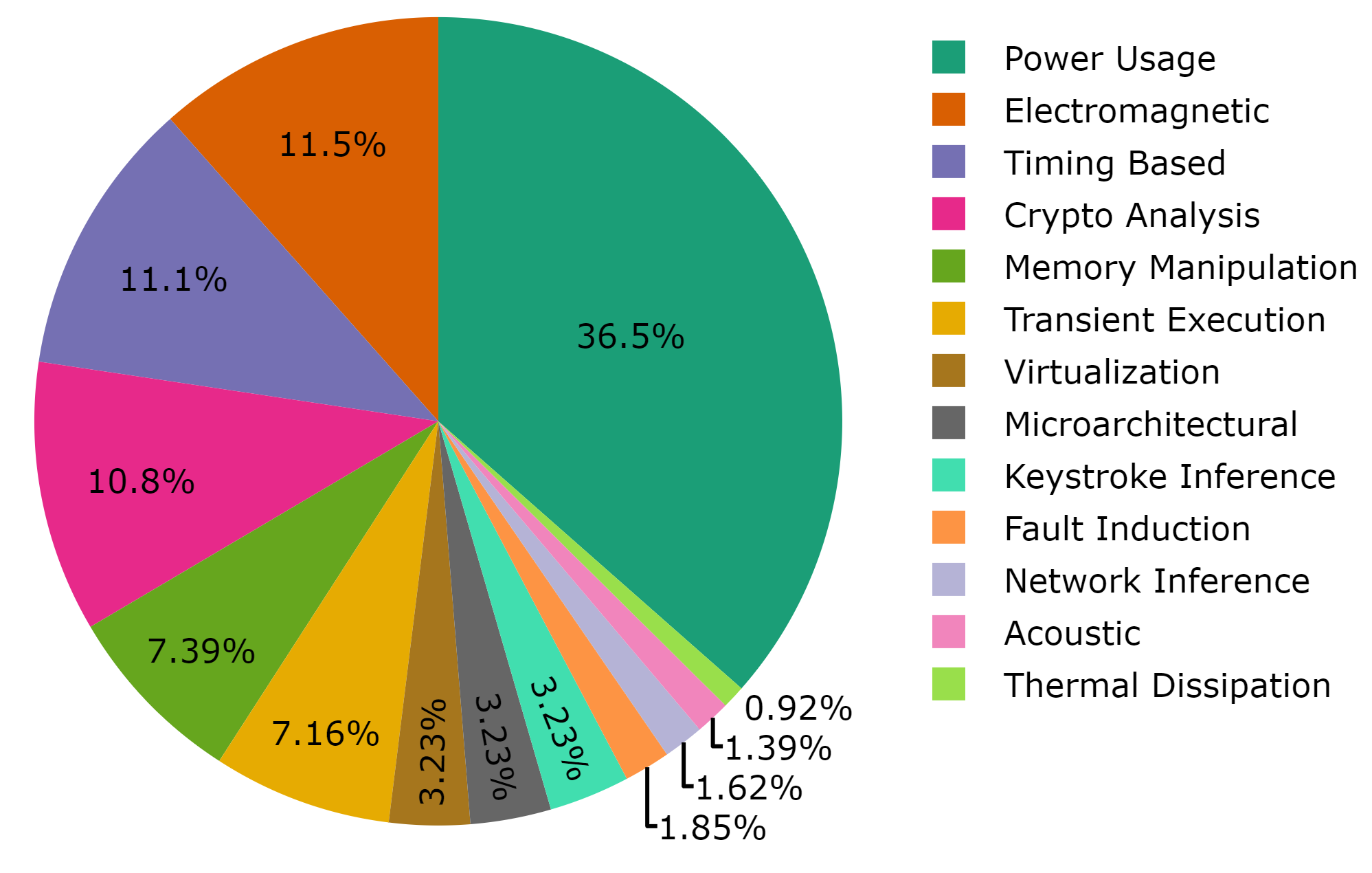}
    \end{center}
      \vspace{-0.2in}
      \caption{Most common side-channel attack types based on published results~\cite{Johnson2021Side}. Power side-channel represents the most researched attack in the literature.}
 
      \label{fig:sca_types}
\end{figure}

\begin{figure*}[htp]
    \begin{center}
        \includegraphics[width=0.75\textwidth]{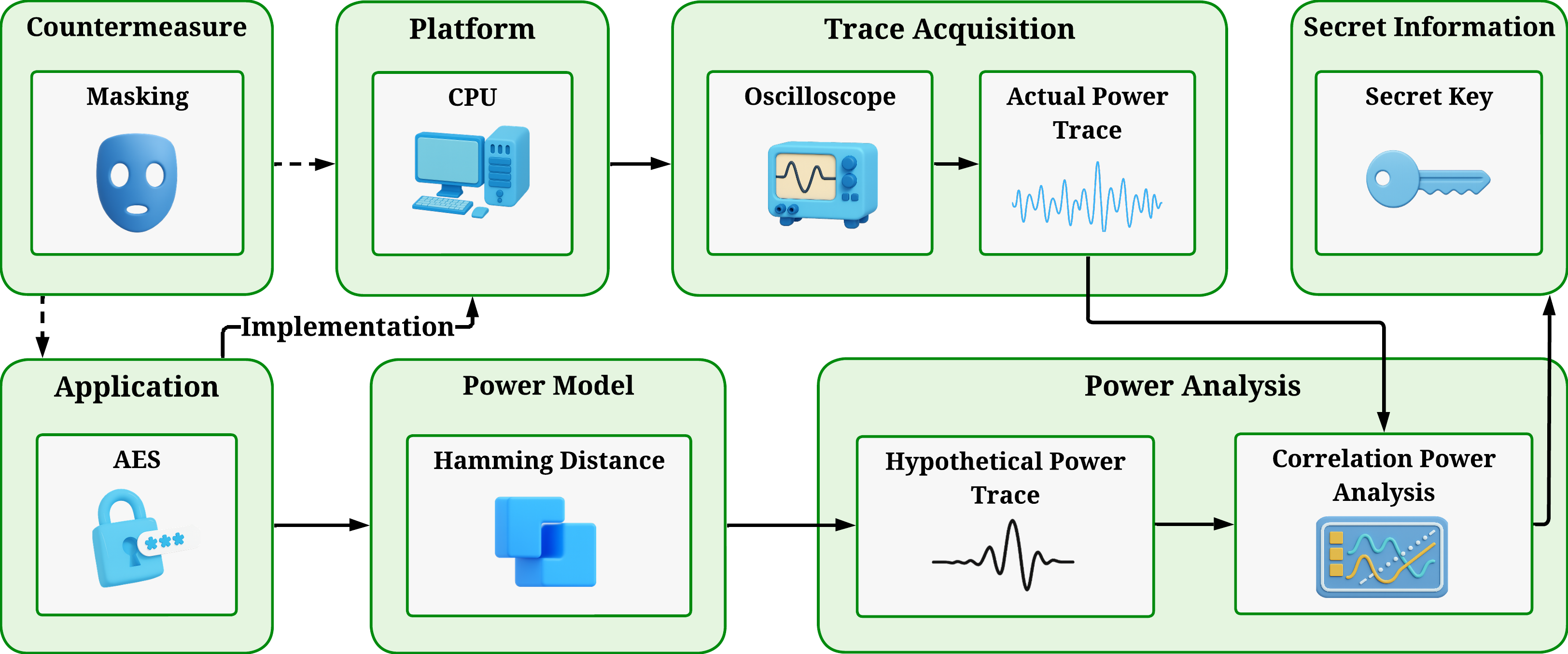}
    \end{center}
      \vspace{-0.2in}
      \caption{Correlation power analysis (CPA) attack flow, illustrating the targeted cryptographic algorithm (AES), the implementation platform (CPU) for power trace collection, the trace acquisition method (oscilloscope), the power model (Hamming distance) used to develop hypothetical traces, and the statistical analysis method (CPA) employed to correlate actual and hypothetical traces to extract the secret key.
      Additionally, the figure highlights the use of countermeasures to mitigate future power side-channel attacks.}
      \label{fig:attack_flow}
\end{figure*}

Power side-channel (PSC) attacks have gained significant attention due to their ability to exploit power consumption patterns to retrieve sensitive information from different application implementations. As illustrated by Figure~\ref{fig:sca_types}, PSC attack represents the most researched side-channel attack type between 2015 and 2020, accounting for 36.5\% of the published work. This highlights the prominent interest among researchers in exploiting vulnerabilities related to power consumption in electronic devices. Since the introduction of Differential Power Analysis (DPA) by  \citeauthor{Kocher1999Differential}
\cite{Kocher1999Differential}, PSC attacks have become a major concern, particularly with the expanding use of cryptographic algorithms in consumer devices. Figure~\ref{fig:attack_flow} illustrates a specific attack scenario utilizing correlation power analysis (CPA) on AES cryptographic implementation running on a CPU-based platform. The attack flow involves the collection of power traces, statistical analysis, and correlation with hypothesized power traces to successfully recover the secret key. Also, designers integrate countermeasure methods into the both platform and application to prevent future attacks.  

\begin{figure*}[htp]
\centering
\includegraphics[width=0.9\textwidth]{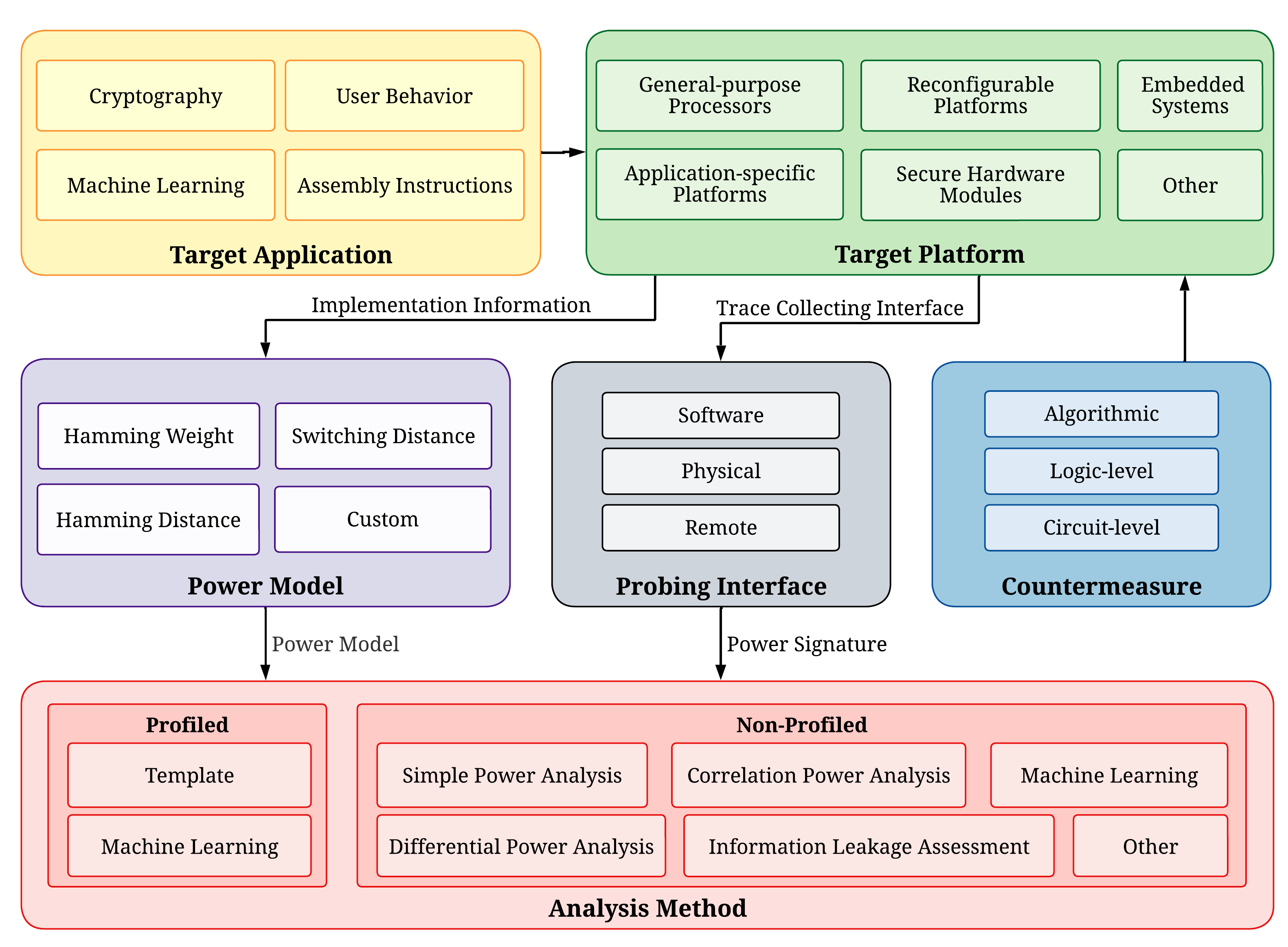}
\caption{Overview of the proposed survey. It covers power side-channel attacks for diverse applications (secret key extraction from cryptographic implementations, extraction of inputs and architectures from machine learning models, derivation of user-specific behavior, and identification of assembly instructions from power traces), trace collecting interfaces (physical, remote and software), power models (Hamming weight. Hamming distance, switching distance, and custom), analysis methods (profiled and non-profiled), and target platforms. It also considers countermeasures at different abstraction levels (algorithm, logic, circuits).}
\label{fig:overview}
\end{figure*}

Figure~\ref{fig:overview} shows that PSC attacks have been demonstrated on a wide range of applications implemented on diverse platforms. Specifically, recent works demonstrate that PSC attacks are successful in extracting a wide variety of information, including secret keys from cryptographic implementations, confidential inputs used in machine-learning models, proprietary machine-learning model architectures, user behavior data, and proprietary instruction-level code. In this paper, we present a survey on application-specific PSC attacks and their countermeasures. 


\subsection{Major Differences with Existing Surveys}
\label{sec:existing}

There are several promising surveys related to PSC attacks~\cite{Standaert2006Overview, Zhang2016Power, Yan2015A, Liptak2022Power, Hasnain2022Power, Taouil2021Power, Randolph2020Power, Hasnain2023Role, Martinez2021SoK, Socha2022Comprehensive, Ravi2024Side, Devi2021Side, Mirzargar2019Physical, Kim2020Gpu, Spreitzer2017Systematic, Shanmugham2018Survey, Ghimire2023Power, Hettwer2020Applications, Picek2023Sok, Panoff2022Review, Maheswari2024Profiling, Chabanne2021Side, Mendez2021Physical, Han2019Side}. However, these surveys are either outdated or limited in scope. For example, there are several surveys that do not consider research efforts from the last decade~\cite{Standaert2006Overview, Zhang2016Power, Yan2015A}. Most of the recent surveys focus exclusively on PSC attacks targeting cryptographic implementations~\cite{Liptak2022Power, Hasnain2022Power, Taouil2021Power, Randolph2020Power, Zhang2016Power, Hasnain2023Role, Martinez2021SoK, Socha2022Comprehensive, Ravi2024Side}. While there are recent surveys that cover side-channel attacks for extraction of neural network architecture~\cite{Chabanne2021Side, Mendez2021Physical} as well as code execution monitoring~\cite{Han2019Side}, they are not focused on power side-channel attacks.

There are also surveys focused on the target implementation platform, including Internet of Things (IoT) devices~\cite{Liptak2022Power, Devi2021Side}, field-programmable gate arrays (FPGA)\cite{Hasnain2022Power, Mirzargar2019Physical}, graphical processing units (GPU)\cite{Kim2020Gpu}, mobile devices~\cite{Spreitzer2017Systematic}, and intelligent sensors~\cite{Shanmugham2018Survey}. There are recent surveys exclusively focused on power analysis techniques, such as machine learning~\cite{Ghimire2023Power, Hasnain2023Role, Hettwer2020Applications}, deep learning~\cite{Picek2023Sok, Panoff2022Review, Maheswari2024Profiling}, and profiled methods~\cite{Hasnain2022Power}. However, these surveys are focused on either hardware platforms or power analysis methods. As a result, they do not capture the interactions between a wide variety of applications, diverse trace collection interfaces, different power models, various hardware platforms, and diverse analysis methods.
\textit{To the best of our knowledge, this survey is the first attempt at a comprehensive and application-oriented review of power side-channel attacks that integrates multiple dimensions into a unified taxonomy, including application types, target platform, trace collection interface, power models, analysis methods, and countermeasures strategies.}
Specifically, this survey makes the following key contributions:
\begin{itemize}
    \item \textbf{Multi-dimensional taxonomy:} We propose a novel six-axis taxonomy that classifies power side-channel (PSC) research based on (i)~target application, (ii)~implementation platform, (iii)~probing interface, (iv)~power model, (v)~analysis method, and (vi)~countermeasure granularity. This taxonomy provides a structured view of the research landscape that unifies previously fragmented literature.
    \item \textbf{Cross-domain coverage:} Unlike prior surveys that focus on a single application domain (e.g., cryptography), we systematically analyze PSC attacks across a wide variety of applications, including machine learning model extraction, user-behavior monitoring, and instruction disassembly, offering the most comprehensive review across various target applications.
    \item \textbf{Platform-specific insights:} We categorize attacks across a diverse range of targeted platforms, including general-purpose processors (e.g., CPUs), reconfigurable hardware (e.g., FPGAs, soft cores), application-specific hardware (e.g., ASICs, GPUs, TPUs), embedded systems (e.g., MCUs, IoT devices), secure hardware platforms (e.g., SGX), and other peripherals (e.g., USB controllers, printers). Our classification highlights platform-specific leakage sources and the unique challenges associated with each platform.
    \item \textbf{Discussion on power trace collection interfaces:} This category encompasses both conventional physical interfaces used in local or remote hardware-based attacks and emerging software-based interfaces such as Intel's RAPL that exploit built-in performance monitoring frameworks. These interfaces facilitate the acquisition of power traces from the targeted platforms.
    \item \textbf{Comparative analysis of power models and analysis methods:} We examine how different power consumption models (e.g., Hamming weight, switching distance) and analysis methods (e.g., template attacks, correlation power analysis) are tailored to specific platforms and applications.
    \item \textbf{Countermeasure classification and evaluation:} We classify countermeasures based on their abstraction levels, such as algorithmic, logic, and circuit, and evaluate their applicability across domains.
    \item \textbf{Literature coverage mapping:} We visualize the density of prior research across the taxonomy dimensions to expose underexplored areas and identify promising directions for future work.
\end{itemize}
\noindent By integrating these contributions into a single coherent framework, this survey not only bridges gaps in existing literature but also serves as a practical reference for researchers and practitioners seeking to understand, evaluate, or develop PSC attacks and defenses across multiple domains.

\subsection{Survey Outline}
\label{sec:methodology}

Figure~\ref{fig:overview} summarizes six dimensions that jointly capture the design space of PSC research. Specifically, this survey seeks to answer the following three fundamental questions:
\begin{itemize}
\item How are power traces collected from various \textit{target applications} implemented on different \textit{target platforms} using diverse \textit{probing interfaces}? This question addresses three key categorization axes: \textbf{Target Application}, \textbf{Target Platform}, and \textbf{Probing Interface}.
\item What are the side-channel  \textit{analysis methods} used to extract sensitive information from the target application using the collected power traces and specific \textit{power models}, which are derived based on the characteristics of the application and platform? This question corresponds to the \textbf{Power Model} and \textbf{Analysis Method} axes.
\item How can we defend against these attacks using \textit{countermeasures}, addressing the final categorization axis: \textbf{Countermeasure}.
\end{itemize}

Each axis reflects a distinct property relevant to power side-channel vulnerability and analysis:

\begin{itemize}
    \item \textbf{Target Application} captures the semantic intent of the computation being attacked, ranging from cryptographic key extraction to neural network inference or instruction monitoring.
    
    \item \textbf{Target Platform} denotes the execution environment, such as general-purpose processors (e.g., CPUs), reconfigurable hardware (e.g., FPGAs, soft cores), application-specific hardware (e.g., ASICs, GPUs, TPUs), embedded systems (e.g., MCUs, IoT devices), secure hardware platforms (e.g., SGX), and other peripherals (e.g., USB controllers, printers). Each platform reflects distinct leakage characteristics and constraints on observability.

    \item \textbf{Probing Interface} captures the adversary's access model, whether leakage is obtained via physical probing (e.g., oscilloscopes), software-based (e.g., utilizing software interface to observe hardware performance counters), or remote channels that convert power variations into indirectly observable metrics (e.g., timing differences). The choice of interface significantly influences the attacker’s capabilities and the required setup complexity.

    \item \textbf{Power Model} defines the abstraction that relates internal transitions to observable power, such as Hamming weight or switching activity.
    
    \item \textbf{Analysis Method} categorizes the adversarial effort to interpret the trace, distinguishing profiled and non-profiled analysis methods.
    
    \item \textbf{Countermeasure} captures the abstraction level at which defenses are applied, such as algorithmic (e.g., masking), logic-level (e.g., dual-rail encoding), or circuit-level (e.g., noise injection).
    
\end{itemize}

\begin{figure}[htp]
    \begin{center}
        \small
        \includegraphics[width=0.48\textwidth]{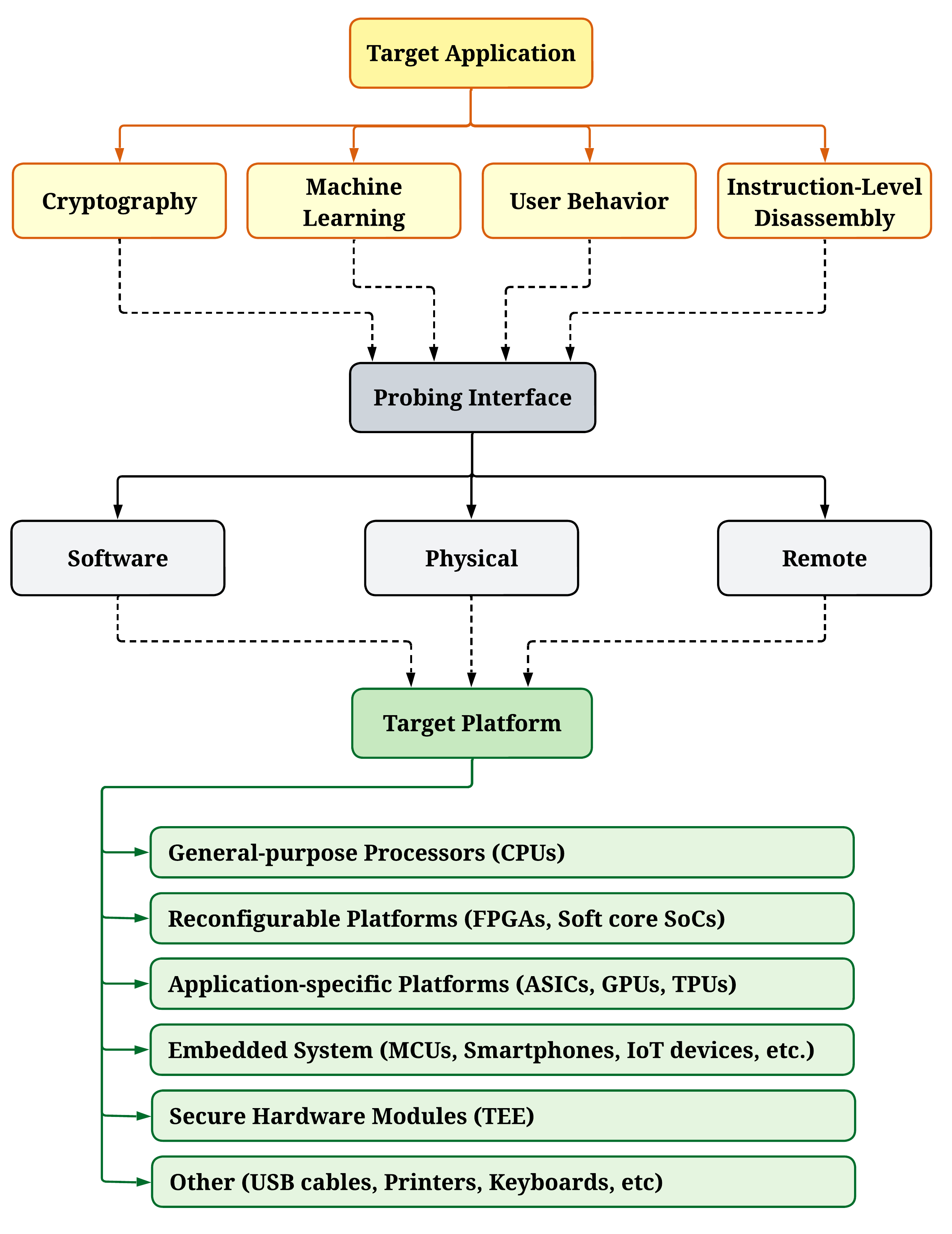}
    \end{center}
    \vspace{-0.1in}
      \caption{Overview of application-specific power side-channel analysis attacks that cover various application domains (e.g., cryptography, machine learning, user behavior, and assembly code) as well as diverse probing interfaces on different devices under attack}
      \label{fig:target_overview}
\end{figure}

The remainder of this section details each axis and explains how the subsequent sections of the survey map onto the taxonomy as follows. Section~\ref{sec:models} surveys various leakage models, including Hamming weight, Hamming distance, switching distance, and custom models. 
Section~\ref{sec:analysis} surveys both profiled and non-profiled PSC analysis methods. From Section~\ref{sec:crypto} onwards, we survey various side-channel attacks based on the target applications, including cryptographic implementations (Section~\ref{sec:crypto}), user behavior data (Section~\ref{sec:user}), assembly instructions (Section~\ref{sec:instruction}), and machine learning models (Section~\ref{sec:ml}). Figure~\ref{fig:target_overview} depicts the categorization of the surveyed literature based on the targeted applications. Each application is further classified by the method of trace collection (software, physical, and remote) as well as the type of device under attack, including CPUs, FPGAs, ASICs, GPUs, TPUs, embedded devices, trusted execution environments (TEEs), and other hardware peripherals. We survey application-specific countermeasures in Section~\ref{sec:countermeasures}. Finally, Section~\ref{sec:summary} concludes this survey with a discussion on future research directions.

\section{Power Side-Channel Models}
\label{sec:models}

In order to perform a PSC attack, it is essential to understand the internal operations of the hardware. Since almost all attacks target devices using semiconductor components, it is necessary to discuss the power consumption model in semiconductor devices. These power models are based on the underlying manufacturing technology. Complementary Metal-Oxide-Semiconductor (CMOS) technology is the predominant technology for constructing application-specific integrated circuits (ASICs), including central processing units (CPUs), field programmable gate arrays (FPGAs), graphics processing units (GPUs), memory chips, and other digital logic circuits. CMOS technology is popular for fabrication due to its low power consumption, high noise immunity, and high density of logic functions.

CMOS power consumption can be primarily divided into dynamic and static components. Dynamic power consumption occurs when the circuit transitions between different states, specifically when transistors switch from off (logic 0) to on (logic 1) or vice versa. This form of power consumption can be expressed by Equation~\ref{eqn:dynamic}.
Where \(\alpha\) represents the activity factor (indicating the fraction of time the transistors are switching), \(C_L\) denotes the load capacitance, \(V_{DD}\) is the supply voltage, and \(f\) is the switching frequency.

\vspace{-0.1in}
\begin{equation} \label{eqn:dynamic}
P_{\text{dynamic}} = \alpha C_L V_{DD}^2 f 
\end{equation}

On the other hand, static power consumption arises from leakage currents even when transistors are not switching. This leakage power illustrated by Equation~\ref{eqn:static} has become more significant in advanced technology nodes. In Equation~\ref{eqn:static}, \(I_{\text{leakage}}\) represents the leakage current.

\vspace{-0.1in}
\begin{equation}\label{eqn:static}
P_{\text{static}} = I_{\text{leakage}} V_{DD}
\end{equation}

In the earlier days of CMOS technology, dynamic power consumption was the dominant component because the switching activity and load capacitance significantly contributed to the overall power usage. However, as technology has scaled down to nanometer dimensions, the leakage currents have increased due to thinner gate oxides and higher electric fields, making static power consumption comparable to dynamic power consumption in many domains, such as embedded devices with large memory. Additionally, with the increase in the number of transistors per chip, the aggregate leakage current has become substantial, elevating the importance of addressing static power in modern CMOS designs.

In order to perform a successful PSC attack, the discussed parameters should be computed from the device under attack. The computation of these parameters varies depending on whether the attack is performed on pre-silicon designs or post-silicon (real hardware) implementations.

\begin{figure}[htp]
    \begin{center}
        \includegraphics[width=0.48\textwidth]{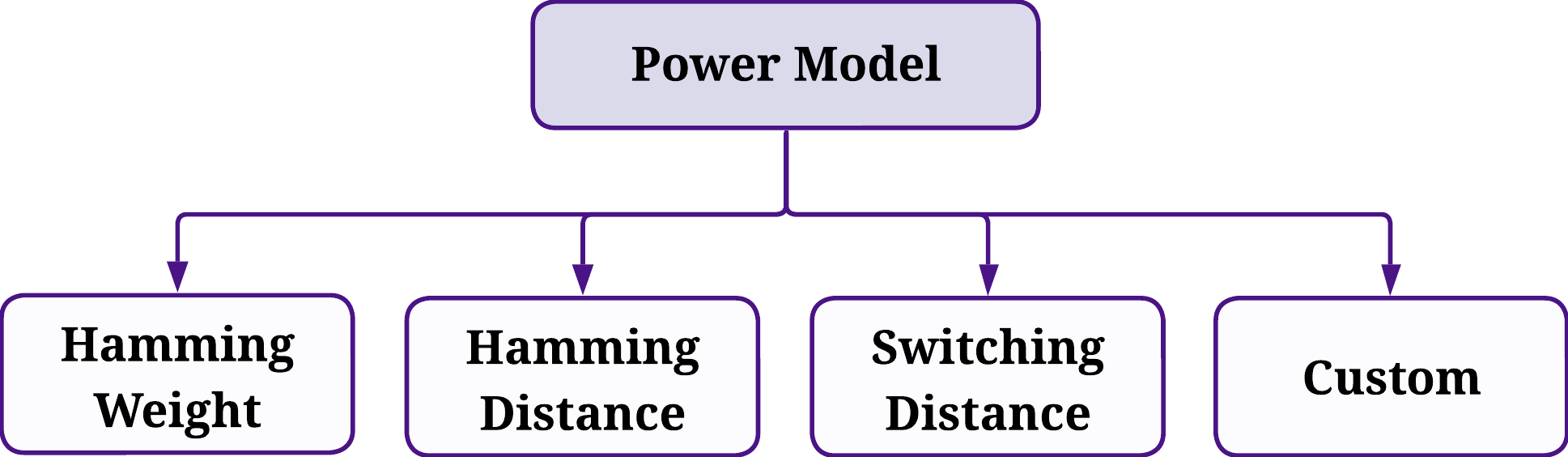}
    \end{center}
     \vspace{-0.1in}
      \caption{Overview of various power models used for power side-channel analysis, including Hamming weight, Hamming distance, switching distance, and custom combinations and models.}
      \label{fig:power_model_overview}
      \vspace{-0.2in}
\end{figure}

\begin{itemize}
    \item During pre-silicon information leakage assessment, these values are estimated from simulation results~\cite{He2019Rtl, Pundir2022Power, jayasena2023tvla}. The pre-silicon designs are simulated with appropriate inputs, obtaining value change dump (VCD) files that capture the behavior of internal signals. By analyzing signal transitions across various power model combinations, the expected power signature of the design is extracted, allowing for an early evaluation of potential PSC leakages before the device's physical implementation. These estimation models are particularly useful for emerging technologies, such as negative capacitance field-effect transistors (NCFETs), which aim to overcome limitations in traditional MOSFETs. Although NCFETs are not yet at the production level, their security applicability can be evaluated using pre-silicon leakage analysis methods~\cite{Knechtel2020Power}. While pre-silicon analysis enables early detection of potential leakages, it relies on simulated switching activity and abstract power models that often overlook real-world effects such as parasitics, IR drop, and process variation. Consequently, its accuracy is generally lower than post-silicon analysis, which captures actual physical leakage but requires fabricated hardware and measurement effort.
    \item For post-silicon power side-channel attacks, the researchers utilize abstract versions of power models that are easier to model, easier to compute, and still provide accurate results~\cite{Liu2022Frequency}. Before the actual power side-channel analysis, the expected power consumption is computed based on these power models. Unlike real power traces, the abstract versions of the power models are computationally efficient. These abstract models have been used to generate the hypothetical power traces \cite{Aysu2018Binary, Luo2015Side, Benhadjyoussef2021Power, Munny2021Power, Iyer2021Systematic} to perform statistical analysis with actual power traces. However, the use of abstract power models may fail to capture subtle fine-grained microarchitectural behaviors, introducing trade-offs between analysis simplicity and modeling accuracy. While, post-silicon analysis benefits from high fidelity due to real power measurements, it requires access to fabricated hardware, precise instrumentation, and careful noise handling, which can be costly and time-consuming.
\end{itemize}
While this section discusses the power models used in PSC attacks, Section~\ref{sec:analysis} explains how these power models are utilized in analysis methods.
Figure~\ref{fig:power_model_overview} shows the four popular power models for side-channel analysis in the literature, including Hamming weight, Hamming distance, and switching distance. Among these, the Hamming weight, Hamming distance, and custom combinations of these two power models are widely used. These power models play a crucial role since the accuracy of the attack depends on the correct representation of the power model. 

\subsection{Hamming Weight} 
\label{sec:model_hwt}

The Hamming weight (HWT) represents the number of ``1'' bits in a binary  number~\cite{Messerges2000Using}. This metric is directly related to the power consumption of the device, as each ``1'' bit contributes to the static power dissipation~\cite{tTehranipoor2011Introduction}. The formula to compute Hamming weight is illustrated by Equation~\ref{eqn:hw}, where \(D_i\) is the i-th bit of the data byte \(D\).

\begin{equation} \label{eqn:hw}
HWT(D) = \sum_{i=0}^{n-1} D_i 
\end{equation}

For example, the Hamming weight of the binary number 10110010 is 4. This model has been successfully applied in attacks on ASIC, CPU, and FPGA implementations of CMOS devices \cite{Bellizia2021Sc, Moos2020Static, Benhadjyoussef2021PowerAES, Dumitru2023Borrowed, Moos2017Static, Vafa2020Efficient}.

\subsection{Hamming Distance}

The Hamming distance (HD) measures the number of bit positions at which two binary numbers differ, providing insight into the power consumed during state transitions. This power model was introduced by \citeauthor{Brier2004Correlation}~ \cite{Brier2004Correlation} in a PSC attack. The Hamming distance is calculated using the formula shown in Equation~\ref{eqn:hd}, where \(D1_i\) and \(D2_i\) are the i-th bits of data bytes \(D1\) and \(D2\), respectively, and \(\oplus\) denotes the XOR operation.

\begin{equation} \label{eqn:hd}
HD(D1, D2) = \sum_{i=0}^{n-1} (D1_i \oplus D2_i)
\end{equation}

 For instance, the Hamming distance between 10110010 and 11011001 is 5. The Hamming distance model was successfully used by various PSC attacks \cite{Zhang2020Memory, Benhadjyoussef2021Power, Wei2018I, Srivastava2024Scar, Luo2015Side, Nevskovic2023Systemc, Lipp2021Platypus}.

\subsection{Switching Distance}

\citeauthor{Peeters2007Power} introduced a refined power model based on the fact that transitions from 0 to 1 and from 1 to 0 consume different power~\cite{Peeters2007Power}. This model aims to provide a more accurate representation of leakage by distinguishing between the power consumed during different state transitions in CMOS gates, compared to the Hamming distance model. The switching distance power consumption model is detailed in Table \ref{tab:power_model}, where transitions from 0 to 1 and from 1 to 0 are assigned specific power values to reflect their differing consumption patterns. Here, $d$ is the normalized difference in transition leakages, as defined in Equation~\ref{eqn:sd}, where $P_{0 \rightarrow 1}$ is the probability of a 0 to 1 transition, and $P_{1 \rightarrow 0}$ is the probability of a 1 to 0 transition.

\begin{equation} \label{eqn:sd}
d = \left( P_{0 \rightarrow 1} - P_{1 \rightarrow 0} \right) / P_{0 \rightarrow 1}
\end{equation}

\citeauthor{Peeters2007Power}~\cite{Peeters2007Power} and \citeauthor{Tran2023Transition}~\cite{Tran2023Transition} used this model to perform attacks on RISC-based processors, while there were promising efforts 
 of applying this model to perform attacks on ASICs and FPGAs \cite{Liu2010Aes,Mestiri2013Comparative, Cao2020Attacking}.

\begin{table}[h]
\caption{Switching distance power consumption model}
\centering
\begin{tabular}{|c|c|}
\hline
\textbf{Transition} & \textbf{Power} \\ \hline
$0 \rightarrow 0$ & 0 \\ \hline
$0 \rightarrow 1$ & 1 \\ \hline
$1 \rightarrow 0$ & $1 - d$ \\ \hline
$1 \rightarrow 1$ & 0 \\ \hline
\end{tabular}
\label{tab:power_model}
\vspace{-0.2in}
\end{table}

\subsection{Custom Power Models}
\label{sec:model_custom}

Besides the above-mentioned power models, there are also custom power models in the literature based on the target application and implemented hardware. The stochastic power model is such a custom model, which was introduced to enhance the efficiency of differential side-channel analysis against block ciphers by approximating the leakage function within a low-dimensional vector space, thereby avoiding the need for full high-dimensional profiling as required in classical template attacks~\cite{schindler2005stochastic}. The stochastic model learns a compact set of parameters using measurements from only a single profiling key, enabled by the equal-images property that keeps the leakage structure invariant across keys. The model expresses each observed power sample as
\[
I_t(x,k) = h_t(x,k) + R_t,
\]
where the deterministic component $h_t(x,k)$ captures key- and data-dependent leakage, while the random component $R_t$ represents key- and data-independent noise. For the AES implementation on an 8-bit microcontroller, the deterministic leakage is approximated by a linear combination of S-box output bit functions,
\[
\hat{h}^*_t(\phi(x,k)) = b_{0,t} + \sum_{i=1}^{8} b_{i,t}\, g_i(\phi(x,k)), \quad \phi(x,k)=x\oplus k,
\]
where each $g_i$ extracts one bit of the S-box output and the coefficients $b_{i,t}$ are obtained via least-squares fitting during profiling. The noise term is modeled as a multivariate Gaussian with an empirically estimated covariance matrix.

\citeauthor{Xiang2020Open} proposed another custom power model for Deep Neural Networks (DNNs)~\cite{Xiang2020Open}. The model calculates the power consumption for each layer of the neural network, considering convolutional, pooling, fully connected, and activation layers. Specifically, Equation~\ref{eqn:p_conc} illustrates the power consumption of a convolutional layer, where $C$ is the number of input channels, $L$ and $W$ are the input dimensions, $S$ is the stride, $N$ is the number of kernels, $F$ is the kernel size, and $p_m$ and $p_a$ are the average power consumptions for multiplication and addition operations, respectively.

\vspace{-0.1in}
\begin{equation} \label{eqn:p_conc}
\small
P_{conv}(C, L, W, S, N, F) = p_m \frac{L W N C F^2}{S^2} + p_a \frac{L W N C F^2}{S^2}
\end{equation}

The power consumption for the pooling layer is given in Equation~\ref{eqn:p_pl}, where $p_c$ is the average comparison operation power consumption.

\vspace{-0.1in}
\begin{equation} \label{eqn:p_pl}
\small
P_{pl}(C, L, W, S, F) = p_c \frac{C L W F^2}{S^2}
\end{equation}

For the fully connected layer, the power consumption is modeled as shown in Equation~\ref{eqn:p_fc},
where $X$ and $Y$ are the number of input and output neurons, respectively.

\vspace{-0.1in}
\begin{equation} \label{eqn:p_fc}
\small
P_{fc}(X, Y) = p_m X Y + p_a X Y
\end{equation}

Equation~\ref{eqn:p_ac} illustrates the formula for the activation function, where \(p_{ac}\) is the power consumed by one operation in the activation function and \(\alpha\) is the operational coefficient determined by the specific type of activation function. 

\vspace{-0.1in}
\begin{equation} \label{eqn:p_ac}
\small
P_{ac}(C, L, W) = p_{ac} \alpha CLW
\end{equation}

To further refine the power model, the authors introduced a Parameter Sparsity Model~\cite{Xiang2020Open}, accounting for the sparsity in the DNN parameters due to pruning techniques. The sparsity is represented by the coefficients \(\lambda_1\) and \(\lambda_2\), modifying the power consumption as shown in Equation~\ref{eqn:p_conv1} and Equation~\ref{eqn:p_fc1}. The overall power consumption model is thus a combination of these equations, allowing for accurate prediction of the power consumption of DNNs based on their architecture and operational parameters.

\vspace{-0.1in}
\begin{equation} 
\label{eqn:p_conv1}
\small
P'_{\text{conv}}(C, L, W, S, N, F) = \lambda_1 \cdot P_{\text{conv}}(C, L, W, S, N, F)
\end{equation}

\vspace{-0.1in}
\begin{equation} 
\label{eqn:p_fc1}
\small
P'_{\text{fc}}(X, Y) = \lambda_2 \cdot P_{\text{fc}}(X, Y)
\end{equation}

Another custom power model was proposed by \citeauthor{Wang2022Hertzbleed} to perform the `Hertzbleed' attack~\cite{Wang2022Hertzbleed}. Even though the authors have primarily considered the Hamming weight and Hamming distance models to explain the power consumption of modern Intel and AMD x86 CPUs, they further explored an additive model combining both Hamming weight and Hamming distance as the power model. This approach also highlights the non-linearity of the Hamming weight model, especially concerning the bit position of `1's, where the leakage is more pronounced when there are `1's in the most significant byte than in the least significant byte.

Another custom power model was proposed by \citeauthor{Kogler2023Collide+} in the Collide+Power attack~\cite{Kogler2023Collide+}. The authors emphasized that simple HWT or HD models fail to capture the leakage arising from interactions between attacker-controlled and victim cache line values. To address this, they introduced a composite power model incorporating both the HD and the HWT of the attacker-controlled value ($G$) and secret victim value ($V$) of different security domains, formally modeled as:
\begin{equation}
P_{cache} \approx a_0 \cdot \mathsf{hd}(G,V) + w_0 \cdot \mathsf{hwt}(G) + w_1 \cdot \mathsf{hwt}(V) + w,
\label{eq:collidepower_model}
\end{equation}

where $\mathsf{hd}(G,V)$ denotes the Hamming distance between $G$ and $V$, $\mathsf{hwt}(\cdot)$ represents the HWT, and $w$ captures independent noise components~\cite{Kogler2023Collide+}. This model reflects the observation that power consumption during cache line replacement is predominantly influenced by bit transitions (captured by HD), but is also affected by the HWT of the involved values.

In summary, power modeling plays a significant role in shaping the effectiveness of PSC attacks. While traditional models such as HWT and HD offer foundational insights into digital switching activity, recent works demonstrate that these models alone are often insufficient to capture the nuanced leakage behaviors of modern computing platforms. Consequently, researchers have proposed hybrid and architecture-specific models that better reflect practical leakage scenarios, including nonlinear effects, bit position dependencies, and inter-value interactions. These evolving models not only enhance attack success rates but also guide the design of more robust countermeasures. In the subsequent section, we delve into the diverse set of analysis techniques that leverage these models to extract sensitive information from power traces.

\section{Power Side-Channel Analysis Methods}
\label{sec:analysis}

\begin{figure}[htp]
    \begin{center}
        \small
        \vspace{-0.1in}
        \includegraphics[width=0.49\textwidth]{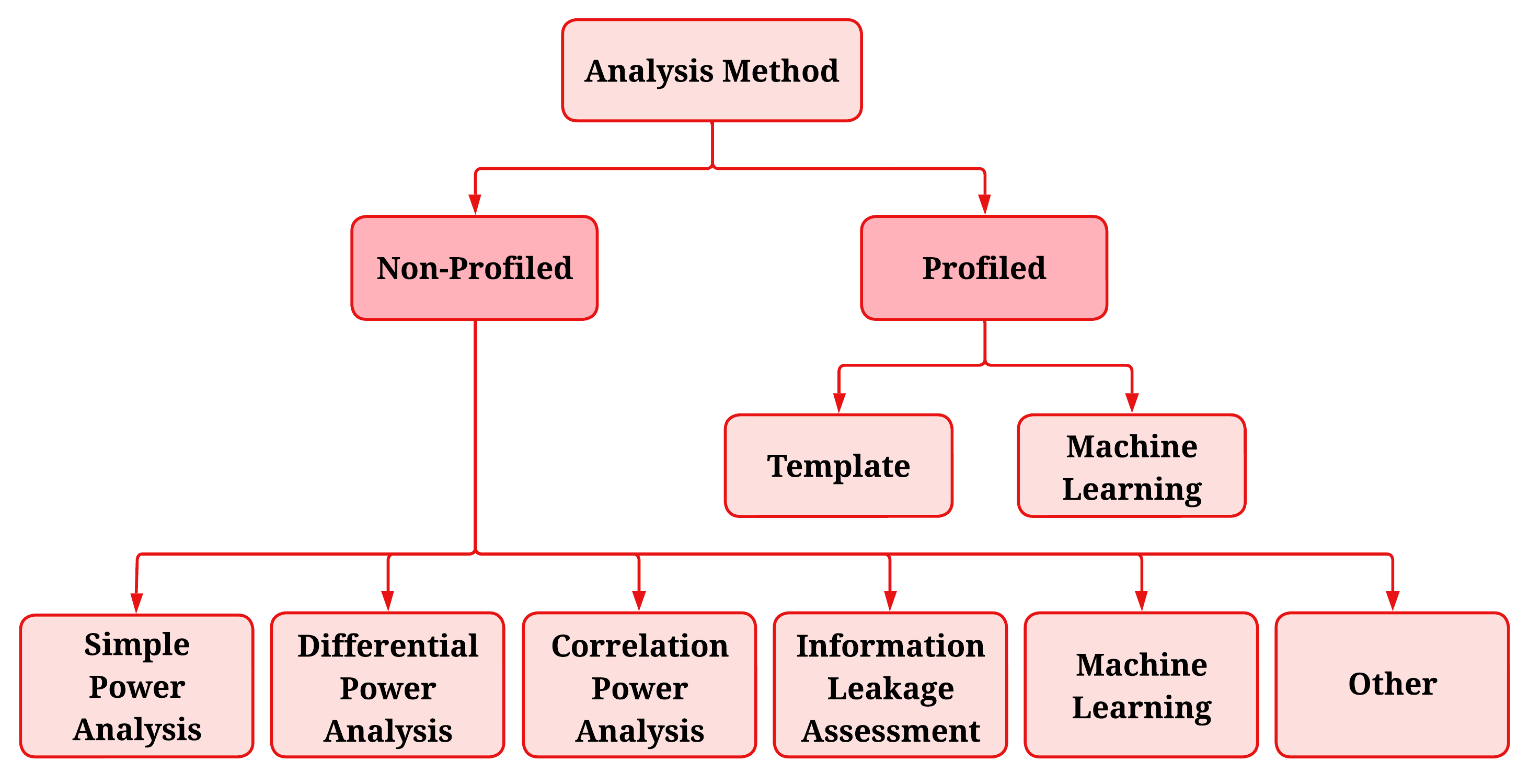}
    \end{center}
      \vspace{-0.1in}
      \caption{Overview of power side-channel analysis methods that can be viewed as profiled or non-profiled based on the procedure used to extract information from the power traces.}
      \label{fig:analysis_overview}
          \vspace{-0.1in}
\end{figure}

Figure~\ref{fig:analysis_overview} shows an overview of PSC analysis methods that can be broadly divided into two categories (profiled and non-profiled) based on the method used to extract information from the power traces.
Both types of attacks rely on collecting power traces based on the power consumption of the platform during the execution of the target applications. By analyzing these power traces, attackers can infer secret information due to the correlation between the power consumption and the operations being performed or data being used in the platform.

\begin{figure}[htp]
    \begin{center}
        \includegraphics[width=0.48\textwidth]{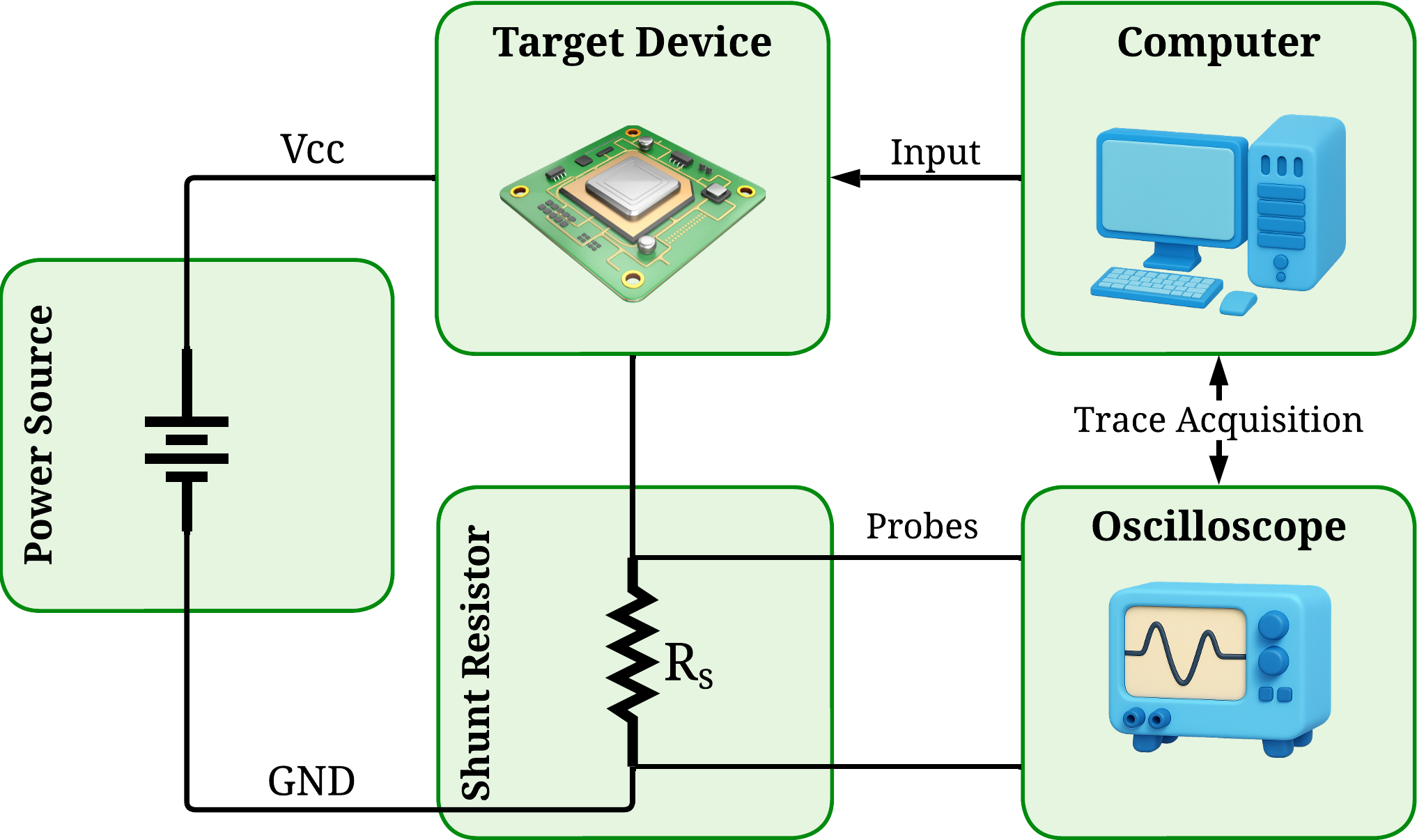}
    \end{center}
      \caption{A typical PSC attack trace acquisition setup includes some mechanism (e.g., oscilloscope) to gather power traces from the device under attack (e.g., a CPU running the target application) as well as a controller to control the device and collect traces for further analysis (e.g., a computer)~\cite{Sanjaya2024Information}.}
      \label{fig:attackSetup}
\end{figure}

As shown in Figure~\ref{fig:attackSetup}, a typical physical power trace acquisition setup involves the following components: a target device (e.g., ASIC, CPU, FPGA, GPU), a power measurement tool (such as an oscilloscope or a dedicated power analysis tool), and a computer to control the device and collect the power traces. The device executes the targeted application while the power measurement tool records its power consumption. These power traces are then analyzed using statistical or machine learning (ML) techniques to extract the secret information.

The power consumption ($ P $) of the device is given by Equation~\ref{eqn:pow}, where $V$ is the supply voltage and $ I $ is the current.

\vspace{-0.2in}
\begin{equation} \label{eqn:pow}
P = IV
\end{equation}

Since $ V $ is a known value by measuring $ I $ we can calculate $ P $. However, to accurately measure $P$, a shunt resistor is often placed in series with the power supply line of the targeted device. Equation~\ref{eqn:ohm} illustrates the voltage drop ($V_{drop}$) across this resistor ${R_s}$.
\begin{equation} \label{eqn:ohm}
{V_{drop}} = I{R_s}
\end{equation}

Therefore $V_{drop}$ across the ${R_s}$ is proportional to $I$. By measuring the voltage drop across the shunt resistor and knowing the resistance, the power consumption can be calculated. 

Beyond traditional physical power trace acquisition using oscilloscopes and shunt resistors, recent advancements have enabled PSC attacks via remote and software-accessible interfaces. Remote PSC attacks leverage implementations such as time-to-digital converters (TDCs), which convert power variations into another observable metric, such as time, to infer power consumption without requiring direct physical access to the victim implementation. Software-based PSC attacks, an emerging class of PSC techniques, such as PLATYPUS~\cite{Lipp2021Platypus}, exploit privileged or publicly accessible hardware performance monitoring frameworks. Notable examples include Intel’s Running Average Power Limit (RAPL) interface and NVIDIA’s NVML interface, both of which provide energy telemetry at fine granularity. These approaches broaden the threat model by eliminating the need for physical proximity, thereby increasing the practicality and scalability of PSC attacks.
The remainder of this section presents two major approaches for PSC analysis that are applied to the power traces collected using the acquisition techniques discussed above. We begin by exploring non-profiled side-channel analysis methods, which operate without prior knowledge of the target's leakage characteristics. Next, we examine profiled side-channel analysis methods, which leverage a trained leakage model obtained from a similar device to improve attack efficiency and accuracy.

\subsection{Non-Profiled Side-channel Analysis Methods}
Non-profiled side-channel analysis attacks are a category of power analysis attacks that do not rely on a pre-established profile of the target device's behavior. These attacks focus on exploiting the power consumption patterns of targeted devices during their operation to extract secret information such as cryptographic keys. The procedure typically involves collecting power consumption traces while the device is executing targeted applications, followed by statistical analysis to identify correlations between the power consumption data and the secret information. Common types of non-profiled side-channel analysis include simple power analysis (SPA), differential power analysis (DPA), and correlation power analysis (CPA). SPA involves visually inspecting power traces to identify patterns directly related to the operations performed. DPA uses statistical methods to compare multiple power traces to filter out noise and highlight the secret information. CPA, a more advanced form of DPA, uses correlation coefficients to match the observed power consumption with hypothetical power consumption models. 

The primary advantage of non-profiled side-channel analysis attacks is that they do not require prior knowledge or access to the target device, making them versatile and applicable to a wide range of devices. In this section, we describe five popular approaches for side-channel analysis: (i) simple power analysis, (ii) differential power analysis, (iii) correlation power analysis, (iv) information leakage assessment, and (v) machine learning.

\subsubsection{\textbf{Simple Power Analysis}}

In the late 1990s, \citeauthor{Kocher1996Timing} introduced simple power analysis (SPA)~\cite{Kocher1996Timing}. SPA involves the direct observation and interpretation of power consumption patterns of a target device while it performs operations using a single or a few inputs on the target application. SPA does not rely on statistical methods but instead focuses on identifying distinct power signatures associated with specific operations. 
During a SPA attack, the attacker captures the power traces of the target device and visually inspects these traces to identify patterns and distinguish between different operations, such as key loading, encryption, and decryption. If the target application is a cryptographic algorithm, the secret key information can sometimes be inferred directly from these visual patterns, particularly if the device performs cryptographic operations in a manner that leaks information through distinct power variations. For instance, the execution of cryptographic algorithms like DES~\cite{Kocher1999Differential} can exhibit power consumption variations depending on the value of the key bits being processed. By carefully analyzing these variations, an attacker can deduce the key bit by bit. 

Existing literature has shown that the standard unsecured hardware implementation of the quantum-secure encryption scheme based on the Binary-Ring-Learning-with-Errors (B-RLWE) problem is particularly vulnerable to SPA due to its use of binary coefficients~\cite{Aysu2018Binary}. During polynomial multiplication, the hardware generates partial products based on the binary values of the secret key bits. If a key bit is `1', the corresponding coefficients are added to the intermediate sum, leading to detectable variations in power consumption. \citeauthor{Aysu2018Binary} illustrated that this phenomenon makes it possible to recover the secret key with a few power measurements by observing these differences in power traces~\cite{Aysu2018Binary}.

\subsubsection{\bf Differential Power Analysis}

Differential power analysis (DPA), introduced by~\citeauthor{Kocher1999Differential}~\cite{Kocher1999Differential}, is a more sophisticated technique than SPA. In DPA, the attacker collects a large number of power traces while the target device processes different known inputs. DPA focuses on the statistical analysis of differences in power consumption to identify secret keys. During a DPA attack on a cryptographic algorithm, the attacker first makes hypotheses about the key bits and computes the expected intermediate values of the cryptographic algorithm. The attacker then partitions the power traces into different sets based on the hypothetical intermediate values. By comparing the average power consumption of these sets, the attacker identifies the presence of significant differences that correlate with specific key bits. 

The first step of a DPA attack is to execute the targeted algorithm \( n \) times to capture \( T_n \) traces. The output ciphertext is also captured and labeled with \( C_i \) corresponding to its \( i^{th} \) trace. The selection function \( S(C_i, K_m) \) is defined, where \( K_m \) is the key guess. The selection function determines whether a specific bit of the intermediate value (derived from the key guess and ciphertext) is 0 or 1, thereby classifying each trace into one of two groups. The differential trace \( T_{\Delta} \) for the guess \( K_m \) is determined by Equation~\ref{eqn:dtr}, where \( T_{\Delta}[j] \) is the differential trace at point \( j \). In Equation~\ref{eqn:dtr}, \( T_i[j] \) is the power consumption at time offset \( j \) in the \( i \)-th trace. The first term in the equation represents the average power consumption when the selection function is 1, and the second term represents the average power consumption when the selection function is 0. The differential trace \( T_{\Delta} \) highlights the differences in power consumption that correlate with the correct key guess. The guesses for \( K_m \) that produce the largest spikes in the differential trace \( T_{\Delta} \) are considered to be the most likely candidates for the correct value. 

\vspace{-0.1in}
\begin{equation} \label{eqn:dtr}
\footnotesize
T_{\Delta}[j] = \frac{\sum_{i=1}^{n} S(C_i, K_m) \cdot T_i[j]}{\sum_{i=1}^{n} S(C_i, K_m)} - \frac{\sum_{i=1}^{n} (1 - S(C_i, K_m)) \cdot T_i[j]}{\sum_{i=1}^{n} (1 - S(C_i, K_m))} \end{equation}
\normalsize

An example of a successful DPA attack is demonstrated on a hardware implementation of B-RLWE~\cite{Aysu2018Binary}. By analyzing the differences in power consumption during polynomial multiplication, they were able to recover the secret key. This attack illustrates the effectiveness of DPA in breaking quantum-secure cryptographic implementations.

\subsubsection{\bf Correlation Power Analysis}

\citeauthor{Brier2004Correlation} introduced correlation power analysis (CPA)~\cite{Brier2004Correlation}. In CPA, the attacker collects multiple power traces while the target device processes different known inputs. The attacker then hypothesizes intermediate values of the cryptographic algorithm based on possible key guesses and uses power models to develop hypothetical power traces. At a specific point in time, the actual power traces are statistically correlated with the hypothetical power traces, typically using Pearson correlation coefficients. The Pearson correlation coefficient, \( r \), is calculated by Equation~\ref{eqn:pea}, where \(X_i\) and \(Y_i\) are the actual and hypothetical power trace values, respectively, \(\bar{X}\) and \(\bar{Y}\) are the means of these values, and \(n\) is the number of samples. The key hypothesis that results in the highest correlation with the actual traces is deemed the correct guess.

\vspace{-0.1in}
\begin{equation} \label{eqn:pea}
\small
r = \frac{\sum_{i=1}^n (X_i - \bar{X})(Y_i - \bar{Y})}{\sqrt{\sum_{i=1}^n (X_i - \bar{X})^2 \sum_{i=1}^n (Y_i - \bar{Y})^2}}
\end{equation}

In recent years, several PSC attacks using CPA have been performed. An application of CPA on an AES implementation running on a state-of-the-art GPU was demonstrated in~\cite{Luo2015Side}. The authors successfully extracted the secret key by analyzing the power consumption patterns of the GPU during AES encryption operations. They developed a Hamming distance-based power model targeting registers used in the last round of AES. \citeauthor{Benhadjyoussef2021Power}~\cite{Benhadjyoussef2021Power} proposed an attack scenario combining both CPA and differential fault analysis. Here, the last round of an AES-128 encryption implementation on an FPGA was selected as the target.

Measurement to Disclosure (MTD) is an indication of the number of power traces required to extract the secret information. \citeauthor{Munny2021Power} evaluated the relationship between the attacker's probing effect in the system and MTD, the time to collect the MTD number of traces, and the time for a successful CPA to extract the full key~\cite{Munny2021Power}. They used an AES-128 implementation on an XMEGA board and targeted the output of the S-Box while using Hamming distance as the power model in their CPA. The results showed that with a sufficiently high resistive probing effect, the MTD is reduced, and the time for MTD collection and the time for the CPA are almost linear with MTD.

\begin{figure}[htp]
    \begin{center}
        \includegraphics[width=0.48\textwidth]{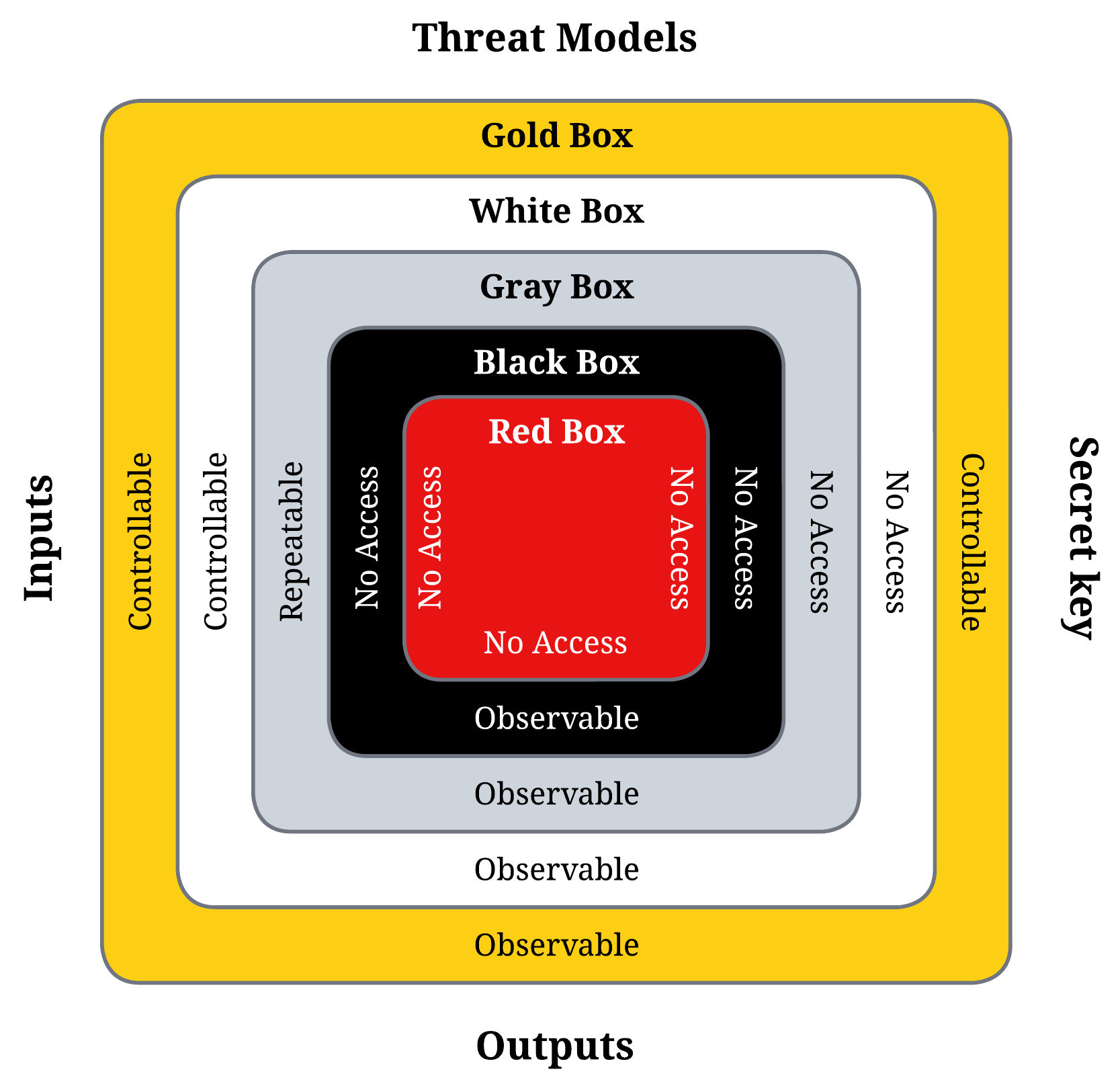}
    \end{center}
     \caption{Different levels of PSC attacks based on model restrictions, ranging from the most flexible Golden Box to the most restrictive Red Box~\cite{Iyer2021Systematic}.}
      \label{fig:restriction_levels}
\end{figure}

Another systematic evaluation of PSC attacks using CPA on AES implementations was explored by~\cite{Iyer2021Systematic}. The authors outlined how to conduct attacks under different restriction levels shown in Figure~\ref{fig:restriction_levels}: the most restrictive “red box” with no additional information, the “black box” with output observability, the “gray box” with input repeatability and output observability, the “white box” with input controllability and output observability, and the “golden box” where both the key and input are controllable and the output is observable. A static PSC attack using CPA on an AES-128 implementation in an FPGA environment was proposed in~\cite{Dumitru2023Borrowed}. The S-box outputs in the final round are the targeted intermediate values, and the Hamming weight power model was used to generate hypothetical power trace values. The CPA attack successfully recovered the entire AES key with 1,500 traces.

\citeauthor{Mozipo2023Residual}~\cite{Mozipo2023Residual} explored the application of CPA attacks on selected lightweight ciphers (LWC), specifically targeting Ascon, Sparkle, PHOTON-Beetle, GIFT-COFB, Grain, Romulus, and TinyJambu. These algorithms are the seven out of ten finalists from the NIST lightweight cryptography competition, which are designed for resource-constrained IoT devices. The authors detailed the attack methodology, including the targeted intermediate values and the necessary conditions for successful CPA~\cite{Mozipo2023Residual}. The findings demonstrated that increasing the number of collected power traces can significantly improve the effectiveness of CPA, thereby highlighting the ongoing risk posed by such attacks even in advanced lightweight cryptographic implementations.

\citeauthor{jayasena2024evilcs} proposed a pre-silicon vulnerability evaluation framework for SoC designs using CPA~\cite{jayasena2024evilcs}. The authors demonstrated a vulnerability (EvilCS) associated with context switching of the operating system kernel which leaks data associated with user applications. This framework was later extended for pre-silicon CPA analysis of cryptographic instruction set extensions (CISE)~\cite{jayasena2024ciseleaks}. The authors proposed a technique to generate firmware images to activate the functional units implementing the CISE components and used the estimated power model to evaluate the leakage of the cryptographic secrets. The authors demonstrated the cryptographic leakage on RISC-V XCRYPTO extension~\cite{jayasena2024ciseleaks}.

Higher-order CPA is one of the promising options for attacking target implementations with countermeasures. \citeauthor{Moos2017Static} demonstrated the use of CPA and third-order CPA to exploit both static and dynamic PSC leakage of the PRESENT-80 cipher implemented on a 150 nm CMOS prototype chip~\cite{Moos2017Static}. The third-order CPA,  collision-based Moments Correlating DPA (MCDPA), is used to attack the PRESENT-80 block cipher with a first-order countermeasure threshold masking.

\begin{figure}[htp]
    \begin{center}
        \includegraphics[width=0.49\textwidth]{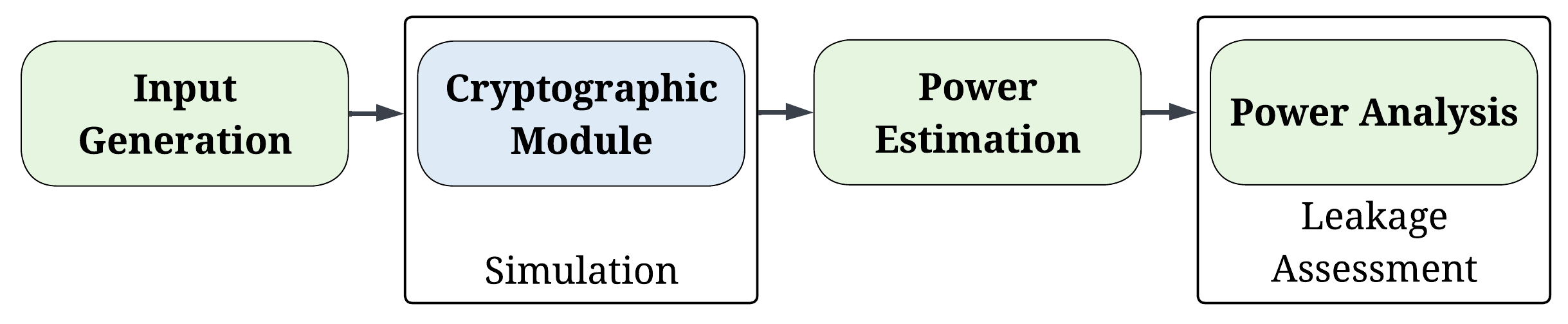}
    \end{center}
     \vspace{-0.2in}
      \caption{Overview of test vector leakage assessment (TVLA) framework.}
      \label{fig:tvla}
\end{figure}

\subsubsection{\bf Information Leakage Assessment}
\label{sec:analysis_non_profiled_information_leakage_assessment}

While SPA, DPA, and CPA are focused on extracting secrets such as key recovery attacks, information leakage assessment focuses on quantifying the amount of information leakage from an implementation. Test Vector Leakage Assessment (TVLA) is a widely used statistical testing methodology used for information leakage assessment~\cite{Gilbert2011Testing}. Test Vector Leakage Assessment (TVLA) is a widely adopted statistical method  for assessing information leakage~\cite{Gilbert2011Testing}. Its primary objective is to detect the presence of side-channel vulnerabilities using a conformance-style testing approach. This allows designers to identify and mitigate leakage early in the development lifecycle, ideally before fabrication for hardware designs and before deployment for software or firmware. TVLA is typically categorized into two types: non-specific and specific. The non-specific (fixed vs. random) leakage assessment technique was proposed to detect information leakage dependent on input data with the power signature. This method statistically compares trace differences between fixed input data and random input data. The specific leakage assessment method was proposed targeting specific intermediate values of the implementation.

Figure~\ref{fig:tvla} illustrates the general steps involved in the process of information leakage assessment with TVLA. Once the type of assessment is determined, the necessary inputs need to be generated and simulated to obtain the expected power signatures using the steps outlined in Section~\ref{sec:models}. 
Once the expected power traces are collected, they have to be analyzed for information leakage. Recent efforts have commonly used statistical techniques such as Welch’s t-test, Pearson’s chi-squared test ($\chi^2$-test), and the Kullback–Leibler (KL) divergence metric~\cite{Kullback1951Information}.

The KL divergence (KLD) metric was utilized for evaluating hardware implementations of symmetric key cryptography implementations~\cite{He2019Rtl, Pundir2022Power}. KL divergence measures the divergence of one probability distribution from a second, expected probability distribution.  Given two probability distributions, $P$ and $Q$, defined over the same probability space, the KL divergence from $Q$ to $P$ in the context of continuous probability distributions is shown by Equation~\ref{eqn:kl}. Where \(p(x)\) and \(q(x)\) are the probability density functions of \(P\) and \(Q\), respectively.

\begin{equation}\label{eqn:kl}
D_{KL}(P \| Q) = \int p(x) \log \frac{p(x)}{q(x)} \, dx
\end{equation}

The RTL-PSC~\cite{He2019Rtl} framework employed a combined KL divergence and success rate metric to evaluate information leakage based on simulation-derived power profiles. This approach facilitated early identification and mitigation of side-channel leakage, proving to be both efficient and cost-effective in securing cryptographic implementations on Galois-Field (GF) and Look-up Table (LUT)-based AES designs.

 \citeauthor{Pundir2022Power} assessed several symmetric and post-quantum cryptographic algorithms using both KL divergence and Welch's t-test~\cite{Pundir2022Power}. A metric was introduced where a high KL divergence indicates distinguishable power leakage probability distributions for two different keys, suggesting that an adversary could easily correlate power consumption between the keys. While this method was noted for its speed, it was less accurate. The pre-silicon assessment was validated through the post-silicon TVLA using the fixed vs. random test and Welch’s t-test. The authors used Welch's t-test to determine the difference between the means of two populations that do not assume equal variances.

  Equation~\ref{eq:ttest} illustrates parameters used by Welch's t-test, where \(\bar{X}_1\) and \(\bar{X}_2\) are the sample means, \(s_1^2\) and \(s_2^2\) are the sample standard deviations, and \(n_1\) and \(n_2\) are the sample sizes of the two groups, respectively. In the context of TVLA, a t-value above 4.5 was used to indicate potential information leakage through side-channel attacks.

\begin{equation}\label{eq:ttest}
t = \frac{\bar{X}_1 - \bar{X}_2}{\sqrt{\frac{s_1^2}{n_1} + \frac{s_2^2}{n_2}}}
\end{equation}

Welch’s t-test was employed for a detailed simulation-based evaluation of how different sampling intervals impact the detection of PSC leakage in AES circuits on an FPGA~\cite{Miura2023Simulation}. Power traces were simulated at various intervals to assess how the sampling rate affects the visibility of leakage and the effectiveness of side-channel attacks.  Welch’s t-test was also used to evaluate how memory-based high-level synthesis (HLS) optimizations affect the security of cryptographic implementations~\cite{Zhang2020Memory}. PSC leakage was assessed under different memory-based HLS optimizations by converting HLS implementations into RTL designs and synthesizing them on an FPGA.

\citeauthor{jayasena2023tvla} proposed a TVLA technique to evaluate hardware implementations of asymmetric 
key cryptographic systems~\cite{jayasena2023tvla}. A major limitation of applying methods used by symmetric key TVLA techniques was the leakage assessment techniques lost the notion of timing during the analysis. In order to solve this, authors have used  Bonferroni correction~\cite{weisstein2004bonferroni} combined with Welch’s t-test. 
This method allowed the authors to partition the power traces along the time axis and perform analysis on bit-serialized algorithms such as Elliptic Curve Cryptography (ECC) and RSA.

\citeauthor{Liu2022Frequency} has explained an example of a use case of TVLA in the post-silicon phase, as it involves evaluating the system's side-channel characteristics after hardware fabrication through experiments on physical systems~\cite{Liu2022Frequency}. Here, TVLA is used to assess potential side-channel leakage from an AES-NI-based AES implementation due to frequency throttling activity. The TVLA methodology utilizes t-scores from Welch’s t-test to compare different data sets and identify statistically distinguishable behaviors, suggesting the presence of side-channel leakage.

In addition to employing TVLA for leakage assessment, promising efforts were made to utilize graph neural networks for analyzing pre-silicon PSC leakages \cite{Srivastava2024Scar}. In this framework, the RTL design was converted into control and data flow graphs (CDFG) to detect vulnerable hardware modules and code lines. The framework was tested on several symmetric and post-quantum cryptographic hardware implementations, highlighting the effectiveness compared to statistical methods.

Since TVLA operates as a conformance-style test and is a strong candidate for use in FIPS certification~\cite{fips}, it cannot quantify the extent of side-channel vulnerability in the same way as evaluation metrics used in Common Criteria (CC)~\cite{ccorg} certification, such as success rate (SR) and guessing entropy (GE). To address this gap between conformance-based (TVLA) and evaluation-based (CC) testing, \citeauthor{roy2018cc} introduced a framework that links the two domains by using TVLA to estimate the signal-to-noise ratio (SNR), which can then be used to compute the expected SR~\cite{roy2018cc}.

Besides the wide use of TVLA for information leakage assessment, \cite{standaert2018not} highlights several limitations of the fixed vs. random Welch's t-test. Specifically, \citeauthor{standaert2018not} demonstrates that this test is unable to reliably detect leakages in implementations protected with higher-order side-channel countermeasures, such as masking. In an $d$-th order masking scheme, the sensitive variable is split into $d{+}1$ shares, and leakage only appears through joint combinations of these shares. As noted in \cite{standaert2018not}, when the noise level increases, the number of traces required to detect such higher-order leakage grows exponentially with the number of shares.


\subsubsection{\bf Machine Learning (N-ML)}
\label{sec:analysis_non_profiled_information_machine_learning}

\citeauthor{Timon2019Non} introduced Differential Deep Learning Analysis (DDLA), a method that applies deep learning to non-profiled PSC attacks by leveraging convolutional neural networks (CNNs) ~\cite{Timon2019Non}. DDLA uses CNNs to handle de-synchronized power traces and extract secret keys from cryptographic devices such as the ChipWhisperer-Lite. The approach includes metrics based on sensitivity analysis to identify the secret key and points of interest like leakage points and mask locations in the traces, making it effective against masked implementations without needing prior knowledge of the masking techniques.

Building on the foundational work of DDLA, \citeauthor{Ahmed2023Optimization} presented optimization techniques to enhance the efficiency of these deep learning methodologies~\cite{Ahmed2023Optimization}. By employing parallel neural network structures and shared layers, this study reduced the time and memory required for training multiple models, thus speeding up the attack process. These optimizations were validated using ASCAD and ChipWhisperer-Lite datasets, showing improved performance and accuracy in extracting secret keys from power traces.

\citeauthor{Ahmed2023Deep} also applied deep learning methods, particularly CNNs, to PSC analysis, automating the process of signal re-alignment and noise reduction~\cite{Ahmed2023Deep}. This approach is validated on ECC by training CNNs to process misaligned power traces.

\subsubsection{\bf Other}
\label{sec:analysis_non_profiled_other}

In recent literature, several PSC attacks have emerged that utilize analysis methods not covered by traditional non-profiled categories. \citeauthor{Wang2022Hertzbleed}~\cite{Wang2022Hertzbleed} introduced a novel attack interface that exploits dynamic voltage and frequency scaling (DVFS) in modern Intel and AMD x86 CPUs, effectively transforming a PSC attack into a remote timing attack. In this approach, the attacker uses the median of a probability distribution to cluster observed trace values, which are then leveraged to recover the secret key from a SIKE implementation. In an extension of this work, \citeauthor{Wang2023Dvfs}~\cite{Wang2023Dvfs} applied a similar probability-distribution-based classification to exploit constant-time implementations of ECDSA and Classic McEliece, and also demonstrated a pixel-stealing side-channel attack. 

\citeauthor{Yu2024Hints} utilized the same DVFS-based vulnerability explored in \cite{Wang2022Hertzbleed} to attack the Number Theoretic Transform in a lattice-based key-encapsulation mechanism. By comparing the mean frequency values, they were able to determine whether the unknown input contained only one or two non-zero coefficients \cite{Yu2024Hints}. {Hot Pixels} proposed by \citeauthor{Taneja2023Hot} read software-accessible internal sensors on Arm SoCs and GPUs while running crafted workloads, showing instruction- and operand-dependent variations in the power telemetry without any external probes~\cite{Taneja2023Hot}. The authors induced heavy, color-dependent computation via an SVG-filter stack and then read out the throttling as timing, enabling pixel stealing, history sniffing, and even website fingerprinting without privileged access using a distribution-based classification.

\citeauthor{Sanjaya2025Sleepwalk}~\cite{Sanjaya2025Sleepwalk} proposed a cryptographic key recovery attack on SIKE and AES by exploiting residual power signatures generated during context switches, specifically when the built-in sleep function induces power spikes. A probability-distribution-based classification method was again employed to infer secret key bit values in SIKE and byte values in AES.

\citeauthor{Wei2018I} introduced a non-traditional {background-detection} analysis on CNN accelerators that exploits the observation that when data inside the convolution unit remains unchanged between cycles, internal transitions are limited and the per-cycle power stays low~\cite{Wei2018I}. Foreground and background have different power signatures because changing convolution windows over foreground pixels trigger more internal transitions, while repeated identical background pixels result in fewer transitions and lower dynamic power.
A per-cycle power histogram is constructed, and a threshold is chosen at the maximal drop in cycle counts to distinguish background cycles from foreground activity. 
Cycles below this threshold are then mapped back to pixel coordinates, producing a binary background mask that reveals the foreground outline of the input image~\cite{Wei2018I}.

\citeauthor{Moini2021Power} utilized a similar analysis method used in ~\cite{Wei2018I} to demonstrate a PSC attack on binarized neural network (BNN) accelerators in multi-tenant FPGA environments by leveraging on-chip time-to-digital converters (TDCs) to sense per-cycle voltage fluctuations~\cite{Moini2021Power}. 
This attack targeted the first convolution layer of the BNN, where foreground and background pixels create distinct power signatures due to differences in switching activity. 
The adversary constructed a histogram of voltage magnitudes and used a threshold to separate foreground from background pixels. 
This process enabled the reconstruction of input images, with denoising techniques further enhancing clarity~\cite{Moini2021Power}.

\citeauthor{Zhang2023Deep} exploited the fact that ReLU implementations consume different amounts of power depending on whether the input is positive or negative, leading to two separable energy distributions observable through Intel’s RAPL interface~\cite{Zhang2023Deep}. 
By repeatedly executing ReLU operations and constructing kernel density estimates of their power consumption, the attacker identifies the branch direction of each neuron and reconstructs the activation pattern of the network. 
With these activation patterns, they search for special input pairs, called flip pairs, that differ only in the activation of a single target neuron, enabling precise localization of the switching point. 
Comparing the input gradients of flip pairs yields direct equations for the associated weights, which are solved layer by layer, while the biases are subsequently derived from input-output relations once the weights are known. 
This density-distribution based analysis turns coarse RAPL power readings into a reliable signal for recovering the complete parameters of DL models.

In~\cite{Gatlin2021Encryption}, the authors captured synchronized multi-axis oscilloscope traces (X, Y, Z, E) from the printer’s stepper motor. The traces are low-pass filtered with a Butterworth filter, and peaks are detected with axis-specific height/prominence thresholds and logic to reject simultaneous or near-duplicate peaks. An algorithm was implemented to learn each axis’s firing order and direction reversals, segment the timeline into valid/invalid regions, and apply a heuristic edit step (peak insertions/deletions) to enforce motor-sequence validity before resuming detection. Finally, a state machine integrated the corrected peaks to estimate head position and extrusion state over time, producing a millimeter-scaled 3D point cloud of the toolpath that reconstructs the printed model.

\begin{figure}[htp]
    \begin{center}
        \includegraphics[width=0.49\textwidth]{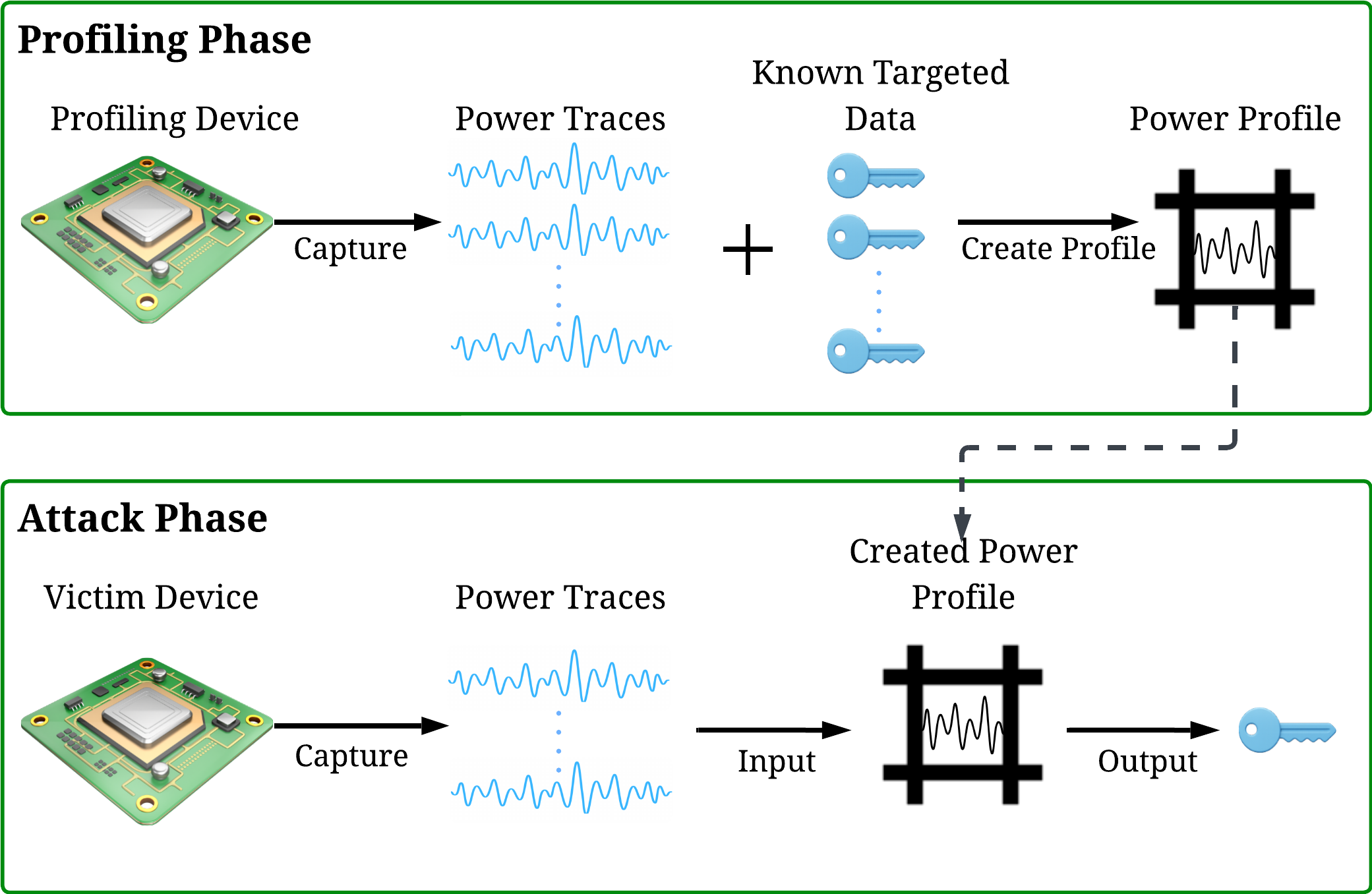}
    \end{center}
     \vspace{-0.1in}
      \caption{An overview of profiled power side-channel analysis that consists of two phases: the profiling phase and the attack phase.}
      \label{fig:profiling_attack_setup}
\end{figure}

\subsection{Profiled Side-Channel Analysis} 

Profiled side-channel analysis attacks leverage a detailed profile of the target device's power consumption behavior, typically obtained through a similar device under controlled conditions. This type of attack is considered more powerful and effective because it uses the pre-established profile to analyze the power traces of the actual target device. Figure \ref{fig:profiling_attack_setup} shows the two major phases for profiled side-channel analysis: the profiling phase and the attack phase. During the profiling phase, a detailed model of the device's power consumption patterns is created, often using statistical or machine learning techniques. During the attack phase, the power traces from the victim device is analyzed to extract the sensitive information. Two prominent types of profiled side-channel analysis are template attacks and ML-based attacks.

\subsubsection{\bf Template-based  Attacks}
\label{sec:analysis_profiled_template}


Template-based attacks were introduced by~\citeauthor{chari2002template}~\cite{chari2002template}. They rely on constructing precise statistical models of a device’s power consumption for each possible value of a secret intermediate, typically using multivariate Gaussian distributions whose mean and covariance matrices are learned during the profiling phase. A key component of template attacks is the identification of points of interest (POIs), which are the time samples most correlated with the secret-dependent leakage. During the attack phase, the attacker compares the observed power traces with the precomputed templates to determine the most likely value of the secret intermediate.

A stochastic-model-based attack enhances the classical DPA framework by replacing simple leakage predictions with a statistically learned approximation of the deterministic leakage and a shared noise model~\cite{schindler2005stochastic, mangard2007power}. During profiling, the attacker fits a low-dimensional model of how intermediate values leak and learns the noise distribution from traces collected under a single key, enabling more accurate hypothesis testing during the attack phase. By combining an approximate deterministic leakage model with a shared noise model, the attack with a stochastic model achieves key-extraction performance close to template attacks while requiring orders of magnitude fewer traces during profiling, making it suitable when profiling measurements are limited or costly.

\citeauthor{Wei2018I} proposed a template attack for active adversaries, where a ``power template'' is constructed by profiling the relationship between power consumption and pixel values across multiple kernels in an input image reconstruction attack~\cite{Wei2018I}. In the profiling stage, each cycle’s associated pixel set $Px_j$ and its corresponding power feature vector 
\[
\rho_j = (\rho_{0,j}, \ldots, \rho_{|K|,j}),
\] 
where $|K|$ denotes the number of kernels used, are collected to form a template:
\[
PT = \{(Px_j : \rho_j) \mid j \in \text{cycles}\}.
\]
The profiling phase is concluded by taking the union set of every input image template as the final power template. 
The image reconstruction in the attack phase selects consistent candidates across multiple cycles by minimizing the sum of the variance of pixel values for each position: 
\[
\min_{\text{Sel}} \sum_{x \in I} \text{var}\{pv(x) \mid pv(x) \in P_{x_{c_t}},\, c_t \in C_x\}.
\] 
,where $I$ is the set of pixel positions, $pv(x)$ denotes a candidate pixel value for position $x$, $P_{x_{c_t}}$ is the set of possible values derived from cycle $c_t$, and $C_x$ is the set of all cycles contributing information about pixel $x$. 
This variance-minimization ensures consistent reconstruction of each pixel, reducing uncertainty and enabling more accurate recovery of fine image details compared to simple threshold-based background detection.

\citeauthor{Nevskovic2023Systemc} also demonstrated a template attack within their SystemC-based evaluation framework for AI accelerators~\cite{Nevskovic2023Systemc}. 
In the profiling phase, the adversary assumes full control of the accelerator and builds templates by modeling the data-dependent power and noise distributions of each processing element, using chosen weights to characterize leakage behavior. 
During the attack phase, these templates are applied to traces from unknown weights, and candidate values are assigned by comparing measured power vectors against the pre-computed templates. 
This method enables recovery of all targeted weights with only a small number of traces and shows that template attacks remain effective even under noisy conditions.

\subsubsection{\bf Machine Learning based (P-ML) Attacks}
\label{sec:analysis_profiled_machine_learning}

ML–based attacks use ML models to classify power traces by learning discriminative patterns from profiling data. Commonly used models include Support Vector Machines (SVM)~\cite{heuser2012intelligent}, Random Forests (RF)~\cite{lerman2015machine}, neural networks~\cite{gilmore2015neural}, and deep neural networks (DNN)~\cite{maghrebi2016breaking}. These models are trained on labeled power traces from a device similar to the target, learning to distinguish between different key values based on subtle variations in the power consumption patterns. In profiled side-channel analysis, inputs, outputs, and secret data are controlled and known during the profiling phase, which helps create accurate models for the attack phase. The key advantage of profiled side-channel analysis attacks is their high success rate and efficiency due to the detailed profiling, making them particularly effective against well-studied and high-value targets.

DeepTheft~\cite{Gao2023Deeptheft} was an ML-based attack designed to extract DNN model architecture, operating in two phases: offline and online. In the offline phase, a set of meta-models was trained using a dataset of energy traces from various neural network architectures. These traces were collected through sampling via the powercap interface on a ML model similar to the victim's, with each sampling point labeled by its corresponding layer type. In the online phase, the target model was queried to capture its energy trace, and a two-step attack was implemented. First, a pre-trained meta-model from the offline phase segmented the energy trace to identify the network structure. Then, using the segmented points, multiple pre-trained meta-models and domain knowledge were applied to infer layer-wise hyperparameters.

A CNN was used to perform a black-box attack against the masking countermeasure implemented in AES for the DPAcontestV4~\cite{Ahmed2023Deep}. X-DeepSCA~\cite{Das2020Electromagnetic} introduced the concept of multi-device training, which uses traces from multiple devices and optimizes hyperparameters for a 256-class DNN. By training the DNN model with data from four different devices, X-DeepSCA achieved an accuracy of over 99.9\% across all four test devices, enabling effective single-trace attacks on an AES implementation.

The study conducted by authors \citeauthor{Ghandali2021Deep} focused on a new approach to profiled PSC attacks, employing a Twin Support Vector Machine (K-TSVM) with a deep kernel method~\cite{Ghandali2021Deep}. This approach enhanced the effectiveness of power analysis by pre-training with Restricted Boltzmann Machine methods and fine-tuning using gradient descent. The attack method was evaluated against both masked and unmasked hardware implementations of AES-128.

An attack involved in creating a database of power consumption templates for various DNN layer and hyper-parameter configurations by profiling the power traces of known DNN models was proposed in~\cite{Zhang2021Stealing}. Once the templates are established, the attacker matches the captured power traces against these templates to identify the corresponding DNN layers and their hyper-parameters. By using this template matching technique, the attack effectively reconstructed the sequence of layers and hyper-parameters of DNN models like MLP, AlexNet, and VGG16, achieving high accuracy.

The profiling phase in the attack proposed in \cite{Meyers2022Reverse} involved training on a known neural network configuration using RF classifiers. The authors used the tsfresh library for feature selection and extraction, obtaining a comprehensive set of features from time-series data. The classifiers were trained to identify folding parameters and neuron counts from power traces by using the top 10 features and top 50 features extracted from the tsfresh library, respectively. Once the model is trained, it is deployed on a target FPGA, where it accurately matches the power traces to the pre-learned profiles, successfully extracting the folding parameters and neuron counts.

A conditional Generative Adversarial Network (cGAN) was used to perform a PSC attack in~\cite{Wang2023Powergan}. The generator in the cGAN aims to reconstruct private input data from the power leakage measurements of the compute-in-memory system. It takes the preprocessed power trace data as input and tries to produce a realistic reconstruction of the input data. The discriminator, on the other hand, attempts to distinguish between the real power traces and the generated ones. The training of the cGAN involves the generator trying to mislead the discriminator, which helps refine the data reconstruction to achieve high-quality results even in the presence of noise. In the attack phase, the trained cGAN model was used to recover the user's private data using newly acquired power traces.

\begin{table*}[t]
\centering
\small
\caption{Summary of Power Analysis Methods.}
\label{tab:power_analysis_summary}
\begin{tabular}{>{\raggedright\arraybackslash}p{2.5cm} >
{\raggedright\arraybackslash}p{3.2cm} >{\raggedright\arraybackslash}p{5.3cm} >{\raggedright\arraybackslash}p{5.1cm}}
\rowcolor[HTML]{BDBDBD} 
\textbf{Method} & \textbf{Figures of Merit} & \textbf{Advantages} & \textbf{Disadvantages / Trade-offs} \\
\hline

SPA & Visual distinctness & Zero statistical overhead. Fast visual feedback. & Fragile to noise \& simple countermeasures. Limited to implementations with strong leakage. \\

\rowcolor{rowgray}
DPA & Difference of means & Effective with large traces; Extract information even from noisy power traces. No precise models required. & Needs multiple trace groups. Sensitive to trace alignment. Susceptible to higher-order countermeasures\\

CPA & Correlation coefficient, Success rate, Guess entropy, MTD & Generic. Widely used in cryptographic evaluations. Efficient. & Assumes an accurate power model. Requires sufficient power traces. \\

\rowcolor{rowgray}
Leakage Assessment & t-value, KL-Divergence & Key-independent leakage detection. Useful for verification. & Cannot directly recover the key. Accuracy depends on trace count and alignment. False-Positive scenarios \\

Non-profiled ML & Accuracy & Learns statistical patterns; does not require a power model & Lower accuracy than DL; needs manual feature extraction \\

\rowcolor{rowgray}
Non-profiled Other & Cluster distance, BCDC & Suited for remote attacks & Attack-specific heuristics. Limited generalizability. \\

Profiled ML & Recall & High success in noisy conditions. Robust to desynchronization and masking & Requires labeled traces. Computationally expensive to train. \\
\hline

\end{tabular}
\end{table*}

TransNet~\cite{Hajra2022Transnet} introduces a novel profiled side-channel analysis method based on a Transformer Network (TN) architecture that is both shift-invariant to trace misalignment and effective against masked implementations. Traditional profiled attacks often rely on well-aligned traces and tend to degrade in performance under high desynchronization. To overcome this limitation, TransNet incorporates a local embedding layer using one-dimensional convolution and a Transformer encoder, enabling the model to capture long-range dependencies and extract shift-invariant features. This architecture allows the network to autonomously learn leakage patterns without requiring manual preprocessing or desynchronization countermeasures. TransNet demonstrates strong performance in profiled attack scenarios where conventional alignment-based techniques struggle, making it a compelling alternative to existing CNN-based profiled attacks.

Building upon TransNet, EstraNet~\cite{Hajra2024Estranet} introduces a computationally efficient TN-based architecture for profiled side-channel attacks. One of the primary challenges addressed by \citeauthor{Hajra2024Estranet} is the scalability issue associated with long power traces in \cite{Hajra2022Transnet}. The proposed method achieves linear memory and time complexity with respect to trace length, while preserving the core principle of shift invariance to handle trace misalignments. EstraNet incorporates a novel attention mechanism that operates with linear computational overhead and introduces a new layer-centering normalization technique to overcome the limitations of standard normalization methods such as batch normalization and layer normalization. Similar to TransNet, EstraNet is trained on labeled power traces, and its learned representations are used to predict key-dependent features in unseen traces. Its computational efficiency and high classification accuracy make EstraNet a practical and effective solution for real-world profiled side-channel attacks targeting embedded cryptographic implementations.

In summary, this section presented a comprehensive overview of the major PSC analysis techniques, categorized into non-profiled and profiled methods. Non-profiled attacks, such as SPA, DPA, and CPA, rely on statistical techniques without prior leakage characterization, making them practical but often less efficient under noisy or desynchronized conditions. In contrast, profiled attacks leverage supervised learning with labeled traces to create templates on device-specific leakage, offering significantly higher accuracy, particularly with the advent of deep learning-based models such as CNNs and transformer architectures. The comparative summary in Table~\ref{tab:power_analysis_summary} highlights the advantages, disadvantages, and key performance metrics of each method. These foundational analysis techniques serve as the core analytical engines behind a broad spectrum of application-specific PSC attacks. In the following sections, we explore how these methods are adapted and applied across four distinct application domains, each presenting unique leakage characteristics, platform constraints, and attacker capabilities.



\section{PSC Attacks on Cryptographic Implementations}
\label{sec:crypto}

\begin{table*}[]
\caption{Summary of power side-channel attacks targeting hardware cryptographic implementations. The table includes details about the probing interface, Software (SW), Remote (RM), and Physical (PH), as well as the target platform, cryptographic algorithm, analysis method (non-profiled: SPA, DPA, CPA, TVLA, N-ML; profiled: P-ML), and the power model used.}
\centering
\small
\begin{tabular}{lccccc}
\rowcolor[HTML]{BDBDBD} 
\multicolumn{1}{c}{\cellcolor[HTML]{BDBDBD}\textbf{Reference}} &
  \textbf{Probing Interface} &
  \textbf{Target Platform} &
  \textbf{Target Algorithm} &
  \textbf{Analysis Method} &
  \textbf{Power Model} \\
\hline
\rowcolor[HTML]{FFFFFF} 
\citeauthor{Jayasankaran2023Securing}~\cite{Jayasankaran2023Securing} &
  RM &
  FPGA &
  AES &
  CPA &
  HD \\
\rowcolor[HTML]{E4E4E4} 
\citeauthor{Schellenberg2021Inside}~\cite{Schellenberg2021Inside} &
  RM &
  FPGA &
  AES &
  CPA &
  N/A \\
\rowcolor[HTML]{FFFFFF} 
\citeauthor{Gnad2020Remote}~\cite{Gnad2020Remote} &
  RM &
  FPGA &
  AES &
  CPA &
  HWT \\
\rowcolor[HTML]{E4E4E4} 
\citeauthor{Gravellier2019High}~\cite{Gravellier2019High} &
  RM &
  FPGA &
  AES &
  CPA &
  HD \\
\rowcolor[HTML]{FFFFFF} 
\citeauthor{Ramesh2018Fpga}~\cite{Ramesh2018Fpga} &
  RM &
  FPGA &
  AES &
  DPA &
  N/A \\
\rowcolor[HTML]{E4E4E4} 
\citeauthor{O2019Device}~\cite{O2019Device} &
  RM &
  \begin{tabular}[c]{@{}c@{}}MCU\\ TEE\end{tabular}  &
  AES &
  CPA &
  HWT \\
\rowcolor[HTML]{FFFFFF} 
\citeauthor{Knechtel2020Power}~\cite{Knechtel2020Power} &
  PH &
  ASIC &
  AES &
  CPA &
  HD \\
\rowcolor[HTML]{E4E4E4} 
\citeauthor{Moos2020Static}~\cite{Moos2020Static} &
  PH &
  ASIC &
  PRESENT &
  CPA &
  HWT \\
\rowcolor[HTML]{FFFFFF} 
\citeauthor{Moos2017Static}~\cite{Moos2017Static} &
  PH &
  ASIC &
  PRESENT-80 &
  CPA &
  HWT \\
\rowcolor[HTML]{E4E4E4} 
\citeauthor{Srivastava2024Scar}~\cite{Srivastava2024Scar} &
  PH &
  \begin{tabular}[c]{@{}c@{}}ASIC\\ FPGA\end{tabular} &
  \begin{tabular}[c]{@{}c@{}}AES\\ RSA\\ PRESENT\\ SABER\\ CRYSTALS-KYBER\end{tabular} &
  TVLA &
  HD \\
\rowcolor[HTML]{FFFFFF} 
\citeauthor{He2019Rtl}~\cite{He2019Rtl} &
  PH &
  \begin{tabular}[c]{@{}c@{}}ASIC\\ FPGA\end{tabular} &
  AES &
  TVLA &
  N/A \\
\rowcolor[HTML]{E4E4E4} 
\citeauthor{Pundir2022Power}~\cite{Pundir2022Power} &
  PH &
  \begin{tabular}[c]{@{}c@{}}ASIC\\ FPGA\end{tabular} &
  \begin{tabular}[c]{@{}c@{}}AES\\ PRESENT\\ SABER\end{tabular} &
  TVLA &
  \begin{tabular}[c]{@{}c@{}}HD\\ HWT\end{tabular} \\
\rowcolor[HTML]{FFFFFF} 
\citeauthor{Benhadjyoussef2021Power}~\cite{Benhadjyoussef2021Power} &
  PH &
  FPGA &
  AES &
  CPA &
  HD \\
\rowcolor[HTML]{E4E4E4} 
\citeauthor{Ghandali2021Deep}~\cite{Ghandali2021Deep} &
  PH &
  FPGA &
  AES &
  P-ML &
  N/A \\
\rowcolor[HTML]{FFFFFF} 
\citeauthor{Dumitru2023Borrowed}~\cite{Dumitru2023Borrowed} &
  PH &
  FPGA &
  AES &
  CPA &
  HWT \\
\rowcolor[HTML]{E4E4E4} 
\citeauthor{Miura2023Simulation}~\cite{Miura2023Simulation} &
  PH &
  FPGA &
  AES &
  TVLA &
  N/A \\
\rowcolor[HTML]{FFFFFF} 
\citeauthor{Zhang2020Memory}~\cite{Zhang2020Memory} &
  PH &
  FPGA &
  \begin{tabular}[c]{@{}c@{}}AES\\ PRESENT\end{tabular} &
  \begin{tabular}[c]{@{}c@{}}TVLA\\ CPA\end{tabular} &
  HD \\
\rowcolor[HTML]{E4E4E4} 
\citeauthor{Aysu2018Binary}~\cite{Aysu2018Binary} &
  PH &
  FPGA &
  Binary Ring-LWE &
  \begin{tabular}[c]{@{}c@{}}SPA\\ DPA\end{tabular} &
  N/A \\
\rowcolor[HTML]{FFFFFF} 
\citeauthor{Ahmadi2023Shield}~\cite{Ahmadi2023Shield} &
  PH &
  FPGA &
  RSA &
  SPA &
  N/A \\
\rowcolor[HTML]{E4E4E4} 
\citeauthor{Iyer2021Systematic}~\cite{Iyer2021Systematic} &
  PH &
  \begin{tabular}[c]{@{}c@{}}FPGA\\ ASIC\end{tabular} &
  AES &
  CPA &
  HD \\
\rowcolor[HTML]{FFFFFF} 
\citeauthor{Kamucheka2021Power}~\cite{Kamucheka2021Power} &
  PH &
  \begin{tabular}[c]{@{}c@{}}FPGA\\ MCU\end{tabular} &
  PQC &
  TVLA &
  N/A \\
\rowcolor[HTML]{E4E4E4} 
\citeauthor{Benhadjyoussef2021PowerAES}~\cite{Benhadjyoussef2021PowerAES} &
  PH &
  \begin{tabular}[c]{@{}c@{}}FPGA\\ MCU\end{tabular} &
  AES &
  CPA &
  HWT \\
\rowcolor[HTML]{FFFFFF} 
\citeauthor{Mozipo2023Residual}~\cite{Mozipo2023Residual} &
  PH &
  IoT &
  LWC &
  CPA &
  HD \\
\rowcolor[HTML]{E4E4E4} 
\citeauthor{Das2020Electromagnetic}~\cite{Das2020Electromagnetic} &
  PH &
  MCU &
  AES &
  P-ML &
  N/A \\ \hline
\end{tabular}
\label{tab:crypto_app_summary_hw}
\end{table*}

\begin{table*}[]
\caption{Summary of power side-channel attacks targeting software cryptographic implementations. The table includes details about the probing interface, Software (SW), Remote (RM), and Physical (PH), as well as the target platform, cryptographic algorithm, analysis method (non-profiled: SPA, DPA, CPA, TVLA, N-ML; profiled: P-ML), and the power model used.}
\centering
\small
\begin{tabular}{lccccc}
\rowcolor[HTML]{BDBDBD} 
\multicolumn{1}{c}{\cellcolor[HTML]{BDBDBD}\textbf{Reference}} &
  \textbf{Probing Interface} &
  \textbf{Target Platform} &
  \textbf{Target Algorithm} &
  \textbf{Analysis Method} &
  \textbf{Power Model} \\
\rowcolor[HTML]{FFFFFF} 
\citeauthor{Liu2022Frequency}~\cite{Liu2022Frequency} &
  SW &
  CPU &
  AES &
  \begin{tabular}[c]{@{}c@{}}CPA\\ TVLA\end{tabular} &
  Custom \\
\rowcolor[HTML]{E4E4E4} 
\citeauthor{Lipp2022Amd}~\cite{Lipp2022Amd} &
  SW &
  CPU &
  KASLR &
  SPA &
  N/A \\
\rowcolor[HTML]{FFFFFF} 
\citeauthor{Lipp2021Platypus}~\cite{Lipp2021Platypus} &
  SW &
  \begin{tabular}[c]{@{}c@{}}CPU\\ TEE\end{tabular} &
  \begin{tabular}[c]{@{}c@{}}AES-NI \\ RSA\\ KASLR\end{tabular} &
  \begin{tabular}[c]{@{}c@{}}SPA\\ CPA\end{tabular} &
  HD \\
\rowcolor[HTML]{E4E4E4} 
\citeauthor{Wang2022Hertzbleed}~\cite{Wang2022Hertzbleed} &
  RM &
  CPU &
  SIKE &
  Other &
  Custom \\
\rowcolor[HTML]{FFFFFF} 
\citeauthor{Wang2023Dvfs}~\cite{Wang2023Dvfs} &
  RM &
  CPU &
  \begin{tabular}[c]{@{}c@{}}ECDS\\ Classic McEliece\end{tabular} &
  Other &
  Custom \\
\rowcolor[HTML]{E4E4E4} 
\citeauthor{Yu2024Hints}~\cite{Yu2024Hints} &
  RM &
  CPU &
  Kyber &
  Other &
  Custom \\
\rowcolor[HTML]{FFFFFF} 
\citeauthor{Lee2020Dlddo}~\cite{Lee2020Dlddo} &
  PH &
  ASIC &
  AES &
  P-ML &
  N/A \\
\rowcolor[HTML]{E4E4E4} 
\citeauthor{Sanjaya2025Sleepwalk}~\cite{Sanjaya2025Sleepwalk} &
  PH &
  CPU &
  \begin{tabular}[c]{@{}c@{}}AES\\ SIKE\end{tabular} &
  Other &
  Custom \\
\rowcolor[HTML]{FFFFFF} 
\citeauthor{Ahmed2023Deep}~\cite{Ahmed2023Deep} &
  PH &
  FPGA &
  AES &
  N-ML &
  N/A \\
\rowcolor[HTML]{E4E4E4} 
\citeauthor{Benhadjyoussef2021PowerAES}~\cite{Benhadjyoussef2021PowerAES} &
  PH &
  \begin{tabular}[c]{@{}c@{}}FPGA\\ MCU\end{tabular} &
  \begin{tabular}[c]{@{}c@{}}AES\\ AES\end{tabular} &
  CPA &
  HWT \\
\rowcolor[HTML]{FFFFFF} 
\citeauthor{Kamucheka2021Power}~\cite{Kamucheka2021Power} &
  PH &
  \begin{tabular}[c]{@{}c@{}}FPGA\\ MCU\end{tabular} &
  PQC &
  TVLA &
  N/A \\
\rowcolor[HTML]{E4E4E4} 
\citeauthor{Luo2015Side}~\cite{Luo2015Side} &
  PH &
  GPU &
  AES &
  CPA &
  HD \\
\rowcolor[HTML]{FFFFFF} 
\citeauthor{Munny2021Power}~\cite{Munny2021Power} &
  PH &
  MCU &
  AES &
  CPA &
  HD \\
\rowcolor[HTML]{E4E4E4} 
\citeauthor{Jevtic2022Side}~\cite{Jevtic2022Side} &
  PH &
  MCU &
  AES &
  \begin{tabular}[c]{@{}c@{}}CPA\\ DPA\end{tabular} &
  N/A \\
\rowcolor[HTML]{FFFFFF} 
\citeauthor{Timon2019Non}~\cite{Timon2019Non} &
  PH &
  MCU &
  AES &
  N-ML &
  N/A \\
\rowcolor[HTML]{E4E4E4} 
\citeauthor{Ahmed2023Optimization}~\cite{Ahmed2023Optimization} &
  PH &
  MCU &
  AES &
  N-ML &
  N/A \\
\rowcolor[HTML]{FFFFFF} 
\citeauthor{Lee2020Security}~\cite{Lee2020Security} &
  PH &
  MCU &
  AES &
  \begin{tabular}[c]{@{}c@{}}CPA\\ DPA\end{tabular} &
  N/A \\
\rowcolor[HTML]{E4E4E4} 
\citeauthor{Saito2022One}~\cite{Saito2022One} &
  PH &
  MCU &
  RSA-CRT &
  P-ML &
  N/A \\ \hline
\end{tabular}
\label{tab:crypto_app_summary_sw}
\end{table*}


Among all the applications targeted by PSC attacks, cryptographic applications are the most prominent targets due to the severe damage that can result from compromising these systems. Since the SPA and DPA attacks conducted by \citeauthor{Kocher1999Differential} on DES implementations in 1999, extensive research has been undertaken to bypass or weaken the confidentiality and privacy provided by cryptographic algorithms~\cite{Kocher1999Differential}. These cryptographic applications can be further categorized into two major subcategories based on their implementation era. These are pre-quantum cryptographic algorithms, which include well-established algorithms such as AES, RSA, and SHA, and post-quantum cryptographic algorithms, which are emerging and aim to provide more secure encryption schemes in the quantum computing era. 

The National Institute of Standards and Technology (NIST) announced a competition aimed at finding encryption methods capable of resisting attacks from future quantum computers in 2016~\cite{moody2018let}. The initial candidates were grouped into two categories based on the algorithm: key-encapsulation mechanisms (KEM) and public-key encryption (PKE), and digital signatures (DS). In 2017, sixty-nine candidates were selected from the submissions for the first round~\cite{alagic2019status}. Then, twenty-six algorithms advanced to the second round, and only fifteen of those from the second round were selected for the third round in 2020~\cite{alagic2020status}. One algorithm (CRYSTALS-Kyber) from the KEM/PKE category and three algorithms (CRYSTALS-Dilithium, Falcon, SPHINCS+) from the DS category were selected for standardization, in 2022~\cite{alagic2022status}. Another four KEM/PKE algorithms (BIKE, Classic McEliece, HQC, SIKE) were advanced to the fourth round of the competition for further evaluation. The selected post-quantum cryptography (PQC) schemes have underlying mathematical formulations that were either lattice-based (CRYSTALS-Kyber, CRYSTALS-Dilithium, Falcon) or hash-based (SPHINCS+). The selected fourth-round candidates were based on code-based (BIKE, Classic McEliece, HQC) and supersingular elliptic curve isogeny-based (SIKE) methods. However, some third-round candidates, like GeMSS were based on multivariate mathematical formulation. Due to their vast use cases, there are several types of hardware and software implementations of the above cryptographic algorithms. Depending on the application, different implementations of the same cryptographic algorithm can be found.

\subsection{\textbf{Software-based Attacks}}

\citeauthor{Lipp2021Platypus} exploited the Intel Running Average Power Limit (RAPL) interface to gather low-resolution power consumption data to perform the side-channel attack PLATYPUS~\cite{Lipp2021Platypus}. RAPL provides an interface for managing the core's frequency and voltage, as well as tracking the power usage of the socket and memory domain. By statistically analyzing variations in power consumption, the PLATYPUS attacks were able to distinguish between different instructions, the hamming weights of operands, the hamming weight of the most significant bit of data values loaded from the cache, and the cache status of a load destination. This enabled the extraction of RSA private keys from mbedTLS, AES-NI keys from Intel SGX enclaves and the Linux kernel using CPA attacks, the breaking of kernel address-space layout randomization (KASLR), and the establishment of covert channels without requiring physical access to the target device.

A frequency throttling side-channel attack that shares a conceptual foundation with the PLATYPUS attack was presented in~\cite{Liu2022Frequency}. However, while PLATYPUS relied on direct power consumption telemetry, accessible via RAPL to perform a CPA, the frequency throttling side-channel attack leveraged the indirect timing information caused by power limit-induced frequency adjustments in the CPU. These adjustments, triggered when power consumption exceeds predefined thresholds, cause variations in execution time, which can then be correlated with the processed data.  The authors were able to successfully retrieve the complete key bytes from both kernel space and user space with a constant-cycle implementation of the AES algorithm, which utilized AES-NI instructions, by analyzing encryption execution times.

The Prefetch+Power~\cite{Lipp2022Amd} attack is a novel side-channel technique that exploits the power consumption patterns associated with the execution of prefetch instructions on AMD CPUs. By leveraging the RAPL interface, the attack measured the power consumption of prefetch instructions to leak information about the page-table level at which the page-table walk is aborted. This method was particularly effective for breaking KASLR by detecting differences in power consumption when accessing different kernel memory locations. The Prefetch+Power attack has the significant advantage of not requiring high-resolution timers, relying instead on the power consumption measurements provided by the RAPL interface, which can be accessed from unprivileged user space. 

\subsection{\textbf{Physical Attacks}}


In the early days of PSC attacks, attackers primarily focused on dynamic power. However, as we explained earlier, the total power consumption of a chip is the sum of both dynamic and static power. One of the main reasons for the initial negligence of static power is its relatively smaller contribution compared to dynamic power. However, with advancements in semiconductor manufacturing processes, the transistor density of a chip has increased exponentially, creating a new target for PSC attackers to exploit static power as a side channel.

Unlike dynamic power-based physical side-channel attacks, performing a physical static power-based side-channel attack presents two primary challenges. The first challenge is the relatively small voltage drop across the shunt resistor, and the second is sensitivity to temperature variations. \citeauthor{Moos2020Static} presented a measurement setup to carry out static PSC attacks by addressing the above challenges~\cite{Moos2020Static}. Additionally, \citeauthor{Moos2020Static} evaluated the effects of temperature, measurement interval, and supply voltage on static PSC attacks, concluding that by increasing the operating temperature and increasing the measurement interval, an attacker can force a device to leak more information~\cite{Moos2020Static}. 

Similarly, \citeauthor{Moos2017Static} focused on comparing static and dynamic PSC analysis of an ASIC implementation targeting a 150 nm CMOS prototype chip implementing the PRESENT-80 lightweight block cipher~\cite{Moos2017Static}. The authors successfully performed CPA and third-order CPA, collision-based Moments Correlating DPA (MCDPA), using the HWT power model on the F-box output for both dynamic and static power traces.

A conventional CPA-based PSC attack on an AES implementation on an 8051 MCU has been conducted~\cite{Benhadjyoussef2021Power}. In this attack, the authors used the HW power model for the AES first-round S-box output values. The authors claim that AES software implementations are more vulnerable to PSC attacks than AES hardware implementations.

\citeauthor{Das2020Electromagnetic} conducted a deep learning (DL)-based profiled attack by developing a cross-device  side-channel attack on cryptographic implementations in edge devices~\cite{Das2020Electromagnetic}. The proposed approach involved training a single deep neural network model using traces from multiple devices, combined with an optimal selection of hyperparameters for the 256-class DNN.

A profiled PSC attack on an 8-bit AVR microcontroller implementing a masked version of AES-128 was shown in~\cite{ Ghandali2021Deep}. By leveraging the DL-based K-TSVM approach, the attack exploits the power consumption patterns of the microcontroller to extract cryptographic keys. 

The residual vulnerabilities of lightweight cryptographic algorithms (LWC) were examined in~\cite{Mozipo2023Residual}. Despite the absence of integrated countermeasures, these LWCs showed vulnerabilities in their power consumption profiles, which can be exploited to leak sensitive information using CPA attacks. For ciphers like Ascon, Sparkle, and PHOTON-Beetle, the study considered scenarios involving chosen ciphertext attacks. On the other hand, the vulnerabilities of GIFT-COFB, Grain, Romulus, and TinyJambu were assessed using the HD power model. 

Interesting vulnerabilities were found in several experiments where the countermeasure exposes new vulnerabilities~\cite{Lee2020Security, Lee2020Dlddo, Saito2022One}. The study \cite{Lee2020Security} showed how attackers can leverage a vulnerability in AES implementations to distinguish between real and dummy operations when using the random insertion of dummy operations and shuffling as software countermeasures. By analyzing power traces with the Bounded Collision Detection Criterion (BCDC), the dummy operations were distinguished from real ones with high accuracy regardless of the dummy operation type. The authors explained that this distinction allows attackers to perform side-channel attacks with reduced complexity, significantly lowering the number of traces required for a successful attack.

\citeauthor{Lee2020Dlddo}~\cite{Lee2020Dlddo}, presented an advanced version of the above attack on AES software implementation that utilized dummy-load and shuffle-based countermeasures. They created a multi-label classification CNN to identify dummy S-box operations and subsequently performed a power analysis, such as DPA. Their results showed that the CNN-based detection method surpassed the accuracy of the approach proposed by \citeauthor{Lee2020Security}~\cite{Lee2020Security}. The same dummy-load vulnerability was utilized by \citeauthor{Saito2022One}~\cite{Saito2022One} to perform a DL-based single-trace PSC attack targeting the RSA–CRT (Chinese Remainder Theorem) implementation with windowed exponentiation. The primary focus was on open-source cryptographic libraries such as Gnu MP, which employ a dummy-load scheme as a countermeasure against side-channel attacks. The attack distinguished between true and dummy loads of the multiplicand using a neural network. This method converted the $2^w$-classification problem of finding the windowed exponentiation where the temporal window value is from 0 to $2^{w-1}$ into a $2^w \times 2$ classification problem.

The vulnerability exploited by \citeauthor{Sanjaya2024Information}~\cite{Sanjaya2024Information} analyzed power variations that propagate through shared voltage domains, allowing data extraction without direct access to the processor where the cryptographic operation occurs. Although this vulnerability was initially exploited in shared FPGAs, the paper showed that this vulnerability can be found in other hardware, including general-purpose MCUs in embedded systems. This novel physical layer supply voltage coupling (PSVC) vulnerability leaks sensitive data without requiring physical modifications to the embedded device, demonstrating PSVC as a viable attack vector for extracting information, even remotely. The authors also explained that this vulnerability can be used to perform attacks in both on-chip and on-board devices by demonstrating several SPA and CPA attacks on AES with different experimental setups.

An analysis of side-channel leakage of post-quantum algorithms was performed by \citeauthor{Kamucheka2021Power}~\cite{Kamucheka2021Power}, where three NIST PQC algorithms selected for standardization (CRYSTALS-KYBER, CRYSTALS DILITHIUM, and FALCON) and two third-round finalists (SABER, NTRU) were analyzed using TVLA. Software implementations of these algorithms, which can be found in the “pqm4” library, were implemented on an STM32F4 Discovery evaluation kit. The authors also proposed a platform for trace collection from software implementations of PQC algorithms.

{SleepWalk} exposes a sleep-induced context-switch power spike as a single-sample leakage source: reverse-engineering the user-space call chain shows that {sleep} ultimately invokes the kernel scheduler, and these calls yield the most prominent, attacker-friendly spikes~\cite{Sanjaya2025Sleepwalk}. Unlike full-trace attacks, the method records only the spike amplitude, eliminating trace alignment and external triggers while an off-the-shelf Raspberry~Pi 4 (Cortex-A72) with oscilloscope capture is used to automate the collection of peak values for analysis~\cite{Sanjaya2025Sleepwalk}. \citeauthor{Sanjaya2025Sleepwalk} showed that, by placing the sleep immediately after the Montgomery three-point ladder in SIKE, yields a clear mean-peak separation due to reduced register HWT when anomalous zeros occur; a practical variant triggers after decapsulation and amplifies residual effects via repeated decaps to recover the key~\cite{Sanjaya2025Sleepwalk}. For AES-128, a chosen-plaintext attack forces HWT~0 at selected final-round bytes and, after large-batch amplification before sleep, the mean-peak differences recover most round-10 key bytes, reducing brute-force effort while using only one point per trace~\cite{Sanjaya2025Sleepwalk}.


FPGA-based pre-quantum cryptographic implementations are well-explored attack targets in the PSC domain. For example, \citeauthor{Benhadjyoussef2021Power} attacked an AES-128 implementation on a Side-channel Attack Standard Evaluation Board (SASEBO) using CPA~\cite{Benhadjyoussef2021Power}. The authors used the last round S-box values as the target intermediate values. In a different setup, the attack proposed by \citeauthor{Ghandali2021Deep} targeted a Virtex-5 FPGA embedded on a SASEBO-GII board, evaluating both masked and unmasked implementations of AES-128~\cite{Ghandali2021Deep}. The deep K-TSVM model was employed to analyze power traces from the FPGA, effectively distinguishing between the masked and unmasked states. This version of the attack underscores the robustness of the deep learning model in profiling and identifying vulnerabilities in more complex hardware platforms, highlighting its capability to bypass both types of cryptographic protections.

A static PSC attack targeting cryptographic implementations on FPGA hardware was proposed~\cite{Dumitru2023Borrowed}. Specifically, it evaluated the effectiveness of a CPA attack on an AES-128 implementation within an FPGA. \citeauthor{Kamucheka2021Power} presented a TVLA analysis of PCQ implementation on an FPGA~\cite{Kamucheka2021Power}. They implemented CRYSTALS-Kyber, the selected lattice-based KEM/PKE PQC scheme for standardization. The authors demonstrated that there are some leakages at specific locations based on their TVLA results.

A hardware implementation of a quantum-secure encryption scheme based on the B-RLWE problem was introduced in~\cite{Aysu2018Binary}. The research detailed the vulnerabilities of the B-RLWE polynomial multiplication process, where partial product generation and intermediate sum updates create exploitable power leakage. During polynomial multiplication, each bit of the secret key polynomial determines whether the corresponding coefficients of the input polynomial are added to the intermediate sum. This step-by-step accumulation of products leads to discernible patterns in power consumption, which can be exploited by SPA and DPA attacks. The SPA attack leveraged the partial product generation and binary nature of B-RLWE coefficients to recover secret keys with minimal measurements. The DPA attack, on the other hand, utilized the values of intermediate sum updates in the B-RLWE implementation.


Performing PSC attacks on GPU implementations is a relatively difficult task compared to performing attacks on conventional CPUs due to the parallel processing nature of GPUs. Therefore, \citeauthor{Luo2015Side} proposed a method for capturing power traces, sampling power traces, and processing power traces to perform a CPA and extract the key from a CUDA AES implementation on an NVIDIA TESLA GPU~\cite{Luo2015Side}. The results highlight the potential risks in deploying cryptographic algorithms on GPUs without proper countermeasures, as the power consumption data can reveal encryption keys.

\subsection{\textbf{Remote Attacks}}

The `Hertzbleed'~\cite{Wang2022Hertzbleed} attack exploits DVFS in modern Intel and AMD x86 CPUs to turn a PSC attack into a remote timing attack. This attack targeted cryptographic software and leveraged the data-dependent frequency adjustments caused by DVFS to create a remote timing attack vector. The Hertzbleed attack was theoretically more powerful than traditional software side channels and has a coarse granularity that makes it challenging to exploit. However, through sophisticated, well-crafted inputs that amplify single key-bit guesses into thousands of high- or low-power operations, the authors were able to achieve full key extraction remotely from two production-ready SIKE implementations. This attack is particularly concerning as it undermines the foundations of constant-time programming, which has been a critical defense against timing attacks. \citeauthor{Wang2023Dvfs} expanded on the Hertzbleed attack, demonstrating its applicability beyond the SIKE cryptosystem to include other cryptosystems such as ECDSA and Classic McEliece running on modern x86 CPUs~\cite{Wang2023Dvfs}. The authors illustrated how even highly optimized constant-time cryptographic implementations can be vulnerable when subjected to Hertzbleed attacks.

\begin{figure}[t]
    \begin{center}
        \includegraphics[width=0.49\textwidth]{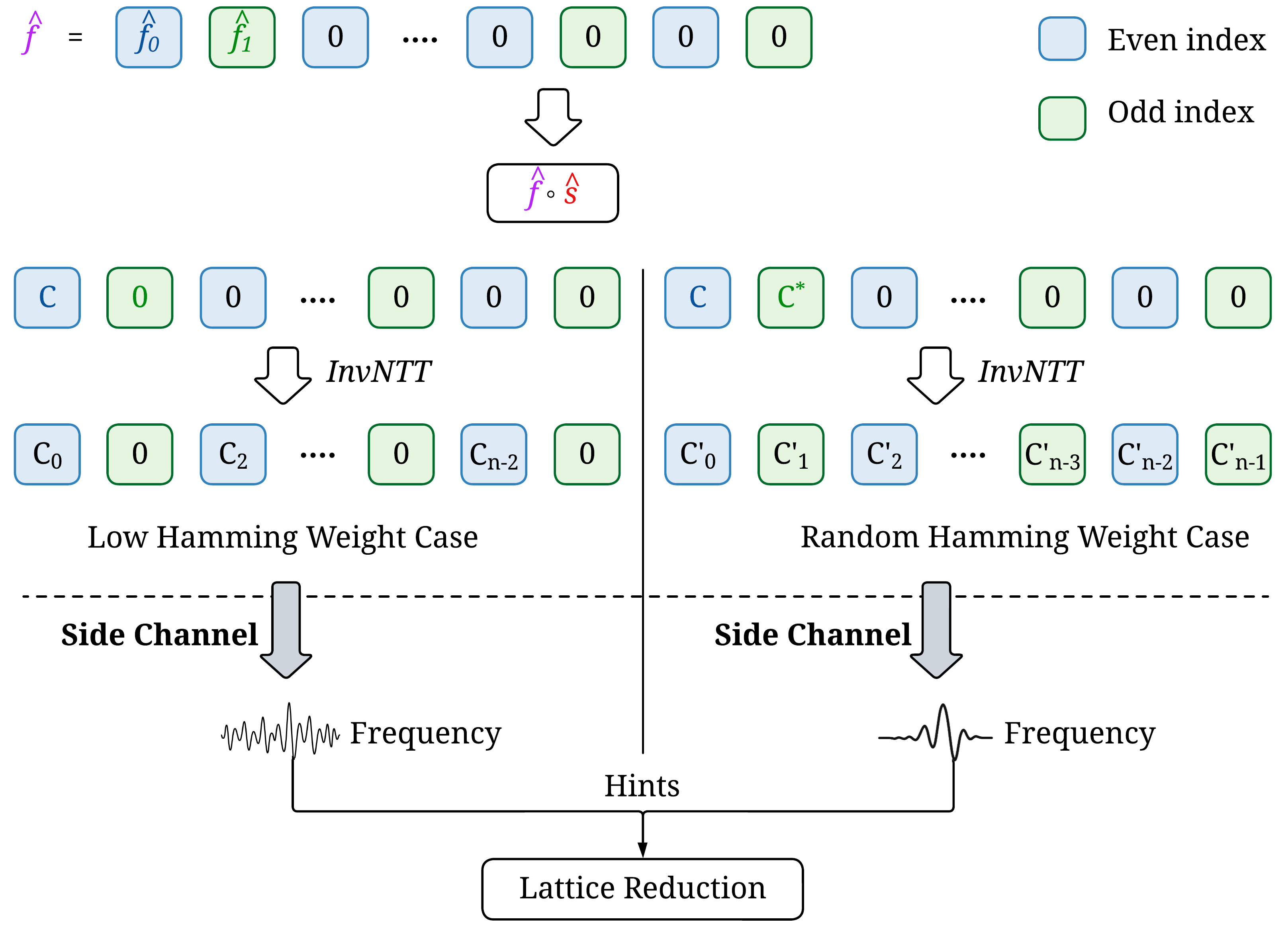}
    \end{center}
      \vspace{-0.1in}
      \caption{Attack idea of the DVFS-based frequency side-channel on lattice-based KEMs in \cite{Yu2024Hints}. The approach relies on the observation that when carefully crafted inputs \(\hat{f}\) are applied to the inverse Number Theoretic Transform $(invNTT)$ function, the resulting transformed outputs \({g} = {InvNTT}(\hat{f} \circ \hat{s})\) exhibit different Hamming weights. Here, \(\hat{s}\) denotes the secret key in the NTT domain, and only the last two coefficients of \(\hat{f}\) are non-zero: \(\hat{f} = (\hat{f}_0, \hat{f}_1, 0, \cdots, 0)\). By exploiting this frequency-dependent timing leakage, the attacker can infer the HWT of \(\hat{s}\) by controlling the \(\hat{f}\).}
      \label{fig:hints_from_hertz}
\end{figure}

\citeauthor{Yu2024Hints} investigated the side-channel leakage induced by DVFS on lattice-based cryptographic schemes, focusing on the Number Theoretic Transform (NTT) operations within CPA-secure Kyber and CCA-secure NTTRU~\cite{Yu2024Hints}. Similar to the~\cite{Wang2022Hertzbleed}, this work demonstrated how frequency scaling effects could be exploited to infer information about modular arithmetic operations central to lattice-based KEMs. As shown in Figure~\ref{fig:hints_from_hertz}, the attack leverages the fact that DVFS-induced frequency variations correlate with the HWT of intermediate NTT coefficients. 
By carefully analyzing execution-time variations correlated with DVFS-induced frequency adjustments, the authors were able to recover hints about intermediate NTT coefficients. This attack highlighted that even post-quantum cryptographic primitives, when implemented on DVFS-enabled CPUs, are vulnerable to software-based PSC attacks. This work extends the threat model beyond classical public-key cryptosystems, underscoring the need for hardware- and software-level countermeasures.

\begin{figure}[t]
    \begin{center}
        \includegraphics[width=0.45\textwidth]{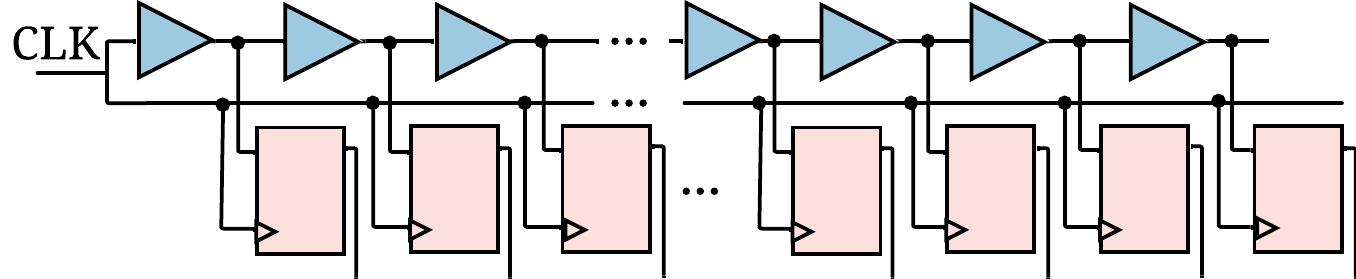}
    \end{center}
      \vspace{-0.1in}
      \caption{Time-to-digital circuit \cite{Jayasankaran2023Securing}.}
      \label{fig:fpga_remote_tdc}
\end{figure}

\begin{figure}[t]
    \centering
    
    


    \subfloat[]{\includegraphics[width=0.3\textwidth]{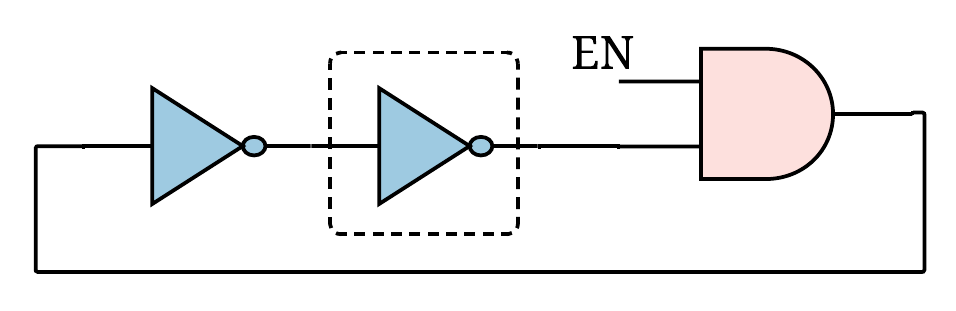}\label{fig:ro_1_fig}}\hskip1ex
    
    \subfloat[]{\includegraphics[width=0.3\textwidth]{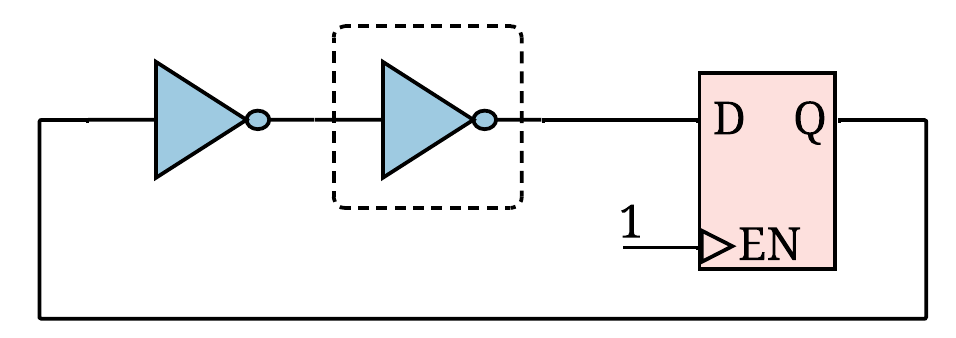}\label{fig:ro_2_fig}}\hskip1ex

    \subfloat[]{\includegraphics[width=0.3\textwidth]{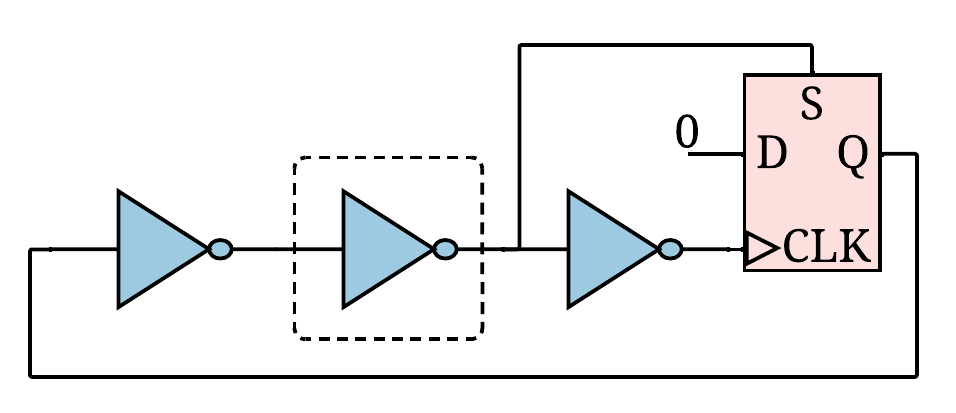}\label{fig:ro_3_fig}}\hskip1ex
    
    \caption{Different types of ring oscillators (RO). The dashed box indicates an even count of inverters. (a) A basic RO utilizing a combinatorial loop, with each inverter represented by a single LUT in the FPGA. (b) An RO design employing latches to prevent the formation of a combinatorial loop. (c) Another RO design that uses flip-flops to prevent combinatorial loops \cite{Jayasankaran2023Securing}.}
    \label{fig:fpga_remote_ro}
\end{figure}

Every remote PSC attack on FPGAs consists of a power monitoring logic design, such as time-to-digital converters (TDC), ring oscillators (RO), and analog-to-digital converters (ADC). Figure~\ref{fig:fpga_remote_tdc} shows an example design of a TDC. Figure~\ref{fig:fpga_remote_ro} depicts three different variants of ROs. These monitors are used to measure the voltage variations caused by cryptographic operations. Such attacks are very common in cloud FPGA environments, also referred to as multi-tenant setups, where the same FPGA  is shared among several users.

Another remote PSC attack was introduced by \citeauthor{Ramesh2018Fpga}, which requires no physical access to the target board \cite{Ramesh2018Fpga}.
They embedded an RO sensor array to measure on-chip voltage variations and align per-byte oscillator counts with the cycles when each ciphertext byte is produced \cite{Ramesh2018Fpga}.
Using the ciphertext and the inverse AES S-box, they built last-round key hypotheses and correlated them with the measured RO counts to recover key bytes.
This work shows that the FPGA fabric itself can serve as the sensing infrastructure, turning multi-tenant devices into self-contained side-channel benches \cite{Ramesh2018Fpga}.

\citeauthor{Schellenberg2021Inside} discussed an attack~\cite{Schellenberg2021Inside} that is capable of performing remote power analysis, which utilized TDC-based voltage sensors presented in~\cite{Zick2013Sensing}. The proposed attack was crafted from FPGA primitives to sense voltage fluctuations in the power distribution network (PDN). A standard CPA attack on AES-128 implementation on the side-channel evaluation platform SAKURA-G was performed using the traces collected using the proposed remote attack.

\citeauthor{Jayasankaran2023Securing} proposed a method to improve the sensitivity of TDC-based monitors by manually placing the TDC monitor in the FPGA~\cite{Jayasankaran2023Securing}. In this work, they analyze the effects of junction temperature and the placement of the victim cryptographic implementation on the attack success rate. They were able to perform a CPA attack on AES-128 with a smaller number of traces using the improved sensitivity of the TDC. Also, the authors demonstrate the repeatability of the attack.

\citeauthor{Gravellier2019High} proposed high-speed RO voltage sensors that enable nanosecond-scale monitoring of supply fluctuations in multi-tenant FPGAs \cite{Gravellier2019High}. Using these sensors, they performed a remote CPA against a co-located 50\,MHz AES hardware core and recover the key, demonstrating the first CPA conducted entirely with RO sensors inside the FPGA fabric \cite{Gravellier2019High}. They compared RO, TDC, and near-field EM setups, showing that on-chip proximity yields EM-comparable results despite lower sampling rates, while RO sensors nearly match TDC accuracy with lower area and simpler deployment \cite{Gravellier2019High}.

\citeauthor{Gnad2020Remote} showed that a tenant can mount a side-channel attack entirely from within the FPGA, undermining digital isolation in multi-tenant deployments \cite{Gnad2020Remote}.
With TDC-based on-chip sensors on a Lattice ECP5 board, they executed a CPA against an AES core and recovered roughly half of the key bytes with up to $10^5$ traces gathered in under three hours, proving the feasibility of remote power analysis \cite{Gnad2020Remote}.
They further demonstrated that deliberately toggling on-chip oscillators at 1–3\,MHz induces timing faults in AES and summarize successful attacks across multiple FPGA families \cite{Gnad2020Remote}.

\citeauthor{O2019Device} mounted an on-device, cross-domain power side-channel attack on TrustZone-M, which is the Trusted Execution Environment (TEE) on ARM Cortex-M microcontrollers, by running non-secure code that triggers secure-world encryptions while the chip’s own ADC samples the core supply voltage, eliminating any need for external probes~\cite{O2019Device}.
The attacker arms the ADC immediately before invoking the secure AES routine, turning standard peripherals into a built-in oscilloscope. The primary target was the hardware AES accelerator in the secure world, and the authors further show that controlling the ADC entirely from the non-secure world still enables key-recovery analysis, demonstrating cross-domain leakage~\cite{O2019Device}.


\subsection{\textbf{Summary of  Attacks and Insights}}
In Table~\ref{tab:crypto_app_summary_hw} and Table~\ref{tab:crypto_app_summary_sw}, we provide an overall summary of the similarities and differences between recent PSC attacks targeting cryptographic applications. This summary highlights the vast array of attack options used in the literature, including targeted hardware, cryptographic algorithms, analysis methods, and power models.

Beyond the traditional focus on physical probing, recent developments in software-based PSC attacks have significantly reshaped the threat landscape for cryptographic implementations. Interfaces such as Intel's RAPL and NVIDIA's NVML enable attackers to obtain power-related telemetry without requiring physical access to the victim device, making software and remote PSC attacks considerably more practical than the classical oscilloscope-based power trace acquisition setup. Their accessibility, scalability, and ability to operate in virtualized or cloud environments broaden the adversarial model and eliminate the laboratory constraints that historically limited physical PSC attacks. 

Meanwhile, the emergence of PQC algorithms has further expanded the scope of PSC research beyond classical primitives such as AES and RSA. Although PQC schemes offer strong mathematical guarantees against quantum adversaries, recent works demonstrate that their implementations still exhibit data-dependent leakage, indicating that PSC vulnerabilities may persist even in the post-quantum era. At the same time, advancements in ML, particularly profiled models capable of extracting shift-invariant features from noisy or misaligned traces, have enhanced the capabilities of modern PSC attacks by improving efficiency and noise tolerance. 

Despite these developments, the literature continues to focus predominantly on symmetric-key algorithms, while asymmetric primitives and PQC implementations remain comparatively underexplored due to their algorithmic complexity, diverse arithmetic structures, and challenging leakage characteristics. Collectively, these trends illustrate how evolving software interfaces, emerging cryptographic standards, and ML-driven analysis methods are expanding both the practicality and the reach of PSC attacks well beyond their classical boundaries.

\section{User-Specific Data Privacy}
\label{sec:user}

Another notable category of PSC attacks targets user behavior data, such as website fingerprinting. These attacks primarily focus on mobile and embedded devices, which frequently involve user interactions. By analyzing the power consumption patterns during user activities, attackers can infer detailed user behavior and patterns. Given the nature of these attacks, most are profiled PSC attacks, where the attacker builds a profile of power consumption patterns to effectively identify and exploit user behaviors. This makes them particularly dangerous, as they can be used to track user activities, infer visited websites, and compromise user privacy without needing direct access to the data being transmitted or stored.

\begin{table*}[]
\caption{Summary of power side-channel attacks targeting user behavior data extraction. The table includes details about the probing interface: Software (SW), Remote (RM), and Physical (PH), as well as the target platform and extracted user behavior data. 
}

\center
\small
\begin{tabular}{lccccc}
\rowcolor[HTML]{BDBDBD} 
\multicolumn{1}{c}{\cellcolor[HTML]{BDBDBD}\textbf{Reference}} &
  \textbf{\begin{tabular}[c]{@{}c@{}}Probing \\ Interface\end{tabular}} &
  \textbf{\begin{tabular}[c]{@{}c@{}}Targeted \\ Platform\end{tabular}} &
  \textbf{Extracted Information} &
  \textbf{\begin{tabular}[c]{@{}c@{}}Preprocessing \\ Method\end{tabular}} &
  \textbf{Attack Model} \\
  \hline
\rowcolor[HTML]{FFFFFF} 
\citeauthor{Zhang2021Red}~\cite{Zhang2021Red} &
  SW &
  CPU &
  Website fingerprinting &
  Trace alignment &
  \begin{tabular}[c]{@{}c@{}}KNN\\ CNN\end{tabular} \\
\rowcolor[HTML]{E4E4E4} 
\citeauthor{Chen2017Powerful}~\cite{Chen2017Powerful} &
  SW &
  Smartphones &
  Mobile app fingerprinting &
  \begin{tabular}[c]{@{}c@{}}Segmentation\\ Power adjustment\\ Min-max search\end{tabular} &
  \begin{tabular}[c]{@{}c@{}}C4.5\\ RF\\ SVM\end{tabular} \\
\rowcolor[HTML]{FFFFFF} 
\citeauthor{Kogler2023Collide+}~\cite{Kogler2023Collide+} &
  \begin{tabular}[c]{@{}c@{}}SW\\ RM\end{tabular} &
  CPU &
  Data in memory hierarchy &
  N/A &
  Non-profiled CPA \\
\rowcolor[HTML]{E4E4E4} 
\citeauthor{Taneja2023Hot}~\cite{Taneja2023Hot} &
  RM &
  \begin{tabular}[c]{@{}c@{}}CPU\\ GPU\end{tabular} &
  \begin{tabular}[c]{@{}c@{}}Website fingerprinting\\ Image pixel value\end{tabular} &
  N/A &
  Non-profiled Other \\
\rowcolor[HTML]{FFFFFF} 
\citeauthor{Wang2023Dvfs}~\cite{Wang2023Dvfs} &
  RM &
  CPU &
  Image pixel value &
  N/A &
  Non-profiled Other \\
\rowcolor[HTML]{E4E4E4} 
\citeauthor{La2021Wireless}~\cite{La2021Wireless} &
  PH &
  Smartphones &
  \begin{tabular}[c]{@{}c@{}}Website fingerprinting\\ Passcode length\\ OLED screen activity\\ Audio Fingerprinting\end{tabular} &
  Segmentation &
  \begin{tabular}[c]{@{}c@{}}1D CNN\\ RF\end{tabular} \\
\rowcolor[HTML]{FFFFFF} 
\citeauthor{Gatlin2021Encryption}~\cite{Gatlin2021Encryption} &
  PH &
  3D Printers &
  3D model reconstruction &
  Noise filtering &
  Non-profiled Other \\
\rowcolor[HTML]{E4E4E4} 
\citeauthor{Yang2016Inferring}~\cite{Yang2016Inferring} &
  PH &
  Smartphones &
  Website fingerprinting &
  Segmentation &
  RF - WEKA \\
\rowcolor[HTML]{FFFFFF} 
\citeauthor{Matovu2020Defensive}~\cite{Matovu2020Defensive} &
  PH &
  Smartphones &
  \begin{tabular}[c]{@{}c@{}}Website fingerprinting\\ Keystroke inference\\ Incoming call detection\\ Camera usage detection\end{tabular} &
  Segmentation &
  RF, SVM \\
\rowcolor[HTML]{E4E4E4} 
\citeauthor{Cronin2021Charger}~\cite{Cronin2021Charger} &
  PH &
  Smartphones &
  \begin{tabular}[c]{@{}c@{}}Touchscreen activities\\ Passcode inference\end{tabular} &
  High pass filter &
  1D CNN \\
\rowcolor[HTML]{FFFFFF} 
\citeauthor{Yucebas2018Power}~\cite{Yucebas2018Power} &
  PH &
  VLC systems &
  Binary data &
  N/A &
  Non-profiled SPA \\
\rowcolor[HTML]{E4E4E4} 
\citeauthor{Spolaor2023Plug}~\cite{Spolaor2023Plug} &
  PH &
  USB peripherals &
  Peripheral type and identity &
  Segmentation &
  RF \\
\rowcolor[HTML]{FFFFFF} 
\citeauthor{Qin2018Website}~\cite{Qin2018Website} &
  PH &
  Smartphones &
  Website fingerprinting &
  \begin{tabular}[c]{@{}c@{}}Trace alignment\\ Resampling\end{tabular} &
  SVM \\
  \hline
\end{tabular}
\label{tab:user_data_app_summary}
\end{table*}

\subsection{\textbf{Software-based Attacks}}

Intel's RAPL interface can be exploited to conduct website fingerprinting attacks. By monitoring the fine-grained power consumption data available through RAPL, attackers can infer which websites a user is visiting. This approach relies on the distinctive energy usage patterns generated during the rendering of different web pages, which can be captured and analyzed to identify specific websites with high accuracy. \citeauthor{Zhang2021Red} demonstrated that using this method, it is possible to achieve up to 99\% accuracy for regular network scenarios and 81\% accuracy when the Tor browser is used~\cite{Zhang2021Red}.

POWERFUL~\cite{Chen2017Powerful} demonstrated a method for fingerprinting mobile applications by analyzing the power consumption profiles on Android devices. By capturing the distinct power consumption patterns associated with different apps, the authors inferred which app a user is currently using. To collect power data, the authors implemented an Android application that periodically reads voltage and current measurements from publicly accessible system files at a frequency of 2 Hz. The collected data was processed and used to build models using three lightweight supervised classifier techniques, which can accurately identify which app was being used based on real-time power consumption data.

\citeauthor{Kogler2023Collide+} introduced {Collide+Power}, a software-based PSC that exploits the mere co-location of attacker- and victim-controlled values in the CPU memory hierarchy to create exploitable, data-dependent power differences~\cite{Kogler2023Collide+}. The attacker primes a target cache set with a chosen guess \(G\), lets the victim access the secret value \(V\), and observes that the energy used when a cache line transitions from \(G\) to \(V\) scales with the \(HD(G,V)\). This leakage is observable either via power interfaces (e.g., RAPL) or indirectly through throttling-induced timing variations, enabling unprivileged attacks. They model the leakage as a linear combination of HD and HWT, and amplify the signal with a differential measurement: measure once with \(G\) and once with the masked inverse \(\tilde{G}=G\oplus m\), then subtract, which cancels constant terms, doubles the HD component, improving the SNR by \(\approx 8.8\times\) in practice~\cite{Kogler2023Collide+}. Instantiations include {MDS-Power} (leaking “data-in-use” from a sibling hyperthread) and {Meltdown-Power} (leaking “data-at-rest” via kernel prefetch gadgets), illustrating that Collide+Power is generic, independent of a specific cryptographic algorithm, and effective with either direct power readings or timing proxies.

\subsection{\textbf{Physical Attacks}}

A profiled attack to identify web browsing activities on smartphones by analyzing power consumption during USB charging was employed in~\cite{Yang2016Inferring}. By training an RF classifier on power traces of the Alexa top 50 websites on multiple smartphones, the authors developed a model that can accurately identify websites visited by the user. The attack involved feature extraction from the frequency domain of power traces, allowing the classifier to recognize distinct power signatures associated with different websites. The attack was evaluated under different environmental conditions, such as varying battery levels, enabling or disabling the browser cache, using Wi-Fi or LTE, and background app activities, to ensure robustness.

Another profiled attack was implemented by \citeauthor{Qin2018Website} to identify websites visited on Android 7 smartphones~\cite{Qin2018Website}. The authors collected CPU load and frequency data while measuring power data to estimate power consumption without requiring user permission. The linear regression-based power estimation model was used to perform the attack to identify web browsing activities. By training an SVM classifier on the estimated power traces, the authors achieved high accuracy in distinguishing between different websites.

\citeauthor{Matovu2020Defensive} explored three distinct PSC attacks on charging smartphones~ \cite{Matovu2020Defensive}: website inference, keystroke inference, and incoming call detection. In these scenarios, the authors leverage the power consumption data captured while the smartphone is charging to infer sensitive information. For instance, the website inference attack aims to identify which websites the user is visiting, while the keystroke inference attack attempts to determine the keys being pressed to extract passwords. The attacks are implemented on two different Samsung Galaxy phones during the charging process, in a public charging station controlled by an adversary.

The Charger-surfing attack~\cite{Cronin2021Charger} exploited PSC leakage to extract sensitive information from smartphones during charging. This attack targeted dynamic content on the smartphone touchscreen by analyzing power consumption patterns, specifically during touchscreen animations such as button presses. By capturing power traces during touchscreen activities, the authors trained neural network models to classify the location and type of animations displayed on the screen. The attack was tested under various environmental conditions, such as different battery levels and screen technologies (LCD and OLED), to ensure robustness and effectiveness across a wide range of scenarios.

A method to exploit PSC attacks during wireless charging, demonstrating that activities such as website loading on smartphones can be inferred by monitoring the power consumption of a Qi wireless charging transmitter was proposed~\cite{La2021Wireless}. The authors captured power traces and successfully identified websites loaded on iPhone 11 and Google Pixel 4 devices by analyzing the current drawn during wireless charging without needing direct device access. This was done using a CNN classifier to analyze these power traces and identify specific websites loaded on the device based on the power consumption patterns during wireless charging.

PowerID was a framework proposed for demonstrating the feasibility of fingerprinting USB peripherals by analyzing their power consumption patterns~\cite{Spolaor2023Plug}. The evaluation was conducted using power traces from 82 USB peripherals, covering 35 models across 8 types, including flash drives, external hard drives, WiFi adapters, Bluetooth adapters, microphones, webcams, mice, and keyboards. Using RF classifiers, the authors developed profiles that could distinguish between different models and types of USB peripherals with high accuracy. These profiles were used to analyze the power traces to accurately identify the type, model, activity, and identity of each USB peripheral.

A method to reconstruct 3D-printed models by exploiting PSC leakages from Fused Deposition Modeling (FDM) 3D printers was demonstrated in~\cite{Gatlin2021Encryption}. By analyzing the power consumption of the printer's stepper motors, the authors developed a model that correlates specific current draw patterns with the movement of the printhead, such as reversal, start, stop, and dwell. Using this profiled data, the toolpath and subsequently the entire 3D model were accurately reconstructed. This attack showed that even with end-to-end encryption, the PSC provides sufficient information to bypass security measures and extract sensitive design data in environments like outsourced additive manufacturing (AM).

Security vulnerabilities of visible light communication (VLC) systems through PSC attacks were investigated by~\cite{Yucebas2018Power}. The authors designed a bidirectional VLC system and implemented a current sensor on the power lines from the power supply to the VLC transmitter. The attack exploited the power variation caused by the LEDs rapidly switching on and off to transmit binary data, resulting in distinct power levels. By analyzing the instantaneous power consumption during data transmission, the transmitted binary data was successfully extracted. 

\subsection{\textbf{Remote Attacks}}

\citeauthor{Wang2023Dvfs} demonstrated how the Hertzbleed attack can be adapted into a pixel-stealing attack on modern Intel and AMD x86 CPUs~\cite{Wang2023Dvfs}. This attack exploits the variation in power consumption associated with rendering different pixel colors, which leads to detectable changes in CPU frequency. These fluctuations are caused by the data-dependent power usage of the integrated GPU (iGPU) when applying SVG filters to graphical data. Here, instead of relying on an approach where the CPU frequency is directly observed, the authors took advantage of how CPU frequency influences the number of JavaScript operations that can be executed within a given time frame. The proof-of-concept demonstrated in the study successfully extracted pixel values at a rate of 1 to 3 pixels per second, with a relatively low error rate, depending on the system configuration in the latest version of Google Chrome.

\citeauthor{Taneja2023Hot} utilized a similar approach to the Hertzbleed attack to perform a pixel-stealing attack that exploits the DVFS mechanism in integrated and discrete GPUs (iGPUs and dGPUs), as well as ARM  SoCs~\cite{Taneja2023Hot}. The Hot Pixels attack was executed through a JavaScript-based approach, where an attacker applies an SVG filter to a targeted pixel in a cross-origin iframe. This attack can recover images from cross-origin iframes and extract browsing history in web browsers like Chrome and Safari, even with all side-channel countermeasures enabled. Additionally, the authors performed a website fingerprinting attack using Apple iGPUs. This attack leveraged the fact that different websites induce distinct patterns of computation on the GPU, leading to unique power and frequency fluctuations that can be measured via internal sensors through unprivileged software.


\subsection{\textbf{Summary of  Attacks and Insights}}
Table~\ref{tab:user_data_app_summary} summarizes key information from our survey of PSC attacks targeting user behavior data extraction. The table includes details such as the probing interface, targeted hardware, extracted user behavior data, trace preprocessing methods, and machine learning models used to perform the profiled attacks. It highlights the similarities and differences among recent approaches. As the majority of the surveyed attacks on user behavior data are profiled, power model information is not included. Instead of categorizing by power analysis methods, this table focuses on preprocessing techniques and the profiled models employed in each study.

A notable trend in PSC attacks targeting user behavior data is the predominance of fingerprinting-based approaches, with website fingerprinting being the most widely explored. These attacks leverage the strong correlation between high-level software activity and power consumption patterns, enabling profiled classifiers to distinguish between websites based on their unique computational and rendering characteristics. This close alignment between fingerprinting tasks and profiled analysis methods stems from the fact that such attacks rely less on explicit power models and more on learned representations that capture subtle statistical differences in power traces. 

While existing works have primarily focused on website identification, similar fingerprinting principles could be extended to a broader range of behavioral contexts, such as application usage states, background service activity, or device-specific interaction patterns. In addition to website identification, surveyed studies show that PSC attacks are capable of extracting even fine-grained touchscreen-related behaviors, including touch positions, swipe patterns, and interaction timing, demonstrating that modern user interfaces expose rich and highly distinguishable side-channel leakage.

\section{Reverse Engineering of Functionality}
\label{sec:instruction}

\begin{table*}[]
\caption{Summary of power side-channel attacks targeting instruction disassembly. The table includes details about the probing interface: Software (SW), Remote (RM), and Hardware (HW), as well as the target platform and the type of extracted instruction-level information. 
}
\center
\small
\begin{tabular}{lcccccc}
\rowcolor[HTML]{BDBDBD} 
\multicolumn{1}{c}{\cellcolor[HTML]{BDBDBD}\textbf{Reference}} &
  \textbf{\begin{tabular}[c]{@{}c@{}}Probing \\ Interface\end{tabular}} &
  \textbf{\begin{tabular}[c]{@{}c@{}}Targeted \\ Platform\end{tabular}} &
  \textbf{\begin{tabular}[c]{@{}c@{}}Extracted \\ Information\end{tabular}} &
  \textbf{\begin{tabular}[c]{@{}c@{}}Preprocessing \\ Method\end{tabular}} &
  \textbf{\begin{tabular}[c]{@{}c@{}c@{}}Feature \\ Extraction \\ Method\end{tabular}} &
  \textbf{Attack Model} \\ \hline
\rowcolor[HTML]{FFFFFF} 
\citeauthor{Lipp2021Platypus}~\cite{Lipp2021Platypus} &
  SW &
  CPU &
  Instructions &
  Median &
  N/A &
  Non-profiled Other \\
\rowcolor[HTML]{E4E4E4} 
\citeauthor{Glamovcanin2023Instruction}~\cite{Glamovcanin2023Instruction} &
  RM &
  Soft-Core CPU &
  Instructions &
  CWT &
  \begin{tabular}[c]{@{}c@{}}PCA\\ LDA\end{tabular} &
  \begin{tabular}[c]{@{}c@{}}LSTM\\ CNN\\ MLP\\ ResNet\end{tabular} \\
\rowcolor[HTML]{FFFFFF} 
\citeauthor{Krishnankutty2020Instruction}~\cite{Krishnankutty2020Instruction} &
  PH &
  \begin{tabular}[c]{@{}c@{}}Soft-Core CPU\\ MCU\end{tabular} &
  Instructions &
  Filtering &
  DP &
  SVM \\
\rowcolor[HTML]{E4E4E4} 
\citeauthor{Fendri2022Deep}~\cite{Fendri2022Deep} &
  PH &
  CPU &
  Instructions &
  Alignment &
  SDL &
  MLP \\
\rowcolor[HTML]{FFFFFF} 
\citeauthor{Han2022Hiding}~\cite{Han2022Hiding} &
  PH &
  MCU &
  Control flow &
  N/A &
  N/A &
  \begin{tabular}[c]{@{}c@{}}BiLSTM\\ BiRNN\\ HMM\end{tabular} \\
\rowcolor[HTML]{E4E4E4} 
\citeauthor{Park2018Power}~\cite{Park2018Power} &
  PH &
  MCU &
  Instructions &
  CWT &
  \begin{tabular}[c]{@{}c@{}}KLD\\ PCA\end{tabular} &
  \begin{tabular}[c]{@{}c@{}}LDA\\ QDA\\ SVM\\ Naive Bayes\end{tabular} \\
\rowcolor[HTML]{FFFFFF} 
\citeauthor{Vafa2020Efficient}~\cite{Vafa2020Efficient} &
  PH &
  MCU &
  Instructions &
  CWT &
  \begin{tabular}[c]{@{}c@{}}KLD\\ PCA\end{tabular} &
  Ensemble of Classifiers \\
\rowcolor[HTML]{E4E4E4} 
\citeauthor{Narimani2021Side}~\cite{Narimani2021Side} &
  PH &
  MCU &
  Instructions &
  N/A &
  \begin{tabular}[c]{@{}c@{}}STFT\\ MFCC\end{tabular} &
  CNN \\
\rowcolor[HTML]{FFFFFF} 
\citeauthor{Van2022Side}~\cite{Van2022Side} &
  PH &
  MCU &
  Instructions &
  Average &
  KLD &
  \begin{tabular}[c]{@{}c@{}}LDA\\ QDA\\ MLP\\ CNN\end{tabular} \\
\rowcolor[HTML]{E4E4E4} 
\citeauthor{Bae2022Implementation}~\cite{Bae2022Implementation} &
  PH &
  MCU &
  Opcode &
  \begin{tabular}[c]{@{}c@{}}CWT\\ Alignment\end{tabular} &
  KLD &
  \begin{tabular}[c]{@{}c@{}}RF\\ KNN\\ MLP\\ CNN\end{tabular} \\ \hline
\end{tabular}
\label{tab:disassembly_app_summary}
\end{table*}

PSC-based disassembly attacks represent a sophisticated and emerging threat in cybersecurity, leveraging the analysis of power consumption patterns to deduce the internal operations of a target device. These attacks focus on extracting assembly-level instructions by closely monitoring the power usage during the execution of software on a device. By analyzing variations in power consumption, attackers can identify the sequence of executed instructions, effectively ``disassembling" the software without needing direct access to the code. This type of attack is particularly concerning for embedded systems and mobile devices, where the integrity and confidentiality of the software are important. PSC-based disassembly attacks can be used to reverse-engineer proprietary algorithms, identify vulnerabilities, and even modify the behavior of the software, posing a significant risk to the security and functionality of affected devices.

\subsection{\textbf{Software-based Attacks}}

In PLATYPUS attack~\cite{Lipp2021Platypus}, even though the main attacks targeted cryptographic applications, those attacks utilized the results of Intel RAPL leakage analysis to distinguish between different instructions, HWT of operands, HWT and the value of the most significant bit of data values loaded from the cache, and the cache status of a load destination. Therefore, the methodology used in the PLATYPUS attack can also be used to perform instruction disassembly.

\subsection{\textbf{Physical Attacks}}

A PSC-based disassembler that identified and disassembled instruction sequences in embedded systems at an instruction-level granularity was introduced by~\cite{Park2018Power, Bae2022Implementation}. The proposed methods involved collecting power traces through an oscilloscope, transforming these traces into the time-frequency domain using continuous wavelet transform (CWT), and then applying feature selection with KL-divergence followed by dimensionality reduction using Principal Component Analysis (PCA). Machine learning classifiers, such as quadratic discriminant analysis (QDA), Linear Discriminant Analysis (LDA), SVM, and Naïve Bayes, were used to create profiles of power consumption patterns for each instruction~\cite{Park2018Power}. Machine learning models, including K-nearest neighbor (KNN), RF, CNN, and Multi-Layer Perceptron (MLP), were used~\cite{Bae2022Implementation} to classify and recover the instructions from the power traces with high accuracy. This hierarchical classification framework allowed for the identification of instructions, including register names.

Reverse engineering the instructions of the 8-bit PIC16F690 microcontroller and the 32-bit ARM Cortex-M3 processor were targeted by the authors \citeauthor{Vafa2020Efficient}~\cite{Vafa2020Efficient}. The method leveraged both dynamic and static power consumption to extract the maximum information leakage, revealing opcodes and operands of real instructions. Classification algorithms and KL feature selection methods were employed to improve the classification rate of instructions into different HW groups.

Another disassembler attack aimed to reverse-engineer the instructions executed on the ARM Cortex-M0 processor-based MCU was proposed in~\cite{Van2022Side}. ARM Cortex-M0 is a 32-bit architecture commonly used in low-power IoT devices. This attack also employed a profiling approach using machine learning algorithms, including LDA, QDA, MLP, and CNN, to classify power traces associated with different instructions. The attack successfully identified instruction sequences and distinguished between instruction groups based on the power consumption patterns.

A disassembly attack aimed to reverse-engineer executed instructions by performing a template attack on power traces of the Atmel 8-bit AVR MCU was proposed in~\cite{Narimani2021Side}. The attack employed a profiling technique using CNNs to classify the power traces corresponding to different instructions executed on the AVR microcontroller. It achieved an accuracy of 98.21\% on real code, significantly outperforming previous methods based on machine learning techniques like KNN and LDA.

\citeauthor{Han2022Hiding} craft assembly-level malware that preserves program functionality while manipulating power traces so that power side-channel control-flow monitors classify it as benign~\cite{Han2022Hiding}.
Their methodology builds a substitute detection setup, partitions the payload into context-independent chunks, performs data-flow analysis to identify each chunk’s live range, and uses Monte Carlo Tree Search to select injection points that maximize the detector’s confidence score~\cite{Han2022Hiding}. The attack is evaluated on ARM Cortex-M programs across multiple detector families (e.g., Bidirectional Long Short-Term Memory neural network (BiLSTM), Bidirectional Recurrent Neural Network (BiRNN), and Hidden Markov Model (HMM)). The authors conclude that such monitors are less secure than commonly assumed and recommend stronger designs, such as multi-cycle windows and fusing power with EM to withstand instruction-level, side-channel-aware injections~\cite{Han2022Hiding}.

\citeauthor{Fendri2022Deep} presented a comprehensive framework designed to evaluate the vulnerability of CPU designs to PSC disassembly attacks at the design phase~\cite{Fendri2022Deep}. The proposed method involved generating simulated power traces from HDL models of RISC-V CPUs at the design stage and employing sparse dictionary learning (SDL) for feature extraction and dimensionality reduction. The deep learning model, specifically a multilayer perceptron (MLP), is trained to classify instructions based on these power traces. This developed framework facilitates the evaluation of vulnerabilities and countermeasures before hardware manufacturing.

\citeauthor{Krishnankutty2020Instruction} presented an advanced technique for identifying and disassembling multi-clock cycle instruction sequences on embedded systems with a pipelined architecture using PSC analysis~\cite{Krishnankutty2020Instruction}. By observing the power variations correlated with instruction execution, the authors were able to identify instruction boundaries using a dynamic programming (DP) algorithm and classify instruction types using a fine-grained classifier with high accuracy. The study also demonstrated that by using multiple power supply pin measurements, the precision and reliability of instruction classification can be significantly improved on both FPGA-based soft processors and discrete microcontrollers.

\subsection{\textbf{Remote Attacks}}

\citeauthor{Glamovcanin2023Instruction} evaluated instruction-level code disassembly using PSC analysis on RISC-V soft-core CPUs (RISCY and PicoRV32) implemented in shared FPGA environments~\cite{Glamovcanin2023Instruction}. The focus is on multitenant FPGA scenarios where attackers can deploy on-chip TDC-based voltage sensors to capture power traces remotely. The study presented a profiled attack employing both traditional machine learning and novel deep learning techniques to classify CPU instructions based on power consumption patterns.


\subsection{\textbf{Summary of  Attacks and Insights}}
PSC-based disassembly attacks exploit power variations to recover executed instructions and program behavior, posing a severe threat to software confidentiality on different platforms. As summarized in Table~\ref{tab:disassembly_app_summary}, these attacks have been demonstrated in diverse settings, ranging from software-based approaches that leverage low-level telemetry (e.g., PLATYPUS) to physical and remote setups using oscilloscope measurements or FPGA-integrated sensors. Attackers applied signal processing methods,  such as CWT, KL-divergence, and PCA, combined with a wide range of ML models, including CNNs, RF, SVMs, and MLPs, to classify instruction opcodes and operands with high accuracy. The surveyed works show that profiled analysis dominates this domain, with attacks successfully reverse-engineering instruction sequences across architectures, such as AVR, ARM Cortex, and RISC-V. Therefore, power model information is not included in the Table since power models are typically used only in non-profiled attacks. Instead of categorizing by power analysis methods, this table focuses on preprocessing techniques, feature extraction methods, and the profiled models employed in each study.

A closer examination of PSC-based disassembly attacks reveals several important insights regarding both their methodological trends and their practical constraints. Nearly all surveyed studies rely on physical probing, reflecting the early dominance of MCU-based evaluations where on-chip telemetry interfaces, such as RAPL, NVML, or FPGA-integrated sensors, are unavailable. This historical focus on MCUs naturally constrained the attack surface to physical measurement setups, but it also provided cleaner and more deterministic leakage signals that facilitated instruction classification. 

The surveyed literature shows that profiled analysis methods overwhelmingly dominate this domain, as machine learning models are highly effective at capturing instruction-dependent variations within localized trace segments, enabling fine-grained opcode and operand recovery. 

However, this strong dependence on profiled approaches and physical access highlights several limitations: these attacks have not yet been widely demonstrated on more complex platforms where multi-core interference, dynamic scheduling, cache effects, and power-management policies introduce substantial noise. Additionally, the reliance on MCU platforms leaves open questions about the generalizability of instruction-disassembly attacks to CPUs, GPUs, and heterogeneous accelerators where richer remote or software-accessible telemetry channels exist. These observations suggest that while PSC-based disassembly has matured rapidly in controlled MCU environments, its broader applicability depends on extending current methodologies beyond physical probing and adapting them to modern architectures with more diverse and less predictable leakage behaviors.

\section{Extracting Machine Learning Models}
\label{sec:ml}

In this section, we survey PSC attacks that leverage variations in device power consumption to extract sensitive information from ML models, such as input data, model parameters, and model architectures. These attacks pose significant risks, especially for ML models used in resource-limited environments, such as smartphones and IoT devices. By analyzing power usage during inference and training, attackers can potentially reverse engineer input data and uncover proprietary model details. The increasing use of ML in sensitive applications underscores the need for effective countermeasures to protect against these attacks.

\begin{table*}[]
\caption{Summary of power side-channel attacks targeting machine learning models. The table includes details about the probing interface: Software (SW), Remote (RM), and Hardware (HW), as well as the target platform, targeted model type, and the extracted model information. 
}
\center
\small
\renewcommand{\arraystretch}{1.1}
\begin{tabular}{lcccccc}
\rowcolor[HTML]{BDBDBD} 
\multicolumn{1}{c}{\cellcolor[HTML]{BDBDBD}\textbf{Reference}} &
  \textbf{\begin{tabular}[c]{@{}c@{}}Probing \\ Interface\end{tabular}} &
  \textbf{\begin{tabular}[c]{@{}c@{}}Target \\ Platform\end{tabular}} &
  \textbf{\begin{tabular}[c]{@{}c@{}c@{}}Target \\ Model /\\ Information\end{tabular}} &
  \textbf{\begin{tabular}[c]{@{}c@{}}Preprocessing \\ Methods\end{tabular}} &
  \textbf{\begin{tabular}[c]{@{}c@{}c@{}}Feature \\ Extraction \\ Method\end{tabular}} &
  \textbf{\begin{tabular}[c]{@{}c@{}}Attack \\ Model\end{tabular}} \\ \hline
\rowcolor[HTML]{FFFFFF} 
\citeauthor{Gao2023Deeptheft}~\cite{Gao2023Deeptheft} &
  SW &
  CPU &
  \begin{tabular}[c]{@{}c@{}}DNN / \\ Hyper-parameters\end{tabular} &
  N/A &
  N/A &
  U-Net + BiLSTM \\
\rowcolor[HTML]{E4E4E4} 
\citeauthor{Ryu2023Gamma}~\cite{Ryu2023Gamma} &
  SW &
  CPU &
  \begin{tabular}[c]{@{}c@{}}CNN / \\ Hyper-parameters\end{tabular} &
  Segmentation &
  Statistical &
  \begin{tabular}[c]{@{}c@{}}KNN\\ Gaussian NB\\ RF\end{tabular} \\
\rowcolor[HTML]{FFFFFF} 
\citeauthor{Jha2020Deeppeep}~\cite{Jha2020Deeppeep} &
  SW &
  GPU &
  \begin{tabular}[c]{@{}c@{}}Compact DNN / \\ Architecture\\ Hyper-parameters\end{tabular} &
  Average &
  N/A &
  Non-profiled Other \\
\rowcolor[HTML]{E4E4E4} 
Arefin et al.~\cite{Arefin2024Dissecting} &
  SW &
  GPU &
  \begin{tabular}[c]{@{}c@{}}CNN / \\ Model type\end{tabular} &
  N/A &
  \begin{tabular}[c]{@{}c@{}}Statistical\\ Spectral\end{tabular} &
  \begin{tabular}[c]{@{}c@{}}RF\\ XGBoost\\ Light GBM\end{tabular} \\
\rowcolor[HTML]{FFFFFF} 
\citeauthor{Zhang2023Deep}~\cite{Zhang2023Deep} &
  SW &
  TEE &
  \begin{tabular}[c]{@{}c@{}}DNN / \\ Parameters\end{tabular} &
  Averaging &
  N/A &
  Non-profiled other \\
\rowcolor[HTML]{E4E4E4} 
\citeauthor{Moini2021Power}~\cite{Moini2021Power} &
  RM &
  FPGA &
  \begin{tabular}[c]{@{}c@{}}BNN / \\ Input data\end{tabular} &
  \begin{tabular}[c]{@{}c@{}}Averaging\\ Filtering\end{tabular} &
  N/A &
  Non-profiled other \\
\rowcolor[HTML]{FFFFFF} 
\citeauthor{Zhang2021Stealing}~\cite{Zhang2021Stealing} &
  RM &
  FPGA &
  \begin{tabular}[c]{@{}c@{}}DNN / \\ Hyper-parameters\end{tabular} &
  Segmentation &
  tsfresh &
  XGBoost \\
\rowcolor[HTML]{E4E4E4} 
\citeauthor{Meyers2022Reverse}~\cite{Meyers2022Reverse} &
  RM &
  FPGA &
  \begin{tabular}[c]{@{}c@{}}MLP, CNN / \\ Folding parameters\\ Number of neurons\end{tabular} &
  N/A &
  tsfresh &
  RF \\
\rowcolor[HTML]{FFFFFF} 
\citeauthor{Tian2023Practical}~\cite{Tian2023Practical} &
  RM &
  FPGA &
  \begin{tabular}[c]{@{}c@{}}N/A / \\ Accelerator operations\\ Matrix shapes\end{tabular} &
  Normalization &
  N/A &
  SPA \\
\rowcolor[HTML]{E4E4E4} 
\citeauthor{Huegle2023Power2picture}~\cite{Huegle2023Power2picture} &
  RM &
  FPGA &
  \begin{tabular}[c]{@{}c@{}}LeNet / \\ Input data\end{tabular} &
  N/A &
  N/A &
  GAN \\
\rowcolor[HTML]{FFFFFF} 
\citeauthor{Nevskovic2023Systemc}~\cite{Nevskovic2023Systemc} &
  PH &
  Accelerator &
  \begin{tabular}[c]{@{}c@{}}N/A / \\ Parameters\end{tabular} &
  N/A &
  N/A &
  \begin{tabular}[c]{@{}c@{}}CPA\\ Template attack\end{tabular} \\
\rowcolor[HTML]{E4E4E4} 
\citeauthor{Wang2023Powergan}~\cite{Wang2023Powergan} &
  PH &
  ASIC &
  \begin{tabular}[c]{@{}c@{}}U-Net / \\ Input data\end{tabular} &
  \begin{tabular}[c]{@{}c@{}}Weighted sum\\ Sliding window\end{tabular} &
  Power matrices &
  GAN \\
\rowcolor[HTML]{FFFFFF} 
\citeauthor{Xiang2020Open}~\cite{Xiang2020Open} &
  PH &
  CPU &
  \begin{tabular}[c]{@{}c@{}}DNN / \\ Model type\\ Parameters\end{tabular} &
  N/A &
  Statistical &
  SVM \\
\rowcolor[HTML]{E4E4E4} 
\citeauthor{Wei2018I}~\cite{Wei2018I} &
  PH &
  FPGA &
  \begin{tabular}[c]{@{}c@{}}CNN / \\ Input data\end{tabular} &
  \begin{tabular}[c]{@{}c@{}}Filtering\\ Curve fitting\\ Alignment\end{tabular} &
  N/A &
  \begin{tabular}[c]{@{}c@{}}Non-profiled other\\ Template attack\end{tabular} \\
\rowcolor[HTML]{FFFFFF} 
\citeauthor{Wolf2021Stealing}~\cite{Wolf2021Stealing} &
  PH &
  MCU &
  \begin{tabular}[c]{@{}c@{}}NN, LR / \\ Model type\\ Architecture\\ Hyper-parameters\end{tabular} &
  N/A &
  N/A &
  KNN \\ \hline
\end{tabular}
\label{tab:ml_app_summary}
\end{table*}

\subsection{\textbf{Software-based Attacks}}

DeepTheft is a software PSC attack developed to steal DNN model architectures by leveraging the RAPL interface~\cite{Gao2023Deeptheft}. DeepTheft employed a hybrid learning framework combining U-Net for spatial features and BiLSTM for temporal features to dissect the energy traces captured through RAPL. This allows the attacker to recover detailed layer-wise hyperparameters, such as kernel number and output feature size, through regression analysis of the computing overhead of different layers. Another software-based PSC attack that uses the RAPL interface to capture power traces on deep learning models~\cite{Zhang2023Deep}. The attack exploited power leakage from the ReLU activation function to recover activation patterns, which are then used to extract full DNN model parameters when the victim model was executed on an Intel SGX enclave. By applying input gradients and PSC information, the authors developed a method that significantly reduces the number of queries needed for model extraction. 

The ``gamma-Knife"~\cite{Ryu2023Gamma} attack also used Intel RAPL to measure power consumption without physical access to the hardware to perform a software-based PSC attack to extract the neural network architecture executed on CPUs. This attack was able to identify key architectural elements such as filter sizes, the number of filters, the activation functions, the weights, and the proportion of network layers with high accuracy. This attack can be categorized as a profiled attack, where the profiling step involves identifying the family of neural networks to which the target network belongs by analyzing the power consumption patterns. The subsequent architecture search step reduces the candidate space by leveraging statistical features of power consumption data.

``DeepPeep'' proposed a black-box, two-stage PSC attack that infers the architectural building blocks of compact DNNs on cloud GPUs by correlating power footprints across batch sizes~\cite{Jha2020Deeppeep}. Using {nvidia-smi} for power telemetry, the attacker first performs inter-group identification and then intra-group refinement to pinpoint modules such as fire blocks, depthwise separable convolutions, channel shuffle, dense blocks, and inception cells. The threat model assumes an untrusted multi-tenant GPU setting where only coarse runtime metrics and power/energy readings are observable, yet the modular design of compact DNNs makes their topology distinguishable. Experiments on Tesla P100 and Quadro P4000 GPUs showed that characteristic trends in metrics and kernel utilization generalize across hardware, enabling architecture recovery without physical access.

\citeauthor{Arefin2024Dissecting} analyzed another software-based PSC attack on GPUs that also leveraged the built-in {nvidia-smi} interface (with no additional privileges) to log power draw and classify CNN architectures~\cite{Arefin2024Dissecting}. The attacker collected high-rate power traces from both on-premise and cloud GPUs and demonstrated that distinctive draw patterns reveal model families and variants, highlighting a practical avenue for architecture inference in multi-tenant environments. Overall, this work further clarifies the sources of DNN power footprints and suggests countermeasures such as restricting or rate-limiting access to {nvidia-smi} and injecting power noise in virtualized cloud environments where feasible.

\subsection{\textbf{Physical Attacks}}

\citeauthor{Nevskovic2023Systemc} demonstrated that a SystemC model can be used to simulate PSC attacks on AI accelerators at the electronic system level (ESL)~\cite{Nevskovic2023Systemc}. The authors proposed a dynamic power consumption model that accounts for switching activities during multiplication operation, addition operations, and during register access. They employed CPA and template attacks using the proposed power model to evaluate the security of AI accelerators. These simulations revealed that both CPA and template attacks can successfully extract secret parameters from AI models by analyzing power traces generated during the inference phase.

\citeauthor{Xiang2020Open}  demonstrated a PSC attack that dissects the structure and parameters of DNN models~\cite{Xiang2020Open}. By analyzing power traces collected during the inference phase, the authors were able to identify different DNN models, including AlexNet, VGG, and ResNet, implemented on embedded platforms like the Raspberry Pi. The study employed power models to understand the power consumption patterns of different DNN layers, including convolutional layers, pooling layers, fully connected layers, and activation functions. The study used a machine learning classifier to analyze power features, achieving high accuracy in distinguishing model architectures and estimating parameter sparsity.

PSC analysis was used to differentiate and extract information from ML models running on MCUs~\cite{Wolf2021Stealing}. Using the ChipWhisperer-Lite, the authors captured power traces during the training of neural networks and linear regression models. The analysis involved using KNN models with dynamic time warping (DTW) and traditional Euclidean distance (ED) metrics to classify power traces. The study demonstrated that power traces can reveal differences between neural network and linear regression models, between neural networks with different hidden layer sizes, and between two nearly identical linear regression models with different hyperparameters.


\citeauthor{Wei2018I} proposed an attack where both passive and active adversaries can recover input images processed by a CNN accelerator~\cite{Wei2018I}. Passive adversaries eavesdrop on power consumption during the inference stage, while active adversaries build a power template to profile the relationship between power consumption and input pixels. This attack used cycle-accurate power traces acquired through oscilloscope measurements.

\citeauthor{Wang2023Powergan} introduced {PowerGAN}, a PSC attack targeting analog compute-in-memory (CIM) accelerators that aims to recover sensitive user inputs~\cite{Wang2023Powergan}. 
By capturing power traces from the first layer of DNN inference and feeding them into a conditional GAN, the framework successfully reconstructs high-quality input data. 
The study emphasizes that CIM, while promising for overcoming the memory bottleneck in digital architectures, inherits unique leakage properties that make it vulnerable to side-channel attacks. 
Moreover, the evaluation demonstrated that the PowerGAN method remains effective even under noisy measurement conditions, underscoring its practicality.

\subsection{\textbf{Remote Attacks}}

\citeauthor{Moini2021Power} presented a detailed remote PSC attack on Binarized Neural Network (BNN) accelerators running on various FPGA platforms, including those provided by cloud services like AWS F1 instances and different Xilinx FPGAs~\cite{Moini2021Power}. The authors successfully detected and exploited power consumption variations by using TDC-based on-chip voltage sensors to extract information about the input data processed by the BNN. The first kernel of the first convolution layer for an input image is used as the attack point. Experimental results indicate that the processing of foreground pixels shows higher dynamic power consumption compared to background pixels.

\citeauthor{Zhang2021Stealing} performed a remote PSC attack on DNN models executed on cloud FPGA instances~\cite{Zhang2021Stealing}. The attack captured voltage fluctuations to infer the structure of DNN models utilizing RO-based power sensors. The attack flow involved segmenting power traces to identify layer boundaries, classifying the type of each layer (e.g., convolutional, fully-connected, and pooling), and then predicting the hyper-parameters, such as the number of neurons, filter sizes, and strides. The evaluation of models like MLP, AlexNet, and VGG16 demonstrated that the attack achieves high accuracy in reconstructing the network architectures. Similar remote PSC attack on folded neural network implemented on FPGAs was proposed in~\cite{Meyers2022Reverse}. The authors used TDCs to capture voltage fluctuations during neural network inference. By employing a profile-based attack, they first recovered the folding parameters, which denote the degree of parallelism and sequential operations in the neural network layers. This information was then used to recover the number of neurons in each layer, achieving significantly higher accuracy compared to prior works. This approach was validated on different neural networks, including MLP and CNN models, under various environmental conditions.

\citeauthor{Tian2023Practical} demonstrated a power trace collection method for practical remote PSC attacks on machine learning accelerators in cloud FPGAs~\cite{Tian2023Practical}. By implementing compressed TDC RTL kernels within the cloud FPGA, the authors captured power traces to create detailed profiles of different ML operations in a realistic environment. Additionally, authors introduced an auto-triggered mechanism that enables power trace collection without requiring hard-wired communication between the attacker and the victim.

\citeauthor{Huegle2023Power2picture} proposed {Power2Picture}, a GAN-based framework designed to extract sensitive user inputs from DNN inference implementations on FPGAs~\cite{Huegle2023Power2picture}. 
The attack leverages on-chip TDC sensors to capture power traces during inference and then employs a generative adversarial network to reconstruct the corresponding input data. 
Their results show that even in resource-constrained FPGA environments, power side-channel information can be effectively transformed into input-level leakage through a learning-based approach.


\subsection{\textbf{Summary of  Attacks and Insights}}
Table~\ref{tab:ml_app_summary} summarizes the growing body of work on PSC attacks targeting machine learning models across software, remote, and physical probing interfaces. These studies highlight how power leakage can be exploited to recover diverse forms of sensitive information, ranging from model architectures, hyperparameters, and parameters to input data. A clear trend is the use of profiled analysis pipelines, where preprocessing and feature extraction methods are coupled with statistical or learning-based classifiers, enabling high accuracy in model reconstruction. However, non-profiled attacks show that even coarse-grained telemetry,  such as RAPL counters or on-chip sensors, can still yield exploitable leakage. 

The rapid advancement of ML has also contributed to the evolution of PSC attacks by both enriching profiled analysis pipelines and expanding the range of viable attack targets. On the analysis side, modern ML techniques provide highly expressive models capable of learning complex leakage patterns from large trace sets, enabling profiled attacks to reconstruct architectural details, layer dimensions, and activation behaviors with greater accuracy and resilience than traditional statistical methods. At the same time, the increasing adoption of ML across domains such as healthcare, finance, autonomous systems, and personalized services has amplified the implications of PSC leakage, since extracting model parameters or input data from these systems can directly compromise privacy, intellectual property, financial assets, and safety-critical decision processes. 

Another key observation is that software-based and remote PSC attacks have become the most prevalent, primarily due to widely available power-telemetry interfaces, including RAPL, NVML, and various on-chip sensors, on modern ML-capable platforms. The accessibility and scalability of these interfaces make such attacks significantly easier to mount than physical probing. However, this balance is likely to shift as edge-AI devices and ML-enabled IoT platforms become more widespread, creating new opportunities for physical PSC attacks on embedded accelerators and lightweight inference engines. 

The surveyed works further demonstrate that the attack surface spans the full spectrum of ML deployment platforms, including CPUs, GPUs, FPGAs, ASICs, and TEEs, each exposing unique leakage properties influenced by hardware primitives, execution environments, and model implementations. 
Overall, Sections~\ref{sec:crypto} to \ref{sec:ml} illustrate that PSC attacks are not confined to cryptographic workloads but also extend to diverse applications, including user data exploitation, program instruction disassembly, and ML model extraction. This reveals a broader class of vulnerabilities that must be addressed by designing sufficient countermeasures, which are discussed in the next section.

\section{Power Side-Channel Countermeasures}
\label{sec:countermeasures}

\begin{figure}[htp]
    \begin{center}
        \small
        \includegraphics[width=0.37\textwidth]{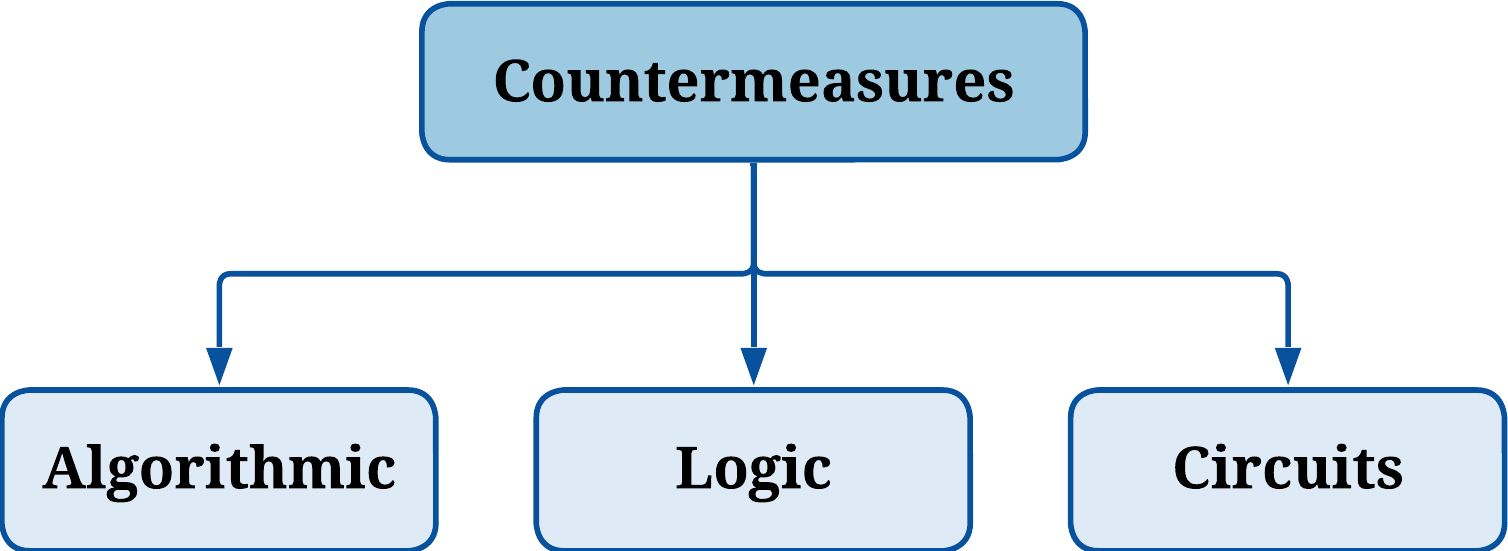}
    \end{center}
     \vspace{-0.1in}
      \caption{Overview of countermeasures to defend against PSC attacks.}
      \label{fig:countermeasure_overview}
\end{figure}

PSC attacks pose a significant threat to user privacy and confidentiality by exploiting their power consumption patterns to extract sensitive information. To mitigate these attacks, various countermeasures have been proposed in the literature. Figure~\ref{fig:countermeasure_overview} illustrates the three main countermeasure categories: design-specific and circuit-level (generic) countermeasures. Design-specific countermeasures can be further divided into two levels: logic level and algorithmic level. The basic idea is to decrease the Signal to Noise Ratio (SNR) by obscuring the power consumption patterns or reducing the correlation between the power consumption and the secret data being processed. This is typically achieved using two key techniques: noise injection and sensitive power signature flattening. 

Noise injection involves adding random or structured noise to the power consumption signal to mask the variations caused by targeted operations, thereby making it more difficult for attackers to extract useful information. There are promising techniques for noise injection, including random power consumption and voltage noise generation. Random power generation utilizes additional operations to introduce noise, whereas noise is directly injected into the power supply for voltage noise generation. Sensitive power signature flattening, on the other hand, aims to flatten the power consumption profile by ensuring that the power consumption remains constant or minimally variable regardless of the data being processed. There are popular approaches for power signature flattening, such as masking, hiding, and dual-rail logic. Masking~\cite{Asfand2021Low, Bhandari2024Lightweight} combines data with random values to hide its true value. Hiding~\cite{Parrilla2022Time, Lee2020Security, Alipour2020Performance} makes some changes to equalize power consumption across different operations. Dual-rail logic can balance the power consumption of each bit operation, and constant time execution ensures that operations take the same amount of time, regardless of the input data. By employing these techniques across different levels of design, it is possible to significantly enhance the resilience of cryptographic devices against PSC attacks.

\begin{figure}[htp]
    \begin{center}
        \includegraphics[width=0.2\textwidth]{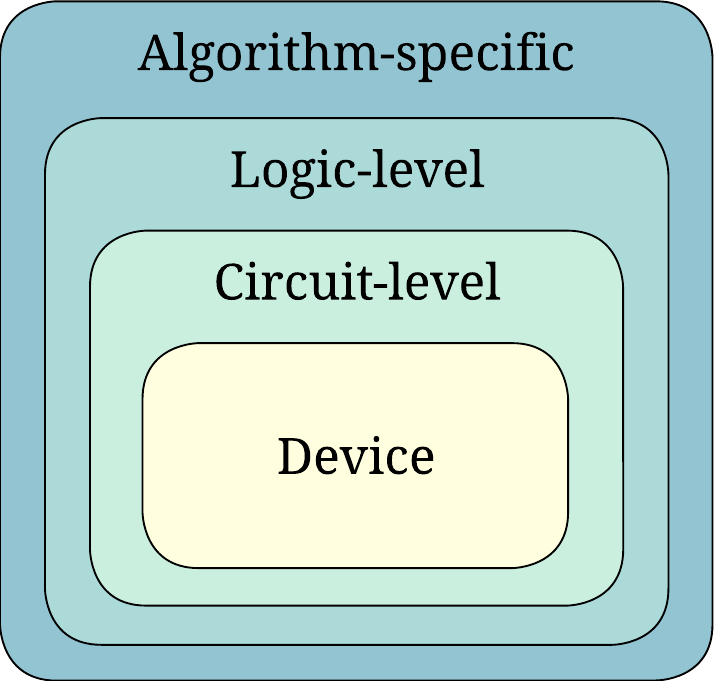}
    \end{center}
      \vspace{-0.2in}
      \caption{Abstraction levels of countermeasures against PSC attacks. Algorithmic- and logic-level countermeasures are design-specific, targeting particular architectures or implementations, whereas circuit-level countermeasures are generic techniques applied at the transistor level.}
    \vspace{-0.2in}
      \label{fig:countermeasures}
\end{figure}

Figure~\ref{fig:countermeasures} categorizes various countermeasures in terms of design abstraction levels. We first survey design-specific countermeasures that are applicable on logic and algorithmic levels. Next, we discuss generic countermeasures that are suitable for circuit-level implementation. 

\subsection{Algorithm-Specific Countermeasures}
\label{sec:countermeasures_algorithmic}

Design-specific masking can be implemented at both the algorithmic and logic levels. Algorithmic-level masking has the advantage of being easy to implement but incurs high latency for higher-order masking. The basic principle of masking is to randomly divide the secret data into multiple shares, which breaks the link between sensitive data and intermediate operations in cryptographic algorithm. If the number of shares is d+1, then the masking scheme theoretically defends against d-order side-channel attacks. 

A hardware-aware algorithmic level masking was proposed for Chaskey, Simon, and AES, while improving power, performance, and area (PPA)~\cite{Yang2024Hardware}. The proposed masking method is well-suited for embedded and IoT domains, which are used in resource-constrained environments. To achieve this improved PPA, the authors implemented a security-aware hardware design with a minimal set of instructions and a simple design inspired by the SubRISC+ embedded processor architecture.

Hiding is another widely used countermeasure against PSC attacks. The goal of hiding is to obscure the sensitive data-dependent power signature with another signature. Methods like noise addition are well-known options for creating additional power signatures. \citeauthor{Benhadjyoussef2021Power} used such a hiding countermeasure against CPA on an FPGA-based AES implementation~\cite{Benhadjyoussef2021Power}. Their attack target was the final round S-box values of AES, and authors employed an additional parallel circuit consisting of AddRoundKey and SubBytes stages with a random key. The power signature from these new stages were able to hide the power signature from the concurrent AddRoundKey and SubBytes stages in the actual AES design.

\citeauthor{Aysu2018Binary} proposed two novel countermeasures againt SPA and DPA attacks on B-RLWE hardware implementation~\cite{Aysu2018Binary}. For SPA, the authors have introduced the ``always-add-with-redundant-store" method, which normalizes power consumption by always performing additions and memory updates, regardless of the binary key value. For DPA, they implemented random initialization and masked threshold decoder. This method begins each decryption process by setting the intermediate sum to a random value, and the final step utilizes a masked threshold decoder to recover the impact of this random initialization. This significantly increased the number of measurements required for a successful DPA attack.

AIAShield~\cite{Yan2023Defense} defense was designed to protect FPGA-based AI accelerators from PSC attacks. Adversarial attack techniques from machine learning were leveraged to generate precise noise that obfuscated unintended power consumption patterns. This approach operated in two phases: offline and online. In the offline phase, a surrogate attack model was trained, and the required adversarial noise was calculated. In the online phase, the adversarial noise was injected using ring oscillators (ROs) into the model's execution to obfuscate side-channel traces.

There are several PSC attack detection mechanisms proposed in the literature. These detection mechanisms help users take protective actions against attacks, which countermeasures alone cannot provide. \citeauthor{Munny2021Power} used a battery impedance monitoring system to detect adversarial probing~\cite{Munny2021Power}. Based on real-world experimental results, this detection method has a detection time of around 22 ms, which is sufficient to prevent a successful CPA-based attack.

When it comes to FPGA-based PSC attack countermeasures, some techniques used in the ASIC domain are ineffective due to the transistor-level inaccessibility in FPGAs. Therefore, new methods should be developed to overcome these challenges and defend against PSC attacks on FPGAs. As discussed earlier, FPGA-based remote PSC attacks use power monitors to collect data-dependent power traces in a shared resource environment. Recent works show that designers can use the same technique to monitor their own power fluctuations, and if information leakage is detected, they can enable an obfuscation circuit in the design. Most of the time, this obfuscation circuit is a noise generation circuit. To monitor power fluctuations, \citeauthor{Glamovcanin2020Cloud} used a TDC-based power sensor~\cite{Glamovcanin2020Cloud}, while  \citeauthor{Ahmadi2023Shield} used an RO-based power monitor~\cite{Ahmadi2023Shield}. TDC-based sensors offer high accuracy and high resolution at the cost of high area overhead. In contrast, RO-based sensors provide a lightweight power sensor solution to defend against remote PSC attacks on FPGAs.

Deviating from traditional methods,  a combination of TDC and RO was used to implement a countermeasure against multi-tenant FPGA-based remote PSC attacks~\cite{Krautter2019Active}. The proposed countermeasure works as an active fence where the fence is made of ROs and the activation sensor is made of TDC. These fences are strategically placed around the victim cryptographic module to inject controlled noise into the power grid, thereby obfuscating the voltage fluctuations caused by the cryptographic operations. The countermeasure aims to equalize the power consumption seen by potential adversaries, significantly reducing the signal-to-noise ratio and making side-channel attacks much more difficult.

\citeauthor{Wang2022Hertzbleed} suggested algorithmic-level measures such as disabling Turbo Boost or hardware P-state management, which force the CPU to operate at fixed base frequencies, thereby eliminating DVFS-induced leakage. While effective, these countermeasures introduce significant performance penalties and may not hold under custom power-limit configurations. For cryptographic libraries like SIKE, additional defenses include input validation to prevent attacker-crafted ciphertexts from triggering anomalous zero values, as deployed by Cloudflare and Microsoft after disclosure~\cite{Wang2022Hertzbleed}.
  
The PLATYPUS attack~\cite{Lipp2021Platypus} highlighted how Intel’s RAPL interface can be abused for software-based PSC attacks and stresses the need for both system-level and algorithmic defenses. \citeauthor{Lipp2021Platypus} recommended restricting or removing unprivileged access to RAPL counters, lowering the resolution of energy measurements, and preventing precise execution control primitives such as zero-stepping in SGX. They also note that traditional software hardening techniques like masking or constant-time coding can reduce but not fully eliminate leakage when privileged access is possible. Intel’s mitigation included microcode and driver updates that degrade RAPL fidelity and enforce stricter access controls, though residual risks remain for enclave security.

\subsection{Logic-Level Countermeasures}
\label{sec:countermeasures_logic}
A logic-level countermeasure for static PSC attacks was proposed in~\cite{Dumitru2023Borrowed}. They proposed a technique called Borrowed Time, an on-chip countermeasure that detects stopped clock signals. This countermeasure eliminates the main assumption in most static PSC attacks, which is the ability to stop the clock over a reasonable period. Results showed that the proposed countermeasure has been able to increase the measurement to disclosure (MTD).

\citeauthor{Padmini2016Calpan} proposed CALPAN, a logic-level countermeasure to protect cryptographic hardware against PSC attacks~\cite{Padmini2016Calpan}. The approach enhances Delay-based Dual-Rail Precharge Logic (DDPL) by introducing two techniques: (i) sleep transistors to manage static leakage currents, and (ii) normalization to balance the leakage across input patterns. While the sleep transistor method increases overall power consumption, it reduces the data dependency of leakage by about 20\%. The normalized DDPL technique achieves superior results, lowering power consumption by 30\% and reducing the coefficient of deviation (CoD) by 89.4\%, making leakage traces nearly data-independent. These results demonstrate that CALPAN significantly strengthens resistance against PSC without prohibitive design overheads~\cite{Padmini2016Calpan}.

\citeauthor{Sengupta2019Logic} evaluated the security of logic locking (LL) techniques against PSC attacks and proposed provably secure strategies that resist both SAT and DPA attacks~\cite{Sengupta2019Logic}. It reveals that conventional LL methods like Random Logic Locking (RLL) and Fault-based Logic Locking (FLL) are vulnerable, but resilience improves through {key-aliasing}, where multiple incorrect keys produce indistinguishable leakage, hampering DPA effectiveness. The authors formally established that LL schemes resilient to SAT attacks are inherently resistant to DPA as well, validating this with experiments on the Stripped Functionality Logic Locking (SFLL) approach. By concentrating key gates within logic cones and leveraging aliasing effects, SFLL achieves strong provable resistance to power analysis while still mitigating piracy and overbuilding threats.

A Standard-Cell Delay-based Dual-Rail Precharge Logic (SC-DDPL) was proposed in \cite{Bellizia2021Sc} as a logic-level countermeasure against static power side-channel attacks. SC-DDPL builds on the Time Enclosed Logic (TEL) protocol, which encodes data in the relative timing of dual-rail transitions rather than their steady-state values, thereby eliminating exploitable static power leakage when the clock is halted~\cite{Bellizia2021Sc}. The authors implemented SC-DDPL on an FPGA with a PRESENT crypto-core and show that it achieves superior security metrics, including a lower SNR ratio, reduced mutual information, and no exploitable leakage under t-test evaluation.

\subsection{Circuit-Level Countermeasures}
\label{sec:countermeasures_circuit}
Circuit-level noise injection is used as a generic, application-independent countermeasure. Prior works have developed noise injection designs to suppress data-dependent power signatures by embedding the targeted application hardware in a signature attenuating hardware (SAH)~\cite{Das2017High, Das2018Asni}. \citeauthor{Das2020Electromagnetic} proposed a similar approach, but specifically targeting the current~\cite{Das2020Electromagnetic}. Their method uses a stable constant current source to maintain a supply current independent of the targeted application current consumption.

\citeauthor{Parrilla2022Time} introduced a novel countermeasure against SPA attacks for FPGA-based IoT devices with the Xored High Consumption Module (XHCM)~\cite{Parrilla2022Time}. XHCM operates by generating controlled power consumption peaks with varying amplitudes, which effectively obscure the power consumption patterns of the target device. The authors developed a Multi-Level XHCM (ML-XHCM) that allows generating different power consumption levels with minimal area overhead, enhancing the security of cryptographic operations on FPGAs by adding crafted power noise in both the amplitude and time domains. \citeauthor{Jevtic2022Side} presented a generic circuit-level hiding method that involves modulating the power supply voltage to obscure the power consumption patterns exploited by attackers~\cite{Jevtic2022Side}. By dynamically varying the power supply, the correlation between power consumption and targeted operations is reduced, making it significantly harder for attackers to perform DPA and CPA attacks.

\citeauthor{Mozipo2024Analysis} investigated three generic circuit-level countermeasures to reduce the efficacy of remote and local PSC attacks~\cite{Mozipo2024Analysis}. Integrated voltage regulators (IVRs) are shown to be effective in scrambling the power traces in local attacks by introducing non-linear transformations that obscure the correlation between the IC's internal current and the external measurements. However, IVRs are less effective against remote attacks. On-package decoupling capacitors (OPDs), while providing some level of filtering, only marginally increase the difficulty of remote attacks. Voltage noise injection, which involves introducing noise into the IC power delivery network, significantly increases the MTD by up to 37 times. The study concludes that noise injection is the most effective countermeasure against remote and local PSC attacks.

Transistors are the fundamental building blocks of any semiconductor device. Therefore, to mitigate PSC attacks, existing works focused on security at the transistor level as well. \citeauthor{Knechtel2020Power} presented such a promising technology named NCFET, that aims to replace traditional MOSFETs~\cite{Knechtel2020Power}.  The authors demonstrated that NCFET-based circuits exhibit greater resistance to CPA attacks due to the significant impact of negative capacitance on switching (dynamic) power. LOCK\&ROLL is a mitigation technique proposed using reconfigurable devices and logic locking to obscure power consumption patterns to counter DL-based PSC attacks~\cite{Kolhe2022Lock}. The proposed Symmetrical MRAM-based LUT (SyM-LUT) showed resilience to machine learning-assisted PSC attacks by achieving nearly zero power variation at the output, thereby significantly reducing the feasibility of such attacks. These countermeasures are suitable to defend against diverse PSC attacks.

\subsection{\textbf{Summary of Countermeasures and Insights}}
Across the algorithmic, logic, and circuit levels, countermeasures for the PSC share a unifying objective: reducing the exploitable correlation between power consumption and sensitive data. Algorithmic and logic-level defenses predominantly rely on techniques such as masking, hiding, noise generation, and dual-rail or timing-equalized logic to suppress distinguishable leakage. These strategies often introduce performance or area overheads, yet they remain highly adaptable to diverse architectures, including block ciphers, lattice-based cryptography, and AI accelerators. Notably, emerging defenses leverage adversarial machine learning, dynamic obfuscation circuits, and hybrid sensor-driven activation mechanisms, demonstrating a shift toward adaptive and context-aware security rather than static protection. There is increasing focus on threat detection because practical attacks often rely on long data-collection periods, allowing detection mechanisms to interrupt or stop malicious activity before it succeeds.

At the lower abstraction levels, circuit and transistor-level countermeasures aim to structurally reshape the power delivery and switching characteristics of hardware, yielding more fundamental suppression of leaked information. Noise injection, IVRs, on-package decoupling, and power-supply modulation have shown varying effectiveness against both local and remote attacks, with voltage noise injection consistently emerging as the most robust protection. Novel technologies such as NCFETs and reconfigurable MRAM-based LUTs highlight a trend toward hardware primitives explicitly designed with side-channel resilience in mind, rather than retrofitted security. These observations collectively suggest that future PSC-resistant systems will benefit most from multilayered designs, where adaptive algorithmic defenses, leakage-equalizing logic styles, and noise-modulating circuit techniques operate synergistically to reduce the SNR while preserving system performance and usability.

\section{Conclusion and Future Directions}
\label{sec:summary}

Power side-channel (PSC) attacks have demonstrated their ability to extract secret information across application domains. In the late 1990s, most PSC attacks targeted the extraction of secret keys from cryptographic implementations. In recent years, there are a wide variety of PSC attacks on diverse application domains, including key extraction from cryptographic implementations, reverse engineering of machine learning models, derivation of user-specific behavior data, and identification of assembly instructions from power traces. In this survey, we have provided a comprehensive overview of these attacks and their countermeasures in different application domains. Specifically, this survey covers PSC attacks and countermeasures across a wide variety of application domains with different power models and analysis methods as well as diverse hardware implementations. 

Based on a thorough analysis of recent PSC attacks across various hardware types and proximity settings, several promising directions for future research have emerged. The following discussion is derived from the insights gathered through this survey in each application domain.

\begin{figure*}[t]
    \centering
    \includegraphics[width=\textwidth]{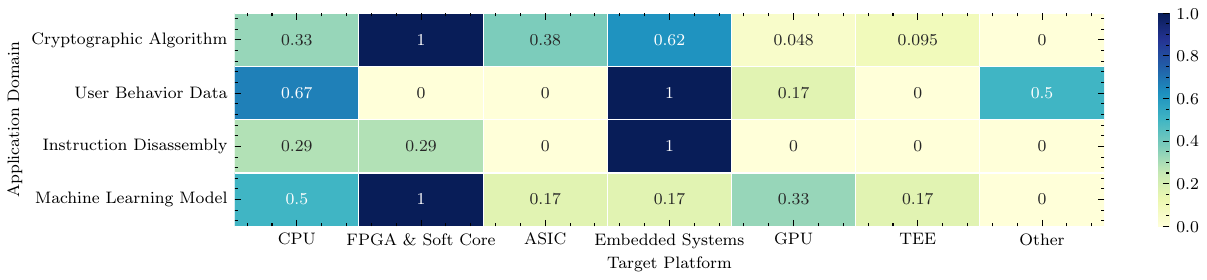}
    \caption{Heatmap of application versus platform coverage for PSC attacks. It shows relative emphasis per application domain, highlighting platform bias within each application domain. For example, there are a lot of research efforts on \textit{instruction disassembly} using \textit{embedded systems}, however, there are no recent efforts on \textit{instruction disassembly} using \textit{GPU} architectures.}
    \label{fig:application_vs_platform_coverage}
\end{figure*}

\subsection{Cryptographic Implementation}

\begin{figure}[htp]
    \centering
    \includegraphics[width=0.48\textwidth]{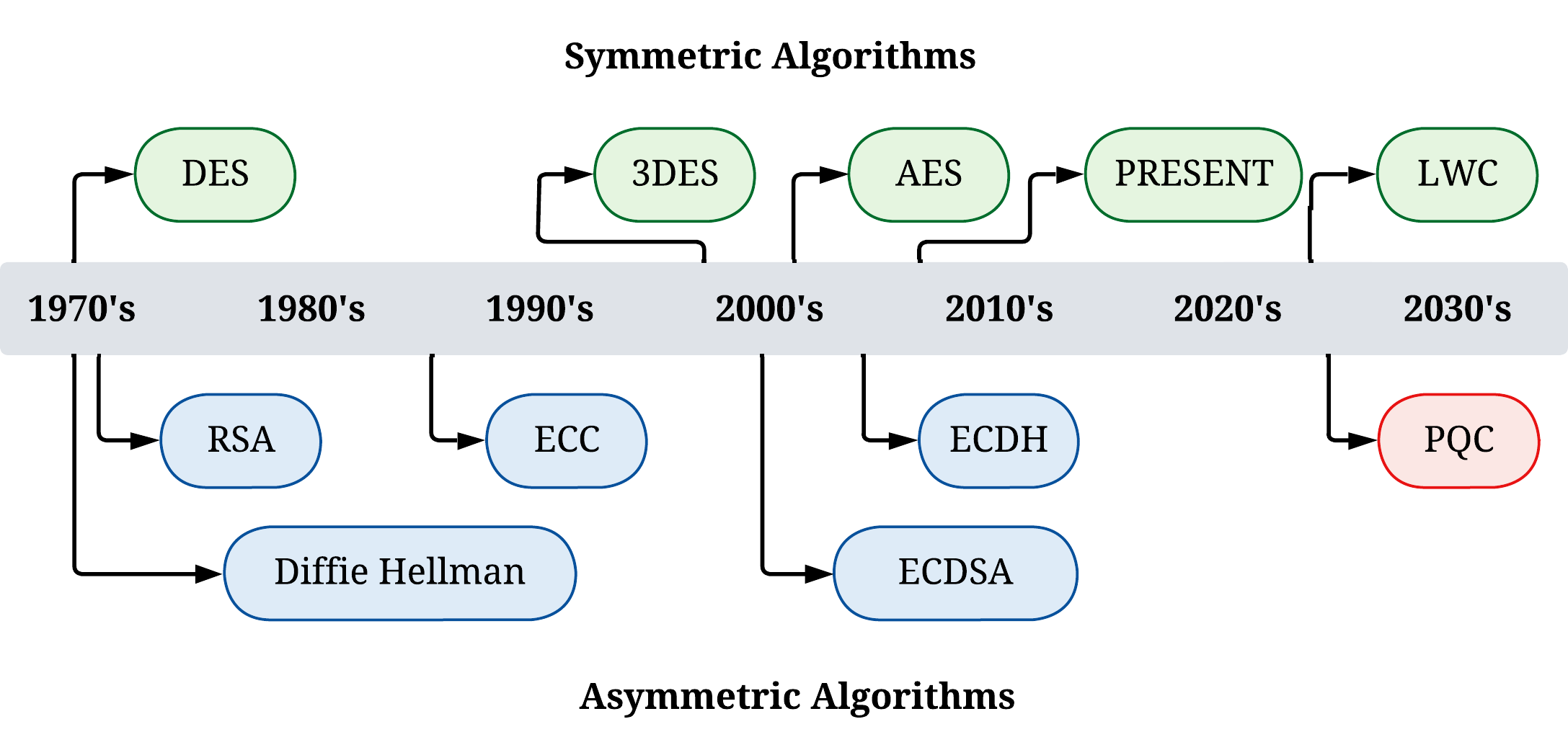}
    \caption{Timeline of PSC attacks targeting cryptographic algorithms, showing progression from DES and AES towards lightweight cryptography (LWC) and post-quantum cryptography (PQC).}
    \label{fig:crypto_timeline}
\end{figure}

Despite decades of research on power-based PSC attacks against cryptographic implementations, several gaps remain. As shown in Table~\ref{tab:crypto_app_summary_hw} and Table~\ref{tab:crypto_app_summary_sw}, most prior works have focused on AES and RSA, while lightweight ciphers and PQC schemes have only recently been evaluated. As illustrated in Figure~\ref{fig:crypto_timeline}, the research trajectory has gradually shifted from classical ciphers, such as DES and AES, toward LWC designed for resource-constrained IoT devices, and more recently to PQC candidates such as CRYSTALS-Kyber, CRYSTALS-Dilithium, and Falcon. Our survey shows that even these NIST-selected PQC algorithms exhibit exploitable leakages under power analysis, raising concerns that algorithmic strength alone is insufficient without robust implementation-level countermeasures. Furthermore, there remains a significant research gap in evaluating PSC vulnerabilities in elliptic curve–based cryptographic algorithms, such as ECDH and ECDSA.

As shown in the first row in Figure~\ref{fig:application_vs_platform_coverage}, while FPGA and embedded platforms have been studied extensively, TEE-based cryptographic implementations remain underexplored, despite their trivial use in security-critical applications. Another gap arises in the countermeasure space: masking and hiding continue to dominate, but attackers have repeatedly demonstrated ways to circumvent these defenses (e.g., distinguishing dummy operations using machine learning). This highlights a pressing need for adaptive and composable countermeasures that can withstand evolving attack methodologies.  

Looking forward, several directions are promising. First, standardized leakage assessment frameworks are needed for PQC algorithms, similar to what has been achieved for AES through TVLA, but extended to account for diverse algorithmic structures and hardware platforms. Second, there is a strong opportunity in leveraging formal security validation tools (pre-silicon and post-silicon) to proactively identify vulnerabilities before deployment. Finally, the community should explore AI-assisted adaptive countermeasures that can monitor leakage in real time and dynamically inject obfuscation, particularly relevant for shared-resource platforms.  

\subsection{User Behavior Data}

PSC attacks targeting user behavior data have revealed the surprising extent to which seemingly low-level power fluctuations can expose high-level user activities such as website visits, keystroke inputs, touchscreen gestures, and even peripheral identities. As shown in Table~\ref{tab:user_data_app_summary}, most works have concentrated on profiled attacks where adversaries build detailed power-consumption signatures to achieve accurate classification of websites or applications. While these attacks demonstrate feasibility across smartphones, USB-powered devices, and wireless charging setups, they are still limited in scope. The majority of research focuses on website fingerprinting, leaving other user interactions, such as continuous voice recognition, health-monitoring wearables, or mixed-reality platforms, largely unexplored. Moreover, most of the evaluated setups assume stable laboratory conditions, whereas real-world scenarios involve far noisier environments with heterogeneous devices, background activity, and dynamic charging states.  

Future directions call for broadening the attack surface as well as advancing defensive design. First, new studies should evaluate PSC leakage across emerging user-centric platforms, including augmented/virtual reality devices, voice assistants, and biometric sensors, to understand the extent of private data exposure. Second, researchers should explore cross-device generalization, where a profile built on one smartphone or wearable can be reliably transferred to another, as this would directly impact the scalability of real-world adversaries. Finally, current countermeasures, such as power noise injection or charging obfuscation, remain reactive. There is strong potential for adaptive and context-aware defenses that dynamically detect abnormal leakage patterns and mitigate them in real time, providing stronger user privacy guarantees in everyday environments.

\subsection{Instruction Disassembly}

PSC-based instruction disassembly attacks highlight how power variations can leak fine-grained program execution details, enabling recovery of opcodes, operands, and control flow. Table~\ref{tab:disassembly_app_summary} shows that most existing works rely on profiled machine learning or deep learning classifiers to map power traces to instruction classes, achieving high accuracy across MCUs, ARM processors, and FPGA soft-cores. However, these studies typically focus on isolated or controlled environments, where synchronization and profiling are feasible. Two notable gaps emerge from these studies. First, the resilience of these attacks in realistic, noisy, multi-threaded, or obfuscated execution environments remains largely unexplored. Second, as shown in the third row in Figure~\ref{fig:application_vs_platform_coverage}, there is limited research on ASIC-based implementations. Additionally, there is an opportunity to develop proactive design-time evaluation frameworks that integrate leakage assessment into the hardware/software co-design cycle, enabling countermeasures before deployment. 

\subsection{Machine Learning Applications}

PSC attacks on ML models demonstrate that power consumption can reveal model parameters, architectures, and even user input data. As presented in Table~\ref{tab:ml_app_summary}, our survey highlights that most attacks have focused on extracting neural network structures and hyperparameters from CPUs, FPGAs, and embedded platforms, often using profiled frameworks to classify power traces. While these works prove feasibility, they also expose critical gaps. First, current efforts primarily target inference-time leakage, whereas training-phase vulnerabilities, which could reveal optimizer states or gradient updates remain underexplored. Second, as depicted in the fourth row in Figure~\ref{fig:application_vs_platform_coverage}, the majority of evaluated platforms are CPUs and FPGAs, leaving GPUs, edge TPUs, and custom accelerators insufficiently studied despite their widespread deployment in modern AI pipelines. Third, existing countermeasures are often borrowed from cryptography and lack adaptation to ML workloads, where model dynamics and data-dependent sparsity create unique leakage patterns.  

Future research should broaden the attack surface to include training-time operations and diverse accelerator architectures, while also developing ML-specific leakage models that capture sparsity, pruning, and quantization effects. There is a strong potential in creating cross-platform attack frameworks capable of generalizing across hardware families, which would mirror realistic adversarial capabilities in cloud and federated learning environments. On the defense side, adaptive countermeasures that combine hardware-level obfuscation with algorithm-aware noise injection or randomized execution strategies can provide robust protection. Finally, proactive assessment tools that integrate side-channel leakage analysis into the ML accelerator design flow are needed to ensure secure deployment of AI in sensitive domains such as healthcare, finance, and defense. 

\subsection{Conclusion}

This survey systematically organized power side-channel attacks using a taxonomy that spans power models, analysis methods, targeted applications, probing interface, target platform, and countermeasures. Our study shows that while classical power models, such as Hamming weight and Hamming distance, remain dominant, customized models are emerging for specialized domains, such as lightweight cryptography and machine learning. Similarly, analysis methods have evolved from simple and differential power analysis toward machine learning and deep learning–based profiled attacks, significantly broadening the attack surface. At the application level, PSC attacks have advanced from classical cryptographic key extraction to user behavior profiling, instruction disassembly, and model extraction in AI accelerators. Finally, countermeasures exist across algorithmic, logic, and circuit levels, yet our taxonomy reveals a persistent gap between proposed defenses and the adaptability of evolving attacks.  

Building on this taxonomy, several future research directions are evident. First, cross-platform analysis across CPUs, GPUs, FPGAs, and heterogeneous accelerators must be prioritized to understand leakage transferability. Second, standardized leakage assessment frameworks are needed for post-quantum cryptographic algorithms and machine learning workloads. Third, adaptive runtime countermeasures that dynamically respond to leakage conditions offer a promising alternative to static design-time protections.  

While our survey has focused on classical and emerging application domains, it is important to recognize that quantum computing may also become a prominent target for PSC attacks in the future~\cite{Xu2023Exploration, Erata2024Quantum}. Early indications suggest that qubit state preparation, measurement operations, and error correction cycles could produce exploitable power or thermal signatures. Although no systematic studies currently exist, extending PSC research toward quantum devices will be critical to ensure that platforms envisioned to secure the post-quantum era are themselves robust against physical leakage.


\begin{thebibliography}{153}
\providecommand{\natexlab}[1]{#1}
\providecommand{\url}[1]{#1}
\csname url@samestyle\endcsname
\providecommand{\newblock}{\relax}
\providecommand{\bibinfo}[2]{#2}
\providecommand{\BIBentrySTDinterwordspacing}{\spaceskip=0pt\relax}
\providecommand{\BIBentryALTinterwordstretchfactor}{4}
\providecommand{\BIBentryALTinterwordspacing}{\spaceskip=\fontdimen2\font plus
\BIBentryALTinterwordstretchfactor\fontdimen3\font minus \fontdimen4\font\relax}
\providecommand{\BIBforeignlanguage}[2]{{%
\expandafter\ifx\csname l@#1\endcsname\relax
\typeout{** WARNING: IEEEtranN.bst: No hyphenation pattern has been}%
\typeout{** loaded for the language `#1'. Using the pattern for}%
\typeout{** the default language instead.}%
\else
\language=\csname l@#1\endcsname
\fi
#2}}
\providecommand{\BIBdecl}{\relax}
\BIBdecl

\bibitem[Kocher et~al.(1999)Kocher, Jaffe, and Jun]{Kocher1999Differential}
P.~Kocher, J.~Jaffe, and B.~Jun, ``Differential power analysis,'' in \emph{Advances in Cryptology—CRYPTO’99: 19th Annual International Cryptology Conference Santa Barbara, California, USA, August 15--19, 1999 Proceedings 19}.\hskip 1em plus 0.5em minus 0.4em\relax Springer, 1999, pp. 388--397.

\bibitem[Johnson(2021)]{Johnson2021Side}
A.~Johnson, ``Side channel attacks and mitigations 2015-2020: A taxonomy of published work,'' in \emph{European Conference on Cyber Warfare and Security}.\hskip 1em plus 0.5em minus 0.4em\relax Academic Conferences International Limited, 2021, pp. 482--XII.

\bibitem[Standaert et~al.(2006)Standaert, Peeters, Rouvroy, and Quisquater]{Standaert2006Overview}
O.-X. Standaert, E.~Peeters, G.~Rouvroy, and J.-J. Quisquater, ``An overview of power analysis attacks against field programmable gate arrays,'' \emph{Proceedings of the IEEE}, vol.~94, no.~2, pp. 383--394, 2006.

\bibitem[Zhang et~al.(2016)Zhang, Vega, and Taylor]{Zhang2016Power}
L.~Zhang, L.~Vega, and M.~Taylor, ``Power side channels in security ics: hardware countermeasures,'' \emph{arXiv preprint arXiv:1605.00681}, 2016.

\bibitem[Yan et~al.(2015)Yan, Guo, Chen, and Mei]{Yan2015A}
L.~Yan, Y.~Guo, X.~Chen, and H.~Mei, ``A study on power side channels on mobile devices,'' vol. 06-November-2015.\hskip 1em plus 0.5em minus 0.4em\relax Association for Computing Machinery, 11 2015, pp. 30--38.

\bibitem[Liptak et~al.(2022)Liptak, Mal-Sarkar, and Kumar]{Liptak2022Power}
C.~Liptak, S.~Mal-Sarkar, and S.~A. Kumar, ``Power analysis side channel attacks and countermeasures for the internet of things.''\hskip 1em plus 0.5em minus 0.4em\relax Institute of Electrical and Electronics Engineers Inc., 2022.

\bibitem[Hasnain et~al.(2022)Hasnain, Asfia, and Khawaja]{Hasnain2022Power}
A.~Hasnain, Y.~Asfia, and S.~G. Khawaja, ``Power profiling-based side-channel attacks on fpga and countermeasures: A survey.''\hskip 1em plus 0.5em minus 0.4em\relax Institute of Electrical and Electronics Engineers Inc., 2022.

\bibitem[Taouil et~al.(2021)Taouil, Aljuffri, and Hamdioui]{Taouil2021Power}
M.~Taouil, A.~Aljuffri, and S.~Hamdioui, ``Power side channel attacks: Where are we standing?''\hskip 1em plus 0.5em minus 0.4em\relax Institute of Electrical and Electronics Engineers Inc., 2021.

\bibitem[Randolph and Diehl(2020)]{Randolph2020Power}
M.~Randolph and W.~Diehl, ``Power side-channel attack analysis: A review of 20 years of study for the layman,'' \emph{Cryptography}, vol.~4, no.~2, p.~15, 2020.

\bibitem[Hasnain et~al.(2023)Hasnain, Asfia, and Khawaja]{Hasnain2023Role}
A.~Hasnain, Y.~Asfia, and S.~G. Khawaja, ``Role of machine learning in power analysis based side channel attacks on fpga,'' in \emph{2023 International Conference on Robotics and Automation in Industry (ICRAI)}.\hskip 1em plus 0.5em minus 0.4em\relax IEEE, 2023, pp. 1--6.

\bibitem[Mart{\'\i}nez-Rodr{\'\i}guez et~al.(2021)Mart{\'\i}nez-Rodr{\'\i}guez, Delgado-Lozano, and Brumley]{Martinez2021SoK}
M.~C. Mart{\'\i}nez-Rodr{\'\i}guez, I.~M. Delgado-Lozano, and B.~B. Brumley, ``Sok: Remote power analysis,'' in \emph{Proceedings of the 16th International Conference on Availability, Reliability and Security}, 2021, pp. 1--12.

\bibitem[Socha et~al.(2022)Socha, Mi{\v{s}}kovsk{\`y}, and Novotn{\`y}]{Socha2022Comprehensive}
P.~Socha, V.~Mi{\v{s}}kovsk{\`y}, and M.~Novotn{\`y}, ``A comprehensive survey on the non-invasive passive side-channel analysis,'' \emph{Sensors}, vol.~22, no.~21, p. 8096, 2022.

\bibitem[Ravi et~al.(2024)Ravi, Chattopadhyay, D’Anvers, and Baksi]{Ravi2024Side}
P.~Ravi, A.~Chattopadhyay, J.~P. D’Anvers, and A.~Baksi, ``Side-channel and fault-injection attacks over lattice-based post-quantum schemes (kyber, dilithium): Survey and new results,'' \emph{ACM Transactions on Embedded Computing Systems}, vol.~23, no.~2, pp. 1--54, 2024.

\bibitem[Devi and Majumder(2021)]{Devi2021Side}
M.~Devi and A.~Majumder, ``Side-channel attack in internet of things: A survey,'' in \emph{Applications of Internet of Things: Proceedings of ICCCIOT 2020}.\hskip 1em plus 0.5em minus 0.4em\relax Springer, 2021, pp. 213--222.

\bibitem[Mirzargar and Stojilovic(2019)]{Mirzargar2019Physical}
S.~S. Mirzargar and M.~Stojilovic, ``Physical side-channel attacks and covert communication on fpgas: A survey.''\hskip 1em plus 0.5em minus 0.4em\relax Institute of Electrical and Electronics Engineers Inc., 9 2019, pp. 202--210.

\bibitem[Kim and Shin(2020)]{Kim2020Gpu}
T.~Kim and Y.~Shin, ``Gpu side-channel attacks are everywhere: A survey,'' in \emph{2020 IEEE International Conference on Consumer Electronics-Asia (ICCE-Asia)}.\hskip 1em plus 0.5em minus 0.4em\relax IEEE, 2020, pp. 1--4.

\bibitem[Spreitzer et~al.(2017)Spreitzer, Moonsamy, Korak, and Mangard]{Spreitzer2017Systematic}
R.~Spreitzer, V.~Moonsamy, T.~Korak, and S.~Mangard, ``Systematic classification of side-channel attacks: A case study for mobile devices,'' \emph{IEEE communications surveys \& tutorials}, vol.~20, no.~1, pp. 465--488, 2017.

\bibitem[Shanmugham and Paramasivam(2018)]{Shanmugham2018Survey}
S.~R. Shanmugham and S.~Paramasivam, ``Survey on power analysis attacks and its impact on intelligent sensor networks,'' \emph{IET Wireless Sensor Systems}, vol.~8, no.~6, pp. 295--304, 2018.

\bibitem[Ghimire et~al.(2023)Ghimire, Baligodugula, and Amsaad]{Ghimire2023Power}
A.~Ghimire, V.~V. Baligodugula, and F.~Amsaad, ``Power analysis side-channel attacks on same and cross-device settings: A survey of machine learning techniques,'' in \emph{IFIP International Internet of Things Conference}.\hskip 1em plus 0.5em minus 0.4em\relax Springer, 2023, pp. 357--367.

\bibitem[Hettwer et~al.(2020)Hettwer, Gehrer, and G{\"u}neysu]{Hettwer2020Applications}
B.~Hettwer, S.~Gehrer, and T.~G{\"u}neysu, ``Applications of machine learning techniques in side-channel attacks: a survey,'' \emph{Journal of Cryptographic Engineering}, vol.~10, no.~2, pp. 135--162, 2020.

\bibitem[Picek et~al.(2023)Picek, Perin, Mariot, Wu, and Batina]{Picek2023Sok}
S.~Picek, G.~Perin, L.~Mariot, L.~Wu, and L.~Batina, ``Sok: Deep learning-based physical side-channel analysis,'' \emph{ACM Computing Surveys}, vol.~55, no.~11, pp. 1--35, 2023.

\bibitem[Panoff et~al.(2022)Panoff, Yu, Shan, and Jin]{Panoff2022Review}
M.~Panoff, H.~Yu, H.~Shan, and Y.~Jin, ``A review and comparison of ai-enhanced side channel analysis,'' \emph{ACM Journal on Emerging Technologies in Computing Systems (JETC)}, vol.~18, no.~3, pp. 1--20, 2022.

\bibitem[Maheswari and Krishnamurthy(2024)]{Maheswari2024Profiling}
R.~Maheswari and M.~Krishnamurthy, ``Profiling and non-profiling key retrieval attacks in programmable object interfaces using deep learning cryptanalytic techniques: A survey,'' in \emph{2024 2nd International Conference on Device Intelligence, Computing and Communication Technologies (DICCT)}.\hskip 1em plus 0.5em minus 0.4em\relax IEEE, 2024, pp. 1--5.

\bibitem[Chabanne et~al.(2021)Chabanne, Danger, Guiga, and K{\"u}hne]{Chabanne2021Side}
H.~Chabanne, J.-L. Danger, L.~Guiga, and U.~K{\"u}hne, ``Side channel attacks for architecture extraction of neural networks,'' \emph{CAAI Transactions on Intelligence Technology}, vol.~6, no.~1, pp. 3--16, 2021.

\bibitem[M{\'e}ndez~Real and Salvador(2021)]{Mendez2021Physical}
M.~M{\'e}ndez~Real and R.~Salvador, ``Physical side-channel attacks on embedded neural networks: A survey,'' \emph{Applied Sciences}, vol.~11, no.~15, p. 6790, 2021.

\bibitem[Han et~al.(2019)Han, Christoudis, Diamantaras, Zonouz, and Petropulu]{Han2019Side}
Y.~Han, I.~Christoudis, K.~I. Diamantaras, S.~Zonouz, and A.~Petropulu, ``Side-channel-based code-execution monitoring systems: a survey,'' \emph{IEEE Signal Processing Magazine}, vol.~36, no.~2, pp. 22--35, 2019.

\bibitem[He et~al.(2019)He, Park, Nahiyan, Vassilev, Jin, and Tehranipoor]{He2019Rtl}
M.~He, J.~Park, A.~Nahiyan, A.~Vassilev, Y.~Jin, and M.~Tehranipoor, ``Rtl-psc: Automated power side-channel leakage assessment at register-transfer level,'' in \emph{2019 IEEE 37th VLSI Test Symposium (VTS)}.\hskip 1em plus 0.5em minus 0.4em\relax IEEE, 2019, pp. 1--6.

\bibitem[Pundir et~al.(2022)Pundir, Park, Farahmandi, and Tehranipoor]{Pundir2022Power}
N.~Pundir, J.~Park, F.~Farahmandi, and M.~Tehranipoor, ``Power side-channel leakage assessment framework at register-transfer level,'' \emph{IEEE Transactions on Very Large Scale Integration (VLSI) Systems}, 2022.

\bibitem[Jayasena et~al.(2023)Jayasena, Andrews, and Mishra]{jayasena2023tvla}
A.~Jayasena, E.~Andrews, and P.~Mishra, ``Tvla*: Test vector leakage assessment on hardware implementations of asymmetric cryptography algorithms,'' \emph{IEEE Transactions on Very Large Scale Integration (VLSI) Systems}, 2023.

\bibitem[Knechtel et~al.(2020)Knechtel, Patnaik, Nabeel, Ashraf, Chauhan, Henkel, Sinanoglu, and Amrouch]{Knechtel2020Power}
J.~Knechtel, S.~Patnaik, M.~Nabeel, M.~Ashraf, Y.~S. Chauhan, J.~Henkel, O.~Sinanoglu, and H.~Amrouch, ``Power side-channel attacks in negative capacitance transistor,'' \emph{IEEE Micro}, vol.~40, pp. 74--84, 11 2020.

\bibitem[Liu et~al.(2022)Liu, Chakraborty, Chawla, and Roggel]{Liu2022Frequency}
C.~Liu, A.~Chakraborty, N.~Chawla, and N.~Roggel, ``Frequency throttling side-channel attack,'' in \emph{Proceedings of the 2022 ACM SIGSAC Conference on Computer and Communications Security}, 2022, pp. 1977--1991.

\bibitem[Aysu et~al.(2018)Aysu, Orshansky, and Tiwari]{Aysu2018Binary}
A.~Aysu, M.~Orshansky, and M.~Tiwari, ``Binary ring-lwe hardware with power side-channel countermeasures,'' in \emph{2018 Design, Automation \& Test in Europe Conference \& Exhibition (DATE)}.\hskip 1em plus 0.5em minus 0.4em\relax IEEE, 2018, pp. 1253--1258.

\bibitem[Luo et~al.(2015)Luo, Fei, Luo, Mukherjee, and Kaeli]{Luo2015Side}
C.~Luo, Y.~Fei, P.~Luo, S.~Mukherjee, and D.~Kaeli, ``Side-channel power analysis of a gpu aes implementation,'' in \emph{2015 33rd IEEE International Conference on Computer Design (ICCD)}.\hskip 1em plus 0.5em minus 0.4em\relax IEEE, 2015, pp. 281--288.

\bibitem[Benhadjyoussef et~al.(2021{\natexlab{a}})Benhadjyoussef, Karmani, and Machhout]{Benhadjyoussef2021Power}
N.~Benhadjyoussef, M.~Karmani, and M.~Machhout, ``Power-based side channel analysis and fault injection: Hacking techniques and combined countermeasure,'' \emph{International Journal of Advanced Computer Science and Applications}, vol.~12, no.~5, 2021.

\bibitem[Munny and Hu(2021)]{Munny2021Power}
R.~Munny and J.~Hu, ``Power side-channel attack detection through battery impedance monitoring,'' vol. 2021-May.\hskip 1em plus 0.5em minus 0.4em\relax Institute of Electrical and Electronics Engineers Inc., 2021.

\bibitem[Iyer et~al.(2021)Iyer, Wang, Kulkarni, and Yilmaz]{Iyer2021Systematic}
V.~Iyer, M.~Wang, J.~Kulkarni, and A.~E. Yilmaz, ``A systematic evaluation of em and power side-channel analysis attacks on aes implementations.''\hskip 1em plus 0.5em minus 0.4em\relax Institute of Electrical and Electronics Engineers Inc., 2021.

\bibitem[Messerges(2000)]{Messerges2000Using}
T.~S. Messerges, ``Using second-order power analysis to attack dpa resistant software,'' in \emph{International Workshop on Cryptographic Hardware and Embedded Systems}.\hskip 1em plus 0.5em minus 0.4em\relax Springer, 2000, pp. 238--251.

\bibitem[Tehranipoor and Wang(2011)]{tTehranipoor2011Introduction}
M.~Tehranipoor and C.~Wang, \emph{Introduction to hardware security and trust}.\hskip 1em plus 0.5em minus 0.4em\relax Springer Science \& Business Media, 2011.

\bibitem[Bellizia et~al.(2021)Bellizia, Sala, and Scotti]{Bellizia2021Sc}
D.~Bellizia, R.~D. Sala, and G.~Scotti, ``Sc-ddpl as a countermeasure against static power side-channel attacks,'' \emph{Cryptography}, vol.~5, 9 2021.

\bibitem[Moos et~al.(2020)Moos, Moradi, and Richter]{Moos2020Static}
T.~Moos, A.~Moradi, and B.~Richter, ``Static power side-channel analysis - an investigation of measurement factors,'' \emph{IEEE Transactions on Very Large Scale Integration (VLSI) Systems}, vol.~28, pp. 376--389, 2 2020.

\bibitem[Benhadjyoussef et~al.(2021{\natexlab{b}})Benhadjyoussef, Karmani, and Machhout]{Benhadjyoussef2021PowerAES}
\BIBentryALTinterwordspacing
N.~Benhadjyoussef, M.~Karmani, and M.~Machhout, ``Power-based side-channel analysis against aes implementations: Evaluation and comparison,'' \emph{IJCSNS International Journal of Computer Science and Network Security}, vol.~21, 2021. [Online]. Available: \url{https://doi.org/10.22937/IJCSNS.2021.21.4.32}
\BIBentrySTDinterwordspacing

\bibitem[Dumitru et~al.(2023)Dumitru, Wabnitz, and Yarom]{Dumitru2023Borrowed}
R.~Dumitru, A.~Wabnitz, and Y.~Yarom, ``On borrowed time--preventing static power side-channel analysis,'' \emph{arXiv preprint arXiv:2307.09001}, 2023.

\bibitem[Moos et~al.(2017)Moos, Moradi, and Richter]{Moos2017Static}
T.~Moos, A.~Moradi, and B.~Richter, ``Static power side-channel analysis of a threshold implementation prototype chip,'' in \emph{Design, Automation \& Test in Europe Conference \& Exhibition (DATE), 2017}.\hskip 1em plus 0.5em minus 0.4em\relax IEEE, 2017, pp. 1324--1329.

\bibitem[Vafa et~al.(2020)Vafa, Masoumi, and Amini]{Vafa2020Efficient}
S.~Vafa, M.~Masoumi, and A.~Amini, ``An efficient profiling attack to real codes of pic16f690 and arm cortex-m3,'' \emph{IEEE Access}, vol.~8, pp. 222\,520--222\,532, 2020.

\bibitem[Brier et~al.(2004)Brier, Clavier, and Olivier]{Brier2004Correlation}
E.~Brier, C.~Clavier, and F.~Olivier, ``Correlation power analysis with a leakage model,'' in \emph{Cryptographic Hardware and Embedded Systems-CHES 2004: 6th International Workshop Cambridge, MA, USA, August 11-13, 2004. Proceedings 6}.\hskip 1em plus 0.5em minus 0.4em\relax Springer, 2004, pp. 16--29.

\bibitem[Zha(2020)]{Zhang2020Memory}
``Memory-based high-level synthesis optimizations security exploration on the power side-channel,'' \emph{IEEE Transactions on Computer-Aided Design of Integrated Circuits and Systems}, vol.~39, pp. 2124--2137, 10 2020.

\bibitem[Wei et~al.(2018)Wei, Luo, Li, Liu, and Xu]{Wei2018I}
L.~Wei, B.~Luo, Y.~Li, Y.~Liu, and Q.~Xu, ``I know what you see: Power side-channel attack on convolutional neural network accelerators.''\hskip 1em plus 0.5em minus 0.4em\relax Association for Computing Machinery, 12 2018, pp. 393--406.

\bibitem[Srivastava et~al.(2024)Srivastava, Das, Choudhury, Psiakis, Silva, Pal, and Basu]{Srivastava2024Scar}
A.~Srivastava, S.~Das, N.~Choudhury, R.~Psiakis, P.~H. Silva, D.~Pal, and K.~Basu, ``Scar: Power side-channel analysis at rtl level,'' \emph{IEEE Transactions on Very Large Scale Integration (VLSI) Systems}, 2024.

\bibitem[Ne{\v{s}}kovi{\'c} et~al.(2023)Ne{\v{s}}kovi{\'c}, Mulhem, Treff, Buchty, Eisenbarth, and Berekovic]{Nevskovic2023Systemc}
A.~Ne{\v{s}}kovi{\'c}, S.~Mulhem, A.~Treff, R.~Buchty, T.~Eisenbarth, and M.~Berekovic, ``Systemc model of power side-channel attacks against ai accelerators: Superstition or not?'' in \emph{2023 IEEE/ACM International Conference on Computer Aided Design (ICCAD)}.\hskip 1em plus 0.5em minus 0.4em\relax IEEE, 2023, pp. 1--8.

\bibitem[Lipp et~al.(2021)Lipp, Kogler, Oswald, Schwarz, Easdon, Canella, and Gruss]{Lipp2021Platypus}
M.~Lipp, A.~Kogler, D.~Oswald, M.~Schwarz, C.~Easdon, C.~Canella, and D.~Gruss, ``Platypus: Software-based power side-channel attacks on x86,'' in \emph{2021 IEEE Symposium on Security and Privacy (SP)}.\hskip 1em plus 0.5em minus 0.4em\relax IEEE, 2021, pp. 355--371.

\bibitem[Peeters et~al.(2007)Peeters, Standaert, and Quisquater]{Peeters2007Power}
E.~Peeters, F.-X. Standaert, and J.-J. Quisquater, ``Power and electromagnetic analysis: Improved model, consequences and comparisons,'' \emph{Integration}, vol.~40, no.~1, pp. 52--60, 2007.

\bibitem[Tran et~al.(2023)Tran, Dao, Hoang, Hoang, and Pham]{Tran2023Transition}
T.-H. Tran, B.-A. Dao, T.-T. Hoang, V.-P. Hoang, and C.-K. Pham, ``Transition factors of power consumption models for cpa attacks on cryptographic risc-v soc,'' \emph{IEEE Transactions on Computers}, vol.~72, no.~9, pp. 2689--2700, 2023.

\bibitem[Liu et~al.(2010)Liu, Qian, Goto, and Tsunoo]{Liu2010Aes}
H.~Liu, G.~Qian, S.~Goto, and Y.~Tsunoo, ``Aes key recovery based on switching distance model,'' in \emph{2010 Third International Symposium on Electronic Commerce and Security}.\hskip 1em plus 0.5em minus 0.4em\relax IEEE, 2010, pp. 218--222.

\bibitem[Mestiri et~al.(2013)Mestiri, Benhadjyoussef, Machhout, and Tourki]{Mestiri2013Comparative}
H.~Mestiri, N.~Benhadjyoussef, M.~Machhout, and R.~Tourki, ``A comparative study of power consumption models for cpa attack,'' \emph{International Journal of Computer Network and Information Security}, vol.~5, no.~3, p.~25, 2013.

\bibitem[Cao et~al.(2020)Cao, Huang, Zheng, and Hu]{Cao2020Attacking}
W.~Cao, F.~Huang, M.~Zheng, and H.~Hu, ``Attacking fpga-based dual complementary aes implementation using hd and sd models,'' in \emph{2020 16th International Conference on Computational Intelligence and Security (CIS)}.\hskip 1em plus 0.5em minus 0.4em\relax IEEE, 2020, pp. 278--282.

\bibitem[Schindler et~al.(2005)Schindler, Lemke, and Paar]{schindler2005stochastic}
W.~Schindler, K.~Lemke, and C.~Paar, ``A stochastic model for differential side channel cryptanalysis,'' in \emph{International Workshop on Cryptographic Hardware and Embedded Systems}.\hskip 1em plus 0.5em minus 0.4em\relax Springer, 2005, pp. 30--46.

\bibitem[Xiang et~al.(2020)Xiang, Chen, Chen, Fang, Hao, Chen, Liu, Wu, Xuan, and Yang]{Xiang2020Open}
Y.~Xiang, Z.~Chen, Z.~Chen, Z.~Fang, H.~Hao, J.~Chen, Y.~Liu, Z.~Wu, Q.~Xuan, and X.~Yang, ``Open dnn box by power side-channel attack,'' \emph{IEEE Transactions on Circuits and Systems II: Express Briefs}, vol.~67, pp. 2717--2721, 11 2020.

\bibitem[Wang et~al.(2022)Wang, Paccagnella, He, Shacham, Fletcher, and Kohlbrenner]{Wang2022Hertzbleed}
Y.~Wang, R.~Paccagnella, E.~T. He, H.~Shacham, C.~W. Fletcher, and D.~Kohlbrenner, ``Hertzbleed: Turning power $\{$Side-Channel$\}$ attacks into remote timing attacks on x86,'' in \emph{31st USENIX Security Symposium (USENIX Security 22)}, 2022, pp. 679--697.

\bibitem[Kogler et~al.(2023)Kogler, Juffinger, Giner, Gerlach, Schwarzl, Schwarz, Gruss, and Mangard]{Kogler2023Collide+}
A.~Kogler, J.~Juffinger, L.~Giner, L.~Gerlach, M.~Schwarzl, M.~Schwarz, D.~Gruss, and S.~Mangard, ``$\{$Collide+ Power$\}$: Leaking inaccessible data with software-based power side channels,'' in \emph{32nd USENIX Security Symposium (USENIX Security 23)}, 2023, pp. 7285--7302.

\bibitem[Sanjaya et~al.(2025{\natexlab{a}})Sanjaya, Jayasena, and Mishra]{Sanjaya2024Information}
S.~Sanjaya, A.~Jayasena, and P.~Mishra, ``Information leakage through physical layer supply voltage coupling vulnerability,'' \emph{IEEE Transactions on Very Large Scale Integration (VLSI) Systems}, 2025.

\bibitem[Kocher(1996)]{Kocher1996Timing}
P.~C. Kocher, ``Timing attacks on implementations of diffie-hellman, rsa, dss, and other systems,'' in \emph{Advances in Cryptology—CRYPTO’96: 16th Annual International Cryptology Conference Santa Barbara, California, USA August 18--22, 1996 Proceedings 16}.\hskip 1em plus 0.5em minus 0.4em\relax Springer, 1996, pp. 104--113.

\bibitem[Mozipo and Acken(2023)]{Mozipo2023Residual}
A.~T. Mozipo and J.~M. Acken, ``Residual vulnerabilities to power side channel attacks of lightweight ciphers cryptography competition finalists,'' pp. 75--88, 7 2023.

\bibitem[Jayasena et~al.(2024)Jayasena, Bachmann, and Mishra]{jayasena2024evilcs}
A.~Jayasena, R.~Bachmann, and P.~Mishra, ``Evilcs: An evaluation of information leakage through context switching on security enclaves,'' in \emph{2024 Design, Automation \& Test in Europe Conference \& Exhibition (DATE)}.\hskip 1em plus 0.5em minus 0.4em\relax IEEE, 2024, pp. 1--6.

\bibitem[Jayasena et~al.(2025)]{jayasena2024ciseleaks}
A.~Jayasena \emph{et~al.}, ``{CISELEAKS}: Information leakage assessment of cryptographic instruction set extension prototypes,'' \emph{IEEE Transactions on Information Forensics and Security}, 2025.

\bibitem[Gilbert~Goodwill et~al.(2011)Gilbert~Goodwill, Jaffe, Rohatgi, et~al.]{Gilbert2011Testing}
B.~J. Gilbert~Goodwill, J.~Jaffe, P.~Rohatgi \emph{et~al.}, ``A testing methodology for side-channel resistance validation,'' in \emph{NIST non-invasive attack testing workshop}, vol.~7, 2011, pp. 115--136.

\bibitem[Kullback and Leibler(1951)]{Kullback1951Information}
S.~Kullback and R.~A. Leibler, ``On information and sufficiency,'' \emph{The annals of mathematical statistics}, vol.~22, no.~1, pp. 79--86, 1951.

\bibitem[Miura et~al.(2023)Miura, Nishikawa, Kong, and Tomiyama]{Miura2023Simulation}
Y.~Miura, H.~Nishikawa, X.~Kong, and H.~Tomiyama, ``Simulation-based analysis of power side-channel leakage at different sampling intervals.''\hskip 1em plus 0.5em minus 0.4em\relax Institute of Electrical and Electronics Engineers Inc., 2023, pp. 364--365.

\bibitem[Weisstein(2004)]{weisstein2004bonferroni}
E.~Weisstein, ``Bonferroni correction,'' \emph{https://mathworld.wolfram.com/}, 2004.

\bibitem[fip()]{fips}
``Fips 140-3 security requirements for cryptographic modules,'' \url{https://csrc.nist.gov/pubs/fips/140-3/}, accessed: 2025-11-20.

\bibitem[cco()]{ccorg}
``The common criteria,'' \url{https://www.commoncriteriaportal.org/}, accessed: 2025-11-20.

\bibitem[Roy et~al.(2018)Roy, Bhasin, Guilley, Heuser, Patranabis, and Mukhopadhyay]{roy2018cc}
D.~B. Roy, S.~Bhasin, S.~Guilley, A.~Heuser, S.~Patranabis, and D.~Mukhopadhyay, ``Cc meets fips: A hybrid test methodology for first order side channel analysis,'' \emph{IEEE Transactions on Computers}, vol.~68, no.~3, pp. 347--361, 2018.

\bibitem[Standaert(2018)]{standaert2018not}
F.-X. Standaert, ``How (not) to use welch’s t-test in side-channel security evaluations,'' in \emph{International conference on smart card research and advanced applications}.\hskip 1em plus 0.5em minus 0.4em\relax Springer, 2018, pp. 65--79.

\bibitem[Timon(2019)]{Timon2019Non}
B.~Timon, ``Non-profiled deep learning-based side-channel attacks with sensitivity analysis,'' \emph{IACR Transactions on Cryptographic Hardware and Embedded Systems}, pp. 107--131, 2019.

\bibitem[Ahmed et~al.(2023{\natexlab{a}})Ahmed, Hasan, Nafi, Aman, Islam, and Nahi]{Ahmed2023Optimization}
A.~A. Ahmed, M.~K. Hasan, N.~S. Nafi, A.~H. Aman, S.~Islam, and M.~S. Nahi, ``Optimization technique for deep learning methodology on power side channel attacks.''\hskip 1em plus 0.5em minus 0.4em\relax Institute of Electrical and Electronics Engineers Inc., 2023, pp. 80--83.

\bibitem[Ahmed et~al.(2023{\natexlab{b}})Ahmed, Salim, and Hasan]{Ahmed2023Deep}
A.~A. Ahmed, R.~A. Salim, and M.~K. Hasan, ``Deep learning method for power side-channel analysis on chip leakages,'' \emph{Elektronika ir Elektrotechnika}, vol.~29, pp. 50--57, 2023.

\bibitem[Wang et~al.(2023{\natexlab{a}})Wang, Paccagnella, Wandke, Gang, Garrett-Grossman, Fletcher, Kohlbrenner, and Shacham]{Wang2023Dvfs}
Y.~Wang, R.~Paccagnella, A.~Wandke, Z.~Gang, G.~Garrett-Grossman, C.~W. Fletcher, D.~Kohlbrenner, and H.~Shacham, ``Dvfs frequently leaks secrets: Hertzbleed attacks beyond sike, cryptography, and cpu-only data,'' in \emph{2023 IEEE Symposium on Security and Privacy (SP)}.\hskip 1em plus 0.5em minus 0.4em\relax IEEE, 2023, pp. 2306--2320.

\bibitem[Yu et~al.(2024)Yu, Cheng, Yang, Wang, Pan, and Weng]{Yu2024Hints}
T.~Yu, C.~Cheng, Z.~Yang, Y.~Wang, Y.~Pan, and J.~Weng, ``Hints from hertz: Dynamic frequency scaling side-channel analysis of number theoretic transform in lattice-based kems,'' \emph{IACR Transactions on Cryptographic Hardware and Embedded Systems}, vol. 2024, no.~3, pp. 200--223, 2024.

\bibitem[Taneja et~al.(2023)Taneja, Kim, Xu, Van~Schaik, Genkin, and Yarom]{Taneja2023Hot}
H.~Taneja, J.~Kim, J.~J. Xu, S.~Van~Schaik, D.~Genkin, and Y.~Yarom, ``Hot pixels: Frequency, power, and temperature attacks on $\{$GPUs$\}$ and arm $\{$SoCs$\}$,'' in \emph{32nd USENIX Security Symposium (USENIX Security 23)}, 2023, pp. 6275--6292.

\bibitem[Sanjaya et~al.(2025{\natexlab{b}})Sanjaya, Jayasena, and Mishra]{Sanjaya2025Sleepwalk}
S.~Sanjaya, A.~Jayasena, and P.~Mishra, ``Sleepwalk: Exploiting context switching and residual power for physical side-channel attacks,'' \emph{arXiv preprint arXiv:2507.22306}, 2025.

\bibitem[Moini et~al.(2021)Moini, Tian, Holcomb, Szefer, and Tessier]{Moini2021Power}
S.~Moini, S.~Tian, D.~Holcomb, J.~Szefer, and R.~Tessier, ``Power side-channel attacks on bnn accelerators in remote fpgas,'' \emph{IEEE Journal on Emerging and Selected Topics in Circuits and Systems}, vol.~11, pp. 357--370, 6 2021.

\bibitem[Zhang et~al.(2023)Zhang, Ding, and Fei]{Zhang2023Deep}
X.~Zhang, A.~A. Ding, and Y.~Fei, ``Deep-learning model extraction through software-based power side-channel.''\hskip 1em plus 0.5em minus 0.4em\relax Institute of Electrical and Electronics Engineers Inc., 2023.

\bibitem[Gatlin et~al.(2021)Gatlin, Belikovetsky, Elovici, Skjellum, Lubell, Witherell, and Yampolskiy]{Gatlin2021Encryption}
J.~Gatlin, S.~Belikovetsky, Y.~Elovici, A.~Skjellum, J.~Lubell, P.~Witherell, and M.~Yampolskiy, ``Encryption is futile: Reconstructing 3d-printed models using the power side-channel.''\hskip 1em plus 0.5em minus 0.4em\relax Association for Computing Machinery, 10 2021, pp. 135--147.

\bibitem[Chari et~al.(2002)Chari, Rao, and Rohatgi]{chari2002template}
S.~Chari, J.~R. Rao, and P.~Rohatgi, ``Template attacks,'' in \emph{International workshop on cryptographic hardware and embedded systems}.\hskip 1em plus 0.5em minus 0.4em\relax Springer, 2002, pp. 13--28.

\bibitem[Mangard et~al.(2007)Mangard, Oswald, and Popp]{mangard2007power}
S.~Mangard, E.~Oswald, and T.~Popp, \emph{Power analysis attacks: Revealing the secrets of smart cards}.\hskip 1em plus 0.5em minus 0.4em\relax Springer, 2007.

\bibitem[Heuser and Zohner(2012)]{heuser2012intelligent}
A.~Heuser and M.~Zohner, ``Intelligent machine homicide: Breaking cryptographic devices using support vector machines,'' in \emph{International Workshop on Constructive Side-Channel Analysis and Secure Design}.\hskip 1em plus 0.5em minus 0.4em\relax Springer, 2012, pp. 249--264.

\bibitem[Lerman et~al.(2015)Lerman, Bontempi, and Markowitch]{lerman2015machine}
L.~Lerman, G.~Bontempi, and O.~Markowitch, ``A machine learning approach against a masked aes: Reaching the limit of side-channel attacks with a learning model,'' \emph{Journal of Cryptographic Engineering}, vol.~5, no.~2, pp. 123--139, 2015.

\bibitem[Gilmore et~al.(2015)Gilmore, Hanley, and O'Neill]{gilmore2015neural}
R.~Gilmore, N.~Hanley, and M.~O'Neill, ``Neural network based attack on a masked implementation of aes,'' in \emph{2015 IEEE International Symposium on Hardware Oriented Security and Trust (HOST)}.\hskip 1em plus 0.5em minus 0.4em\relax IEEE, 2015, pp. 106--111.

\bibitem[Maghrebi et~al.(2016)Maghrebi, Portigliatti, and Prouff]{maghrebi2016breaking}
H.~Maghrebi, T.~Portigliatti, and E.~Prouff, ``Breaking cryptographic implementations using deep learning techniques,'' in \emph{International Conference on Security, Privacy, and Applied Cryptography Engineering}.\hskip 1em plus 0.5em minus 0.4em\relax Springer, 2016, pp. 3--26.

\bibitem[Gao et~al.(2023)Gao, Qiu, Zhang, Wang, Ma, Abuadbba, Xue, Fu, and Nepal]{Gao2023Deeptheft}
Y.~Gao, H.~Qiu, Z.~Zhang, B.~Wang, H.~Ma, A.~Abuadbba, M.~Xue, A.~Fu, and S.~Nepal, ``Deeptheft: Stealing dnn model architectures through power side channel,'' \emph{arXiv preprint arXiv:2309.11894}, 2023.

\bibitem[Das and Sen(2020)]{Das2020Electromagnetic}
D.~Das and S.~Sen, ``Electromagnetic and power side-channel analysis: Advanced attacks and low-overhead generic countermeasures through white-box approach,'' \emph{Cryptography}, vol.~4, pp. 1--19, 12 2020.

\bibitem[Ghandali et~al.(2021)Ghandali, Ghandali, and Tehranipoor]{Ghandali2021Deep}
S.~Ghandali, S.~Ghandali, and S.~Tehranipoor, ``Deep k-tsvm: A novel profiled power side-channel attack on aes-128,'' \emph{IEEE Access}, vol.~9, pp. 136\,448--136\,458, 2021.

\bibitem[Zhang et~al.(2021{\natexlab{a}})Zhang, Yasaei, Chen, Li, and Al~Faruque]{Zhang2021Stealing}
Y.~Zhang, R.~Yasaei, H.~Chen, Z.~Li, and M.~A. Al~Faruque, ``Stealing neural network structure through remote fpga side-channel analysis,'' \emph{IEEE Transactions on Information Forensics and Security}, vol.~16, pp. 4377--4388, 2021.

\bibitem[Meyers et~al.(2022)Meyers, Gnad, and Tahoori]{Meyers2022Reverse}
V.~Meyers, D.~Gnad, and M.~Tahoori, ``Reverse engineering neural network folding with remote fpga power analysis,'' in \emph{2022 IEEE 30th Annual International Symposium on Field-Programmable Custom Computing Machines (FCCM)}.\hskip 1em plus 0.5em minus 0.4em\relax IEEE, 2022, pp. 1--10.

\bibitem[Wang et~al.(2023{\natexlab{b}})Wang, Wu, Park, Yoo, Wang, Eshraghian, and Lu]{Wang2023Powergan}
Z.~Wang, Y.~Wu, Y.~Park, S.~Yoo, X.~Wang, J.~K. Eshraghian, and W.~D. Lu, ``Powergan: A machine learning approach for power side-channel attack on compute-in-memory accelerators,'' \emph{Advanced Intelligent Systems}, vol.~5, no.~12, p. 2300313, 2023.

\bibitem[Hajra et~al.(2022)Hajra, Saha, Alam, and Mukhopadhyay]{Hajra2022Transnet}
S.~Hajra, S.~Saha, M.~Alam, and D.~Mukhopadhyay, ``Transnet: Shift invariant transformer network for side channel analysis,'' in \emph{International Conference on Cryptology in Africa}.\hskip 1em plus 0.5em minus 0.4em\relax Springer, 2022, pp. 371--396.

\bibitem[Hajra et~al.(2024)Hajra, Chowdhury, and Mukhopadhyay]{Hajra2024Estranet}
S.~Hajra, S.~Chowdhury, and D.~Mukhopadhyay, ``Estranet: An efficient shift-invariant transformer network for side-channel analysis,'' \emph{IACR Transactions on Cryptographic Hardware and Embedded Systems}, vol. 2024, no.~1, pp. 336--374, 2024.

\bibitem[Jayasankaran et~al.(2023)Jayasankaran, Guo, Patnaik, Jeyavijayan, Rajendran, and Hu]{Jayasankaran2023Securing}
\BIBentryALTinterwordspacing
N.~G. Jayasankaran, H.~Guo, S.~Patnaik, Jeyavijayan, Rajendran, and J.~Hu, ``Securing cloud fpgas against power side-channel attacks: A case study on iterative aes,'' 7 2023. [Online]. Available: \url{http://arxiv.org/abs/2307.02569}
\BIBentrySTDinterwordspacing

\bibitem[Schellenberg et~al.(2021)Schellenberg, Gnad, Moradi, and Tahoori]{Schellenberg2021Inside}
F.~Schellenberg, D.~R. Gnad, A.~Moradi, and M.~B. Tahoori, ``An inside job: Remote power analysis attacks on fpgas,'' \emph{IEEE Design \& Test}, vol.~38, no.~3, pp. 58--66, 2021.

\bibitem[Gnad et~al.(2020)Gnad, Krautter, Tahoori, Schellenberg, and Moradi]{Gnad2020Remote}
D.~R. Gnad, J.~Krautter, M.~B. Tahoori, F.~Schellenberg, and A.~Moradi, ``Remote electrical-level security threats to multi-tenant fpgas,'' \emph{IEEE Design \& Test}, vol.~37, no.~2, pp. 111--119, 2020.

\bibitem[Gravellier et~al.(2019)Gravellier, Dutertre, Teglia, and Loubet-Moundi]{Gravellier2019High}
J.~Gravellier, J.-M. Dutertre, Y.~Teglia, and P.~Loubet-Moundi, ``High-speed ring oscillator based sensors for remote side-channel attacks on fpgas,'' in \emph{2019 International conference on ReConFigurable computing and FPGAs (ReConFig)}.\hskip 1em plus 0.5em minus 0.4em\relax IEEE, 2019, pp. 1--8.

\bibitem[Ramesh et~al.(2018)Ramesh, Patil, Dhanuskodi, Provelengios, Pillement, Holcomb, and Tessier]{Ramesh2018Fpga}
C.~Ramesh, S.~B. Patil, S.~N. Dhanuskodi, G.~Provelengios, S.~Pillement, D.~Holcomb, and R.~Tessier, ``Fpga side channel attacks without physical access,'' in \emph{2018 IEEE 26th Annual International Symposium on Field-Programmable Custom Computing Machines (FCCM)}.\hskip 1em plus 0.5em minus 0.4em\relax IEEE, 2018, pp. 45--52.

\bibitem[O’Flynn and Dewar(2019)]{O2019Device}
C.~O’Flynn and A.~Dewar, ``On-device power analysis across hardware security domains.: Stop hitting yourself.'' \emph{IACR Transactions on Cryptographic Hardware and Embedded Systems}, pp. 126--153, 2019.

\bibitem[Ahmadi et~al.(2023)Ahmadi, Khalid, Vaidya, Kriebel, Steininger, and Shafique]{Ahmadi2023Shield}
\BIBentryALTinterwordspacing
M.~M. Ahmadi, F.~Khalid, R.~Vaidya, F.~Kriebel, A.~Steininger, and M.~Shafique, ``Shield: An adaptive and lightweight defense against the remote power side-channel attacks on multi-tenant fpgas,'' 3 2023. [Online]. Available: \url{http://arxiv.org/abs/2303.06486}
\BIBentrySTDinterwordspacing

\bibitem[Kamucheka et~al.(2021)Kamucheka, Fahr, Teague, Nelson, Andrews, and Huang]{Kamucheka2021Power}
T.~Kamucheka, M.~Fahr, T.~Teague, A.~Nelson, D.~Andrews, and M.~Huang, ``Power-based side channel attack analysis on pqc algorithms,'' \emph{Cryptology ePrint Archive}, 2021.

\bibitem[Lipp et~al.(2022)Lipp, Gruss, and Schwarz]{Lipp2022Amd}
M.~Lipp, D.~Gruss, and M.~Schwarz, ``$\{$AMD$\}$ prefetch attacks through power and time,'' in \emph{31st USENIX Security Symposium (USENIX Security 22)}, 2022, pp. 643--660.

\bibitem[Lee and Han(2020{\natexlab{a}})]{Lee2020Dlddo}
J.~Lee and D.-G. Han, ``Dlddo: deep learning to detect dummy operations,'' in \emph{International Conference on Information Security Applications}.\hskip 1em plus 0.5em minus 0.4em\relax Springer, 2020, pp. 73--85.

\bibitem[Jevtic et~al.(2022)Jevtic, Perez-Tirador, Cabezaolias, Carnero, and Caffarena]{Jevtic2022Side}
R.~Jevtic, P.~Perez-Tirador, C.~Cabezaolias, P.~Carnero, and G.~Caffarena, ``Side-channel attack countermeasure based on power supply modulation,'' in \emph{2022 30th European Signal Processing Conference (EUSIPCO)}.\hskip 1em plus 0.5em minus 0.4em\relax IEEE, 2022, pp. 618--622.

\bibitem[Lee and Han(2020{\natexlab{b}})]{Lee2020Security}
J.~Lee and D.-G. Han, ``Security analysis on dummy based side-channel countermeasures—case study: Aes with dummy and shuffling,'' \emph{Applied Soft Computing}, vol.~93, p. 106352, 2020.

\bibitem[Saito et~al.(2022)Saito, Ito, Ueno, and Homma]{Saito2022One}
K.~Saito, A.~Ito, R.~Ueno, and N.~Homma, ``One truth prevails: A deep-learning based single-trace power analysis on rsa--crt with windowed exponentiation,'' \emph{IACR Transactions on Cryptographic Hardware and Embedded Systems}, pp. 490--526, 2022.

\bibitem[Moody(2018)]{moody2018let}
D.~Moody, ``Let’s get ready to rumble. the nist pqc “competition”,'' in \emph{Proc. of First PQC Standardization Conference}, 2018, pp. 11--13.

\bibitem[Alagic et~al.(2019)Alagic, Alagic, Alperin-Sheriff, Apon, Cooper, Dang, Liu, Miller, Moody, Peralta, et~al.]{alagic2019status}
G.~Alagic, G.~Alagic, J.~Alperin-Sheriff, D.~Apon, D.~Cooper, Q.~Dang, Y.-K. Liu, C.~Miller, D.~Moody, R.~Peralta \emph{et~al.}, ``Status report on the first round of the nist post-quantum cryptography standardization process,'' 2019.

\bibitem[Alagic et~al.(2020)Alagic, Alperin-Sheriff, Apon, Cooper, Dang, Kelsey, Liu, Miller, Moody, Peralta, et~al.]{alagic2020status}
G.~Alagic, J.~Alperin-Sheriff, D.~Apon, D.~Cooper, Q.~Dang, J.~Kelsey, Y.-K. Liu, C.~Miller, D.~Moody, R.~Peralta \emph{et~al.}, ``Status report on the second round of the nist post-quantum cryptography standardization process,'' \emph{US Department of Commerce, NIST}, vol.~2, p.~69, 2020.

\bibitem[Alagic et~al.(2022)Alagic, Alagic, Apon, Cooper, Dang, Dang, Kelsey, Lichtinger, Liu, Miller, et~al.]{alagic2022status}
G.~Alagic, G.~Alagic, D.~Apon, D.~Cooper, Q.~Dang, T.~Dang, J.~Kelsey, J.~Lichtinger, Y.-K. Liu, C.~Miller \emph{et~al.}, ``Status report on the third round of the nist post-quantum cryptography standardization process,'' 2022.

\bibitem[Zick et~al.(2013)Zick, Srivastav, Zhang, and French]{Zick2013Sensing}
K.~M. Zick, M.~Srivastav, W.~Zhang, and M.~French, ``Sensing nanosecond-scale voltage attacks and natural transients in fpgas,'' in \emph{Proceedings of the ACM/SIGDA international symposium on Field programmable gate arrays}, 2013, pp. 101--104.

\bibitem[Zhang et~al.(2021{\natexlab{b}})Zhang, Liang, Yao, and Gao]{Zhang2021Red}
Z.~Zhang, S.~Liang, F.~Yao, and X.~Gao, ``Red alert for power leakage: Exploiting intel rapl-induced side channels,'' in \emph{Proceedings of the 2021 ACM Asia Conference on Computer and Communications Security}, 2021, pp. 162--175.

\bibitem[Chen et~al.(2017)Chen, Jin, Sun, Zhang, and Zhang]{Chen2017Powerful}
Y.~Chen, X.~Jin, J.~Sun, R.~Zhang, and Y.~Zhang, ``Powerful: Mobile app fingerprinting via power analysis,'' in \emph{IEEE INFOCOM 2017-IEEE Conference on Computer Communications}.\hskip 1em plus 0.5em minus 0.4em\relax IEEE, 2017, pp. 1--9.

\bibitem[La~Cour et~al.(2021)La~Cour, Afridi, and Suh]{La2021Wireless}
A.~S. La~Cour, K.~K. Afridi, and G.~E. Suh, ``Wireless charging power side-channel attacks,'' in \emph{Proceedings of the 2021 ACM SIGSAC Conference on Computer and Communications Security}, 2021, pp. 651--665.

\bibitem[Yang et~al.(2016)Yang, Gasti, Zhou, Farajidavar, and Balagani]{Yang2016Inferring}
Q.~Yang, P.~Gasti, G.~Zhou, A.~Farajidavar, and K.~S. Balagani, ``On inferring browsing activity on smartphones via usb power analysis side-channel,'' \emph{IEEE Transactions on Information Forensics and Security}, vol.~12, no.~5, pp. 1056--1066, 2016.

\bibitem[Matovu et~al.(2020)Matovu, Serwadda, Bilbao, and Griswold-Steiner]{Matovu2020Defensive}
R.~Matovu, A.~Serwadda, A.~V. Bilbao, and I.~Griswold-Steiner, ``Defensive charging: Mitigating power side-channel attacks on charging smartphones.''\hskip 1em plus 0.5em minus 0.4em\relax Association for Computing Machinery, Inc, 3 2020, pp. 179--190.

\bibitem[Cronin et~al.(2021)Cronin, Gao, Yang, and Wang]{Cronin2021Charger}
P.~Cronin, X.~Gao, C.~Yang, and H.~Wang, ``$\{$Charger-Surfing$\}$: Exploiting a power line $\{$Side-Channel$\}$ for smartphone information leakage,'' in \emph{30th USENIX Security Symposium (USENIX Security 21)}, 2021, pp. 681--698.

\bibitem[Yucebas and Yuksel(2018)]{Yucebas2018Power}
D.~Yucebas and H.~Yuksel, ``Power analysis based side-channel attack on visible light communication,'' \emph{Physical Communication}, vol.~31, pp. 196--202, 12 2018.

\bibitem[Spolaor et~al.(2023)Spolaor, Liu, Turrin, Conti, and Cheng]{Spolaor2023Plug}
R.~Spolaor, H.~Liu, F.~Turrin, M.~Conti, and X.~Cheng, ``Plug and power: Fingerprinting usb powered peripherals via power side-channel,'' in \emph{IEEE INFOCOM 2023-IEEE Conference on Computer Communications}.\hskip 1em plus 0.5em minus 0.4em\relax IEEE, 2023, pp. 1--10.

\bibitem[Qin and Yue(2018)]{Qin2018Website}
Y.~Qin and C.~Yue, ``Website fingerprinting by power estimation based side-channel attacks on android 7,'' in \emph{2018 17th IEEE International Conference On Trust, Security And Privacy In Computing And Communications/12th IEEE International Conference On Big Data Science And Engineering (TrustCom/BigDataSE)}.\hskip 1em plus 0.5em minus 0.4em\relax IEEE, 2018, pp. 1030--1039.

\bibitem[Glamo{\v{c}}anin et~al.(2023)Glamo{\v{c}}anin, Shrivastava, Yao, Ardo, Payer, and Stojilovi{\'c}]{Glamovcanin2023Instruction}
O.~Glamo{\v{c}}anin, S.~Shrivastava, J.~Yao, N.~Ardo, M.~Payer, and M.~Stojilovi{\'c}, ``Instruction-level power side-channel leakage evaluation of soft-core cpus on shared fpgas,'' \emph{Journal of Hardware and Systems Security}, vol.~7, no.~2, pp. 72--99, 2023.

\bibitem[Krishnankutty et~al.(2020)Krishnankutty, Li, Robucci, Banerjee, and Patel]{Krishnankutty2020Instruction}
D.~Krishnankutty, Z.~Li, R.~Robucci, N.~Banerjee, and C.~Patel, ``Instruction sequence identification and disassembly using power supply side-channel analysis,'' \emph{IEEE Transactions on Computers}, vol.~69, no.~11, pp. 1639--1653, 2020.

\bibitem[Fendri et~al.(2022)Fendri, Macchetti, Perrine, and Stojilovi{\'c}]{Fendri2022Deep}
H.~Fendri, M.~Macchetti, J.~Perrine, and M.~Stojilovi{\'c}, ``A deep-learning approach to side-channel based cpu disassembly at design time,'' in \emph{2022 Design, Automation \& Test in Europe Conference \& Exhibition (DATE)}.\hskip 1em plus 0.5em minus 0.4em\relax IEEE, 2022, pp. 670--675.

\bibitem[Han et~al.(2022)Han, Chan, Aref, Tippenhauer, and Zonouz]{Han2022Hiding}
\BIBentryALTinterwordspacing
Y.~Han, M.~Chan, Z.~Aref, N.~O. Tippenhauer, and S.~Zonouz, ``Hiding in plain sight? on the efficacy of power side {Channel-Based} control flow monitoring,'' in \emph{31st USENIX Security Symposium (USENIX Security 22)}.\hskip 1em plus 0.5em minus 0.4em\relax Boston, MA: USENIX Association, Aug. 2022, pp. 661--678. [Online]. Available: \url{https://www.usenix.org/conference/usenixsecurity22/presentation/han}
\BIBentrySTDinterwordspacing

\bibitem[Park et~al.(2018)Park, Xu, Jin, Forte, and Tehranipoor]{Park2018Power}
J.~Park, X.~Xu, Y.~Jin, D.~Forte, and M.~Tehranipoor, ``Power-based side-channel instruction-level disassembler,'' vol. Part F137710.\hskip 1em plus 0.5em minus 0.4em\relax Institute of Electrical and Electronics Engineers Inc., 6 2018.

\bibitem[Narimani et~al.(2021)Narimani, Akhaee, and Habibi]{Narimani2021Side}
P.~Narimani, M.~A. Akhaee, and S.~A. Habibi, ``Side-channel based disassembler for avr micro-controllers using convolutional neural networks,'' in \emph{2021 18th International ISC Conference on Information Security and Cryptology (ISCISC)}.\hskip 1em plus 0.5em minus 0.4em\relax IEEE, 2021, pp. 75--80.

\bibitem[Van~Geest and Buhan(2022)]{Van2022Side}
J.~Van~Geest and I.~Buhan, ``A side-channel based disassembler for the arm-cortex m0,'' in \emph{International Conference on Applied Cryptography and Network Security}.\hskip 1em plus 0.5em minus 0.4em\relax Springer, 2022, pp. 183--199.

\bibitem[Bae and Ha(2022)]{Bae2022Implementation}
D.~Bae and J.~Ha, ``Implementation of disassembler on microcontroller using side-channel power consumption leakage,'' \emph{Sensors}, vol.~22, no.~15, p. 5900, 2022.

\bibitem[Ryu et~al.(2023)Ryu, Kim, and Hur]{Ryu2023Gamma}
D.~Ryu, Y.~Kim, and J.~Hur, ``gamma-knife: Extracting neural network architecture through software-based power side-channel,'' \emph{IEEE Transactions on Dependable and Secure Computing}, 2023.

\bibitem[Jha et~al.(2020)Jha, Mittal, Kumar, and Mattela]{Jha2020Deeppeep}
N.~K. Jha, S.~Mittal, B.~Kumar, and G.~Mattela, ``Deeppeep: Exploiting design ramifications to decipher the architecture of compact dnns,'' \emph{ACM Journal on Emerging Technologies in Computing Systems (JETC)}, vol.~17, no.~1, pp. 1--25, 2020.

\bibitem[Arefin and Serwadda(2024)]{Arefin2024Dissecting}
S.~E. Arefin and A.~Serwadda, ``Dissecting the origins of the power side-channel in deep neural networks,'' in \emph{2024 IEEE World AI IoT Congress (AIIoT)}.\hskip 1em plus 0.5em minus 0.4em\relax IEEE, 2024, pp. 414--420.

\bibitem[Tian et~al.(2023)Tian, Moini, Holcomb, Tessier, and Szefer]{Tian2023Practical}
S.~Tian, S.~Moini, D.~Holcomb, R.~Tessier, and J.~Szefer, ``A practical remote power attack on machine learning accelerators in cloud fpgas,'' in \emph{2023 Design, Automation \& Test in Europe Conference \& Exhibition (DATE)}.\hskip 1em plus 0.5em minus 0.4em\relax IEEE, 2023, pp. 1--6.

\bibitem[Huegle et~al.(2023)Huegle, Gotthard, Meyers, Krautter, Gnad, and Tahoori]{Huegle2023Power2picture}
L.~Huegle, M.~Gotthard, V.~Meyers, J.~Krautter, D.~R. Gnad, and M.~B. Tahoori, ``Power2picture: Using generative cnns for input recovery of neural network accelerators through power side-channels on fpgas,'' in \emph{2023 IEEE 31st Annual International Symposium on Field-Programmable Custom Computing Machines (FCCM)}.\hskip 1em plus 0.5em minus 0.4em\relax IEEE, 2023, pp. 155--161.

\bibitem[Wolf et~al.(2021)Wolf, Hu, Cooley, and Borowczak]{Wolf2021Stealing}
S.~Wolf, H.~Hu, R.~Cooley, and M.~Borowczak, ``Stealing machine learning parameters via side channel power attacks,'' vol. 2021-July.\hskip 1em plus 0.5em minus 0.4em\relax IEEE Computer Society, 7 2021, pp. 242--247.

\bibitem[Asfand~Hafeez et~al.(2021)Asfand~Hafeez, Mazyad~Hazzazi, Tariq, Aljaedi, Javed, and Alharbi]{Asfand2021Low}
M.~Asfand~Hafeez, M.~Mazyad~Hazzazi, H.~Tariq, A.~Aljaedi, A.~Javed, and A.~R. Alharbi, ``A low-overhead countermeasure against differential power analysis for aes block cipher,'' \emph{Applied Sciences}, vol.~11, no.~21, p. 10314, 2021.

\bibitem[Bhandari et~al.(2024)Bhandari, Nabeel, Mankali, Sinanoglu, Karri, and Knechtel]{Bhandari2024Lightweight}
J.~Bhandari, M.~Nabeel, L.~Mankali, O.~Sinanoglu, R.~Karri, and J.~Knechtel, ``Lightweight masking against static power side-channel attacks,'' \emph{arXiv preprint arXiv:2402.03196}, 2024.

\bibitem[Parrilla et~al.(2022)Parrilla, Garc{\'\i}a, Castillo, Rodr{\'\i}guez-Bol{\'\i}var, and L{\'o}pez-Villanueva]{Parrilla2022Time}
L.~Parrilla, A.~Garc{\'\i}a, E.~Castillo, S.~Rodr{\'\i}guez-Bol{\'\i}var, and J.~A. L{\'o}pez-Villanueva, ``Time-and amplitude-controlled power noise generator against spa attacks for fpga-based iot devices,'' \emph{Journal of Low Power Electronics and Applications}, vol.~12, no.~3, p.~48, 2022.

\bibitem[Alipour et~al.(2020)Alipour, Papadimitriou, Beroulle, Aerabi, and H{\'e}ly]{Alipour2020Performance}
A.~Alipour, A.~Papadimitriou, V.~Beroulle, E.~Aerabi, and D.~H{\'e}ly, ``On the performance of non-profiled differential deep learning attacks against an aes encryption algorithm protected using a correlated noise generation based hiding countermeasure,'' in \emph{2020 Design, Automation \& Test in Europe Conference \& Exhibition (DATE)}.\hskip 1em plus 0.5em minus 0.4em\relax IEEE, 2020, pp. 614--617.

\bibitem[Yang et~al.(2024)Yang, Ahmed, Inagaki, Sakiyama, Li, and Hara-Azumi]{Yang2024Hardware}
M.~Yang, T.~Ahmed, S.~Inagaki, K.~Sakiyama, Y.~Li, and Y.~Hara-Azumi, ``Hardware/software cooperative design against power side-channel attacks on iot devices,'' \emph{IEEE Internet of Things Journal}, 2024.

\bibitem[Yan et~al.(2023)Yan, Chang, and Zhang]{Yan2023Defense}
X.~Yan, C.~H. Chang, and T.~Zhang, ``Defense against ml-based power side-channel attacks on dnn accelerators with adversarial attacks,'' \emph{arXiv preprint arXiv:2312.04035}, 2023.

\bibitem[Glamo{\v{c}}anin et~al.(2020)Glamo{\v{c}}anin, Coulon, Regazzoni, and Stojilovi{\'c}]{Glamovcanin2020Cloud}
O.~Glamo{\v{c}}anin, L.~Coulon, F.~Regazzoni, and M.~Stojilovi{\'c}, ``Are cloud fpgas really vulnerable to power analysis attacks?'' in \emph{2020 Design, Automation \& Test in Europe Conference \& Exhibition (DATE)}.\hskip 1em plus 0.5em minus 0.4em\relax IEEE, 2020, pp. 1007--1010.

\bibitem[Krautter et~al.(2019)Krautter, Gnad, Schellenberg, Moradi, and Tahoori]{Krautter2019Active}
J.~Krautter, D.~R. Gnad, F.~Schellenberg, A.~Moradi, and M.~B. Tahoori, ``Active fences against voltage-based side channels in multi-tenant fpgas,'' in \emph{2019 IEEE/ACM International Conference on Computer-Aided Design (ICCAD)}.\hskip 1em plus 0.5em minus 0.4em\relax IEEE, 2019, pp. 1--8.

\bibitem[Padmini and Ravindra(2016)]{Padmini2016Calpan}
C.~Padmini and J.~Ravindra, ``Calpan: Countermeasure against leakage power analysis attack by normalized ddpl,'' in \emph{2016 International Conference on Circuit, Power and Computing Technologies (ICCPCT)}.\hskip 1em plus 0.5em minus 0.4em\relax IEEE, 2016, pp. 1--7.

\bibitem[Sengupta et~al.(2019)Sengupta, Mazumdar, Yasin, and Sinanoglu]{Sengupta2019Logic}
A.~Sengupta, B.~Mazumdar, M.~Yasin, and O.~Sinanoglu, ``Logic locking with provable security against power analysis attacks,'' \emph{IEEE Transactions on Computer-Aided Design of Integrated Circuits and Systems}, vol.~39, no.~4, pp. 766--778, 2019.

\bibitem[Das et~al.(2017)Das, Maity, Nasir, Ghosh, Raychowdhury, and Sen]{Das2017High}
D.~Das, S.~Maity, S.~B. Nasir, S.~Ghosh, A.~Raychowdhury, and S.~Sen, ``High efficiency power side-channel attack immunity using noise injection in attenuated signature domain,'' in \emph{2017 IEEE International Symposium on Hardware Oriented Security and Trust (HOST)}.\hskip 1em plus 0.5em minus 0.4em\relax IEEE, 2017, pp. 62--67.

\bibitem[Das et~al.(2018)Das, Maity, Nasir, Ghosh, Raychowdhury, and Sen]{Das2018Asni}
------, ``Asni: Attenuated signature noise injection for low-overhead power side-channel attack immunity,'' \emph{IEEE Transactions on Circuits and Systems I: Regular Papers}, vol.~65, no.~10, pp. 3300--3311, 2018.

\bibitem[Mozipo and Acken(2024)]{Mozipo2024Analysis}
A.~T. Mozipo and J.~M. Acken, ``Analysis of countermeasures against remote and local power side channel attacks using correlation power analysis,'' \emph{IEEE Transactions on Dependable and Secure Computing}, 2024.

\bibitem[Kolhe et~al.(2022)Kolhe, Sheaves, Gubbi, Salehi, Rafatirad, Manoj, Sasan, and Homayoun]{Kolhe2022Lock}
G.~Kolhe, T.~Sheaves, K.~I. Gubbi, S.~Salehi, S.~Rafatirad, P.~D. Manoj, A.~Sasan, and H.~Homayoun, ``Lock{\&}roll: Deep-learning power side-channel attack mitigation using emerging reconfigurable devices and logic locking.''\hskip 1em plus 0.5em minus 0.4em\relax Institute of Electrical and Electronics Engineers Inc., 7 2022, pp. 85--90.

\bibitem[Xu et~al.(2023)Xu, Erata, and Szefer]{Xu2023Exploration}
C.~Xu, F.~Erata, and J.~Szefer, ``Exploration of power side-channel vulnerabilities in quantum computer controllers,'' in \emph{Proceedings of the 2023 ACM SIGSAC Conference on Computer and Communications Security}, 2023, pp. 579--593.

\bibitem[Erata et~al.(2024)Erata, Xu, Piskac, and Szefer]{Erata2024Quantum}
F.~Erata, C.~Xu, R.~Piskac, and J.~Szefer, ``Quantum circuit reconstruction from power side-channel attacks on quantum computer controllers,'' \emph{arXiv preprint arXiv:2401.15869}, 2024.

\end{thebibliography}



\begin{IEEEbiography}
[{\includegraphics[width=1in,clip, keepaspectratio]{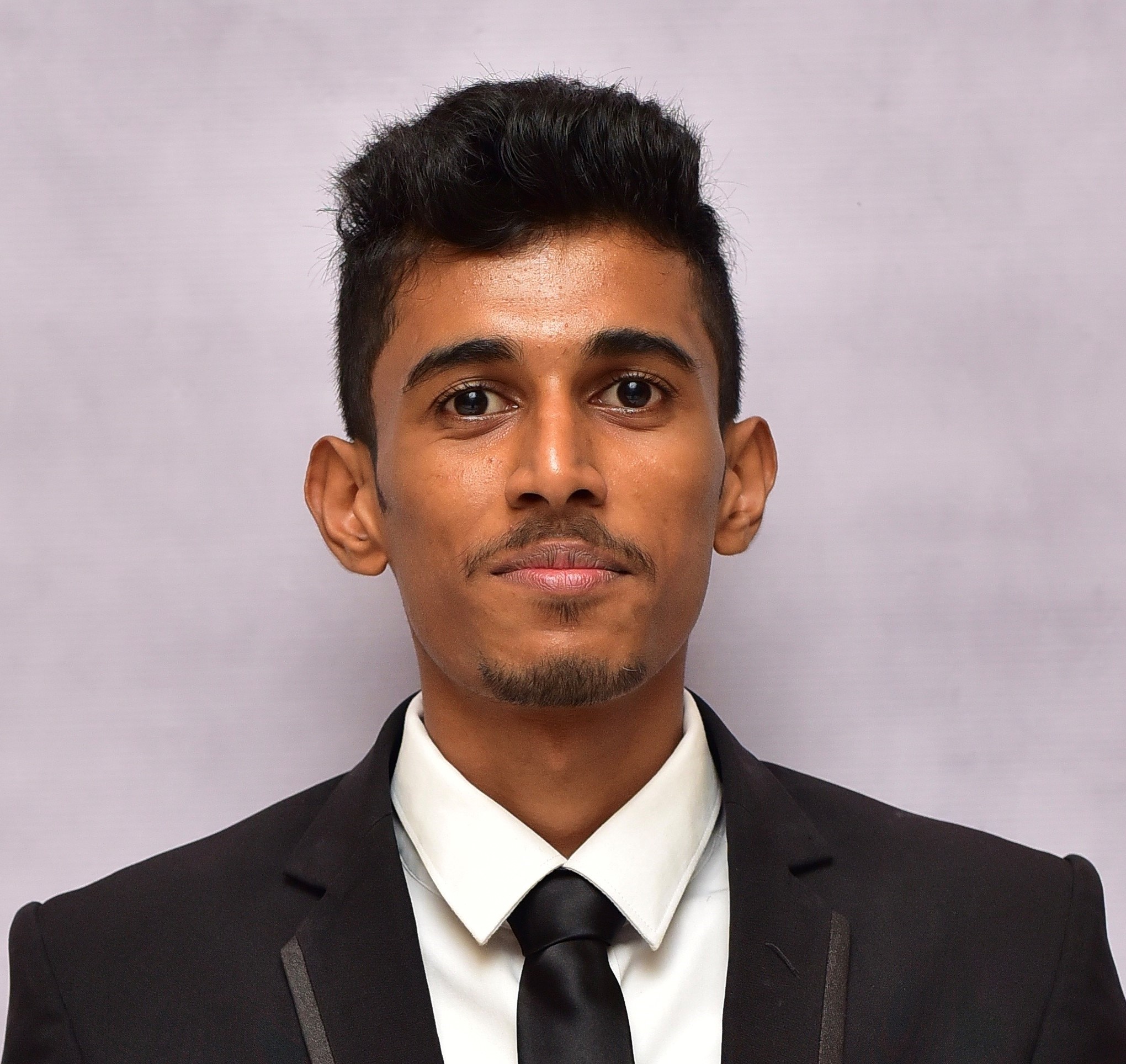}}]{Sahan Sanjaya} is a Ph.D student in the Department of Computer \& Information Science \& Engineering at the University of Florida. In 2022, he completed his B.Sc. in the Department of Electronic and Telecommunication Engineering at the University of Moratuwa, Sri Lanka. His research interests include side-channel attacks, hardware security, pre-silicon validation, and quantum computing.
\end{IEEEbiography}

\begin{IEEEbiography}[{\includegraphics[width=1in,clip, keepaspectratio]{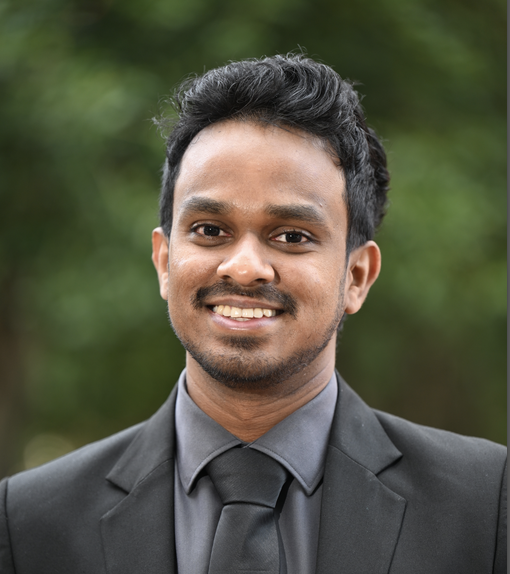}}]{Aruna Jayasena} is an Assistant Professor in the Department of Computer Science and Engineering at the University of Tennessee, Chattanooga. He received his Ph.D. from the University of Florida in 2025. His research focuses on systems security and secure computing architectures. His work lies at the intersection of Computer and Electrical Engineering, with a focus on heterogeneous system design, applied cryptography, trusted execution, and hardware-firmware co-validation.
\end{IEEEbiography}

\begin{IEEEbiography}[{\includegraphics[width=1in,clip,keepaspectratio]{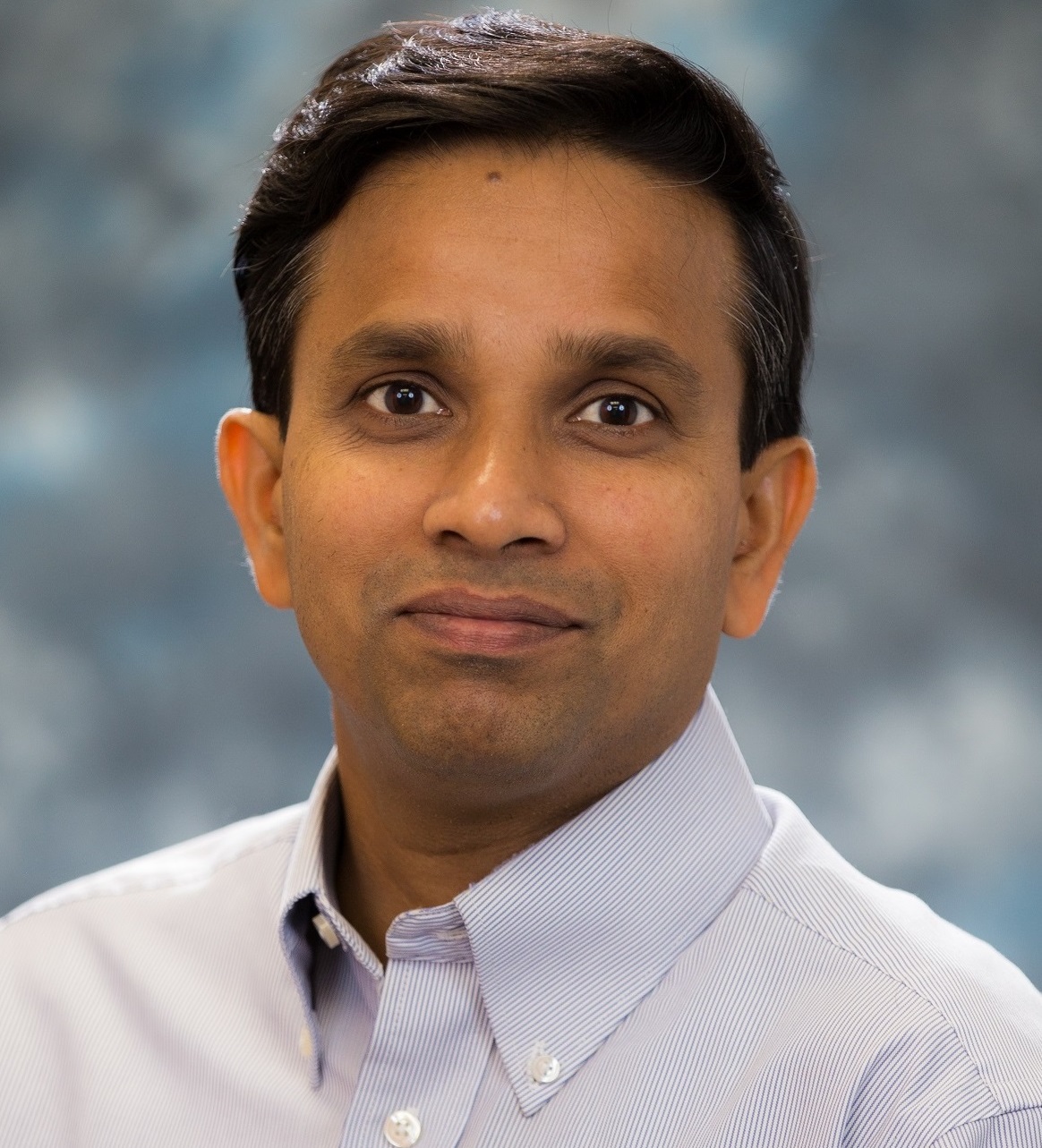}}]{Prabhat Mishra}
is a Professor in the Department of Computer and Information Science and Engineering at the University of Florida. He received his Ph.D. in Computer Science from the University of California at Irvine in 2004. His research interests include hardware security and trust, hardware validation and debug, and embedded systems. He currently serves as an Associate Editor of IEEE Transactions on VLSI Systems and ACM Transactions on Embedded Computing Systems. He is an IEEE Fellow, an AAAS Fellow, and an ACM Distinguished Scientist.
\end{IEEEbiography}


\vfill

\end{document}